\documentclass[english,aps,prl, twocolumn, notitlepage, superscriptaddress,showpacs, amsmath]{revtex4-2}
\pdfoutput=1

\usepackage{bm}% bold math
\usepackage[normalem]{ulem}
\usepackage[T1]{fontenc}
\usepackage[latin9]{inputenc}
\usepackage{color}
\usepackage{units}
\usepackage{amssymb}
\usepackage{graphicx}
\usepackage{esint}
\usepackage{bm}

\usepackage{natbib}

\usepackage{ulem}
\usepackage[colorlinks=true, citecolor=red]{hyperref}

\setcitestyle{journalcolor= blue}

\usepackage{babel}

\usepackage{comment}

\usepackage{caption}
\usepackage{epstopdf}
\usepackage{soul}
\usepackage[dvipsnames]{xcolor}

\usepackage{textcomp}
\begin{document}

\title{ A CMOS compatible platform for high impedance superconducting quantum circuits}

\author{Kazi Rafsanjani Amin}
\affiliation{Univ. Grenoble Alpes, CNRS, Grenoble INP, Institut Ne\'el, 38000 Grenoble, France}
\affiliation{Univ. Grenoble Alpes, CEA, LETI, 38000 Grenoble, France}
\author{Carine Ladner}
\affiliation{Univ. Grenoble Alpes, CEA, LETI, 38000 Grenoble, France}
\author{Guillaume Jourdan}
\affiliation{Univ. Grenoble Alpes, CEA, LETI, 38000 Grenoble, France}
\author{Sebastien Hentz}
\affiliation{Univ. Grenoble Alpes, CEA, LETI, 38000 Grenoble, France}
\author{Nicolas Roch}
\affiliation{Univ. Grenoble Alpes, CNRS, Grenoble INP, Institut Ne\'el, 38000 Grenoble, France}
\author{Julien Renard}
\affiliation{Univ. Grenoble Alpes, CNRS, Grenoble INP, Institut Ne\'el, 38000 Grenoble, France}

\begin{abstract}

  Aluminium based platforms have allowed to reach major milestones for superconducting quantum circuits. For the next generation of devices, materials that are able to maintain low microwave losses while providing new functionalities, such as large kinetic inductance or compatibility with CMOS platform are sought for. Here we report on a combined direct current (DC) and microwave investigation of titanium nitride films of different thicknesses grown using CMOS compatible methods.   For microwave resonators made of TiN film of thickness $\sim$3~nm, we measured large kinetic inductance $L_K~\sim240$~pH/sq, high mode impedance of $\sim4.2~$k$\Omega$  while maintaining microwave quality factor $\sim10^5$ in the single photon limit.   We present an in-depth study of the microwave loss mechanisms in these devices that indicates the importance of quasiparticles and provide insights for further improvement.

\end{abstract}	

\maketitle

%%%%%%%%%%%%%%%%%%%%% main text %%%%%%%%%%%%%%%%%%%%%%%%%%%%%%%

\section{Introduction}

After two decades of intense research, major developments have been made in quantum technologies. Superconducting circuits  is one of the leading platform for quantum computing to date, with major milestone of computational advantage being reported recently\cite{Arute2019, wu2021strong}. In this perspective, while fabrication of long-coherence quantum two-level systems is  a hot pursuit, superconducting microwave devices comprising  qubits and microwave resonators are already being used to study a wide variety of fundamental problems such as ultra-strong light-matter coupling~\cite{FornDiaz2019,FriskKockum2019}, many-body quantum physics~\cite{Leger2019,Kuzmin2019}, quantum simulation of interactions in a lattice~\cite{Roushan2017,Ma2019,Carusotto2020}, or topological protection~\cite{PRXQuantum.2.010339,gyenis2021moving}. For such studies, a prerequesite is the use of a material presenting a minimal amount of microwave losses. In addition, one of the building blocks for the recent developments of superconducting quantum circuits is a lossless, high inductance element, called superinductance~\cite{Manucharyanthesis}.  Its impedance  becomes comparable to the resistance quantum $R_Q = h/4e^2 \approx 6.5$~k$\Omega$ at microwave frequencies. This  is a key feature to realize qubits that are protected from decoherence~\cite{Pop2014,Manucharyan113, Brooks2013} or to enhance the coupling between a qubit and a microwave resonator~\cite{Stockklauser2017,PuertasMartinez2019}.

The superconducting quantum technologies have so far flurished using aluminium based technology that naturally provides a material with very low microwave losses and an excellent oxide for tunnel jonctions. In this platform superinductances have been achieved with large  arrays of Al-AlO$_X$  based Josephson junctions (JJ)s~\cite{Bell2012,Masluk2012,PhysRevB.98.094516}.  However, as the complexity of a real-life problem being addressed by a quantum processor increases, integration of a large number of qubits and resonators becomes inevitable.  Developing  suitable technology achieving  scalablity of  quantum devices  for quantum computing is now  seen as one of the next big challenge. For superinductances, current fabrication technology is incompatible with scalable platforms, and becomes increasingly challenging upon increasing the number of JJs in the array.

This has recently motivated the study of alternative new materials for fabrication of scalable superconducting quantum device. Several materials have been used to fabricate low losses microwave resonators and high-coherence qubits~\cite{Gruenhaupt2019,Place2021,wang2021transmon}.  Fabrication of superinductances using kinetic inductance of disordered superconductor thin films became a promising alternative to Al, and different materials   such as TiN~\cite{6933905,  doi:10.1063/1.3517252, doi:10.1063/1.3480420, doi:10.1063/1.5053461}, NbTiN~\cite{PhysRevApplied.5.044004},  NbN~\cite{PhysRevApplied.11.044014,Yu2021}, AlO$_x$~\cite{Gruenhaupt2019, PhysRevLett.121.117001, PhysRevApplied.11.011003}, InO$_x$~\cite{Astafiev2012,Dupre2017}, doped Si~\cite{bonnet2021strongly}, Tungsten~\cite{Basset2019} have been investigated.    Amongst these materials, TiN and NbTiN have also been studied as  substitutes of Al  to fabricate low-loss microwave quantum circuits~\cite{6933905,  doi:10.1063/1.3517252, doi:10.1063/1.3480420, doi:10.1063/1.5053461}.

Microwave electrodynamics of disordered TiN~\cite{PhysRevLett.109.107003, PhysRevB.88.180505}, as well as study of direct current (DC) resistivity have been reported in TiN and NbTiN. The effect of magnetic disorders on the superconducting behavior of TiN has also been studied~\cite{doi:10.1063/1.4729623}.
Very  high quality resonators, with internal quality factor at high electric field as large as $10^7$~\cite{doi:10.1063/1.3517252, doi:10.1063/1.3480420} have been fabricated using thick films of TiN. In addition, NbTiN-based nanowire superinductances have been  used to realize a  fluxonium qubit~\cite{PhysRevLett.122.010504}. 

Different deposition, film growth techniques~\cite{Ohya_2013, doi:10.1063/1.5053461} and etch methods~\cite{verhaverbeke_parker_1997, doi:10.1063/1.4729623}  towards fabrication of microwave superconducting circuits have been  investigated in details.
 However, with reducing thickness and increasing amount of disorder, the microwave losses and sample variability increase, compromising  future integration in large scale devices. Hence, a detailed understanding of microwave loss mechanisms in these emerging materials is highly needed~\cite{murray2021material}.

%-------------------------------------------------- DC resistivity --------------------------------------------------
\begin {figure*}[!ht]
\begin{center}
    \includegraphics[width=0.9\textwidth, keepaspectratio]{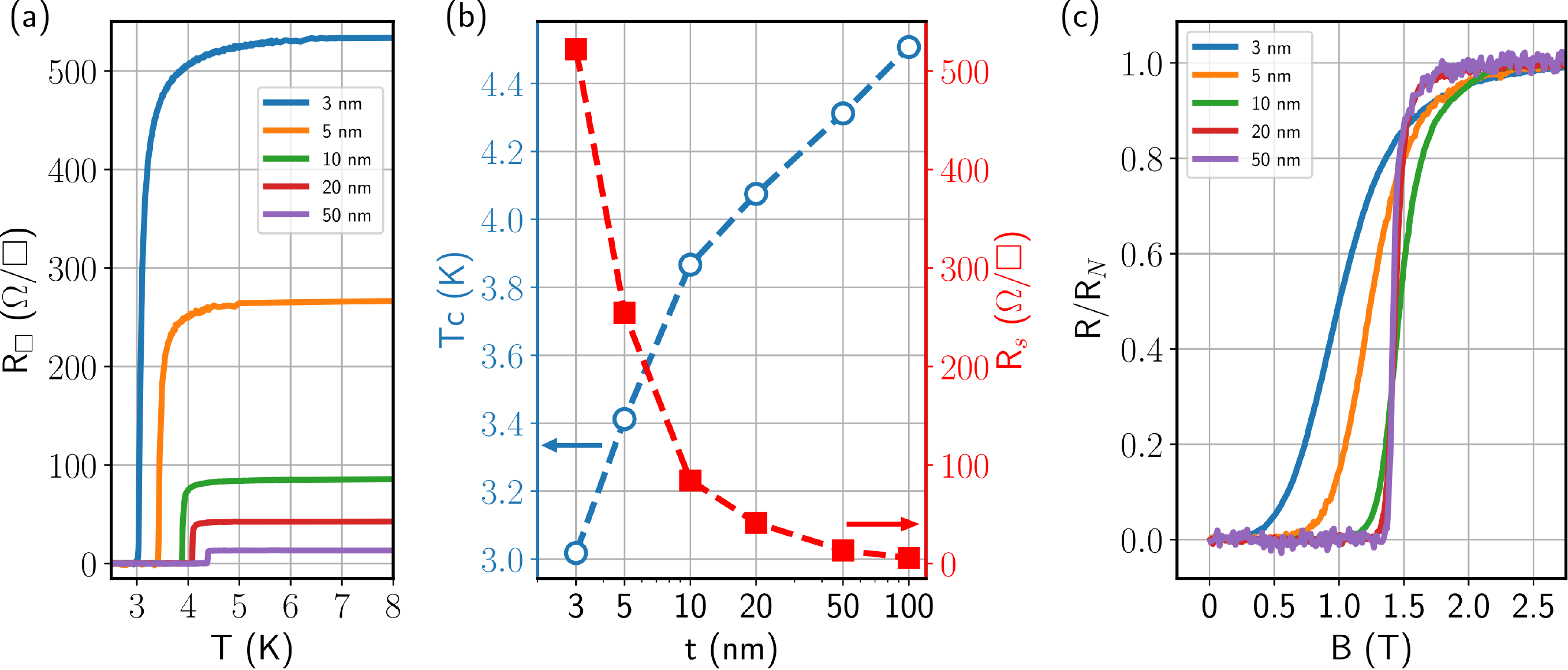} 
    \caption[]{ Superconducting transitions in TiN thin films.
      (a) Sheet resistance $R_\square$  versus temperature $T$ for TiN films of different thicknesses $t$,  zoomed in near the superconducting transitions. Data for 100~nm thin film is shown in Appendix-B, Fig.~\ref{fig:dc_res_full} for clarity.
      (b)  $T_C$ versus $t$, with blue open circles in left axis,  and  $R_S$ versus $t$ in red filled squares in right axis. The dashed lines are guides to the eye. (c) Normalized resistance versus perpendicular magnetic field $B$  measured at $T=2.0$~K for the same TiN film devices as presented in (a). Width of the superconducting to normal state transition increases with decreasing film thickness. 
\label{fig:dc_res} } 
\end{center}
\end {figure*}
%------------------------------------------------------------------------------------------------------------------------

 In this work we demonstrate  TiN superconducting circuits   fabricated with a VLSI, CMOS-compatible process featuring, at the same time  high film quality, ultra-low microwave losses down to the quantum regime and high kinetic inductance. We study in details the microwave loss mechanisms in high kinetic inductance TiN film microwave resonators, and find that the quality factor of very-large kinetic inductance resonators might be limited by quasiparticles.

\section{Organization of this manuscript}

 We begin with presenting  low frequency resistivity measurements from 300~K down to 1.9~K, and magnetic field 0-8~T. Film quality and  superconductivity were characterized from the data. Measurement of 2D microstrip microwave resonators  made out of 3, 5 and 10~nm thin TiN films are presented. Sheet kinetic inductance $L_K$ were computed from both DC and microwave measurements, and their comparison is discussed.   Then we investigate possible loss mechanisms of the thin TiN resonators. For this study, we use a 3D rectangular waveguide, and microstrip resonators are coupled to evanescent microwave fields inside the waveguide. Investigation of stability and aging of the microwave resonators are studied. Finally we discuss high impedance aspect of the microwave resonators. Details of the thin film deposition and resonator device fabrication are presented in details in Appendix-A.

\section{Results}

\subsection*{ Resistivity measurements}

The TiN films of different thicknesses used in this work were deposited, using VLSI CMOS-compatible physical vapour deposition (PVD) method  on high resistive, 725~$\mu$m thick  intrinsic high-resistive silicon (100)  wafer of 200~mm diameters.  Excellent homogeniety and uniformity of film characteristics were achieved as a result of VLSI process.
We first characterize the TiN thin films of different thicknesses by measuring  sheet resistance $R_\square$  as a function of temperature $T$ [Fig.~\ref{fig:dc_res}(a)].
 Starting from room temperature, when T decreases, $R_\square$ decreases first and then tends to   reach a plateau for $T\lesssim$50~K  (see Appendix-B). Except for the $3$~nm film, where the $R_\square$ increases with further cooling, $R_\square$ remains constant with $T$  over a wide range  upon further cooling, until the superconducting transitions are achieved. Such resistivity saturation  below 50~K  is  a signature of disorders present in the films, and is  similar to previous reports in TiN~\cite{PhysRevApplied.12.054001}.  
From these measurements, we   extract the  normal-state sheet resistance $R_s$ [Fig.~\ref{fig:dc_res}(b) right axis, in red filled squares], which is the value of $R_\square$ obtained just before the superconducting transition.  
$R_S$  exhibits an increase  by nearly two orders of magnitude as the film thickness reduces;  from 5.7~$\Omega/\square$ observed for 100~nm film to 522.0~$\Omega/\square$ in 3~nm film. 
The residual-resistance ratios  (RRR) calculated from the sheet resistance at room temperature $R_\square(300~K)$ and $R_s$ as $RRR =  R_\square(300~K)/R_s$ is a measure of disorder present in thin films and characterizes film quality. 
The largest value of RRR=1.3 observed for the 100~nm film is similar to RRR typically reported in  high quality TiN thin films~\cite{Torgovkin2018,PhysRevApplied.12.054001}.
%% ============================== Tc
We then obtain the superconducting transition temperature $T_c$  [Fig.~\ref{fig:dc_res}(b) left axis, blue open circles], defined as the temperature at which  $R_\square$ drops below $1.0~\%$  of $R_s$. 
The $T_c$ decreases from 4.5~K measured  for 100~nm film to 3.3~K for 3 nm film. 
Such decrease in $T_C$ with decreasing film thickness  is   commonly observed  in  disordered thin film superconductors \cite{Sacepe:2008jx}.

%-------------------------------------------------- DC resistivity scaling ----------------------------------------------
\begin {figure}[!ht]
\begin{center}
    \includegraphics[width=0.48\textwidth, keepaspectratio]{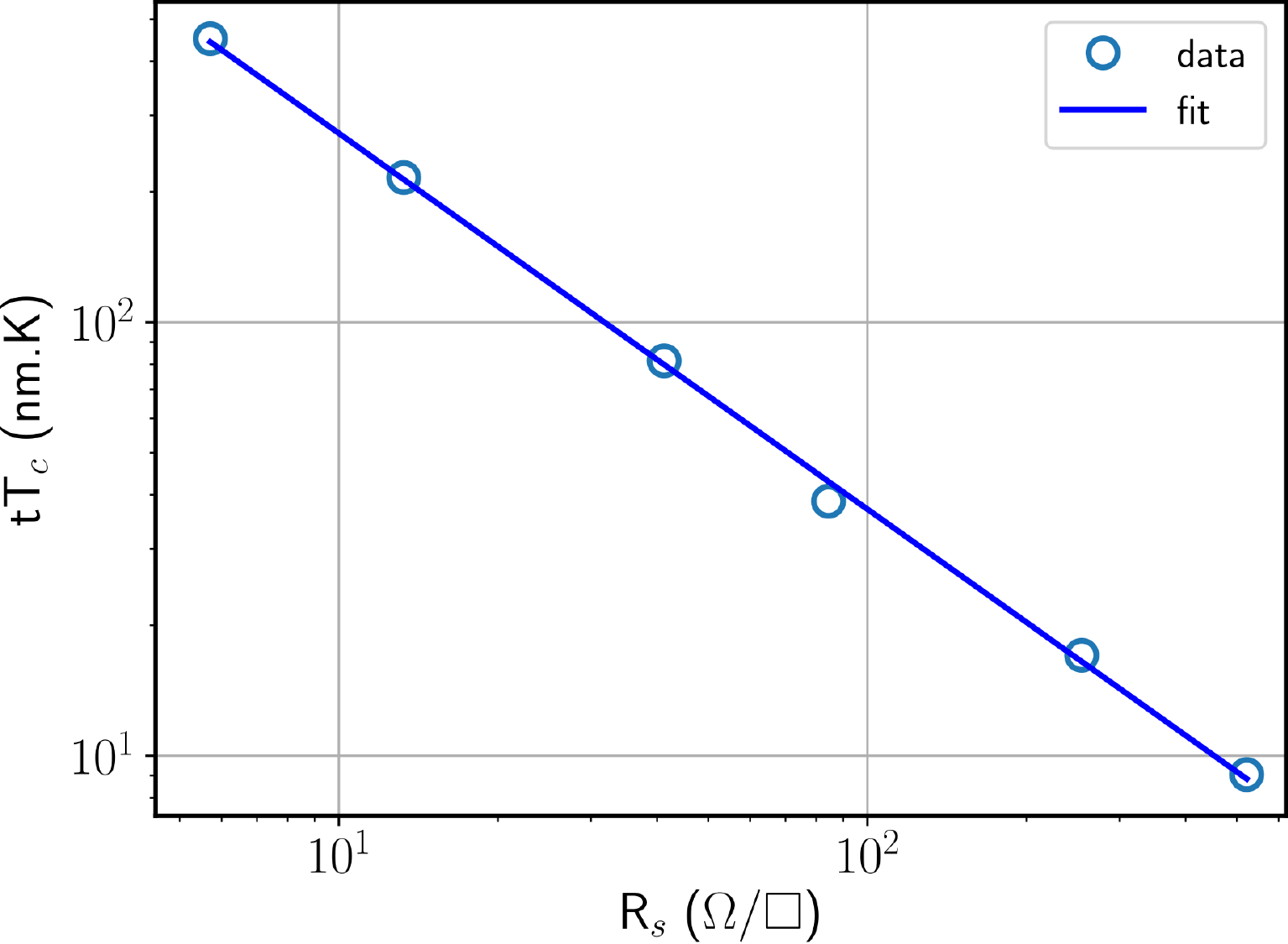}
        \caption[]{Universality and scaling of the superconducting transition in TiN thin films. The blue open circles are plots of $t T_C$ versus $R_S$, in log-log scale. The blue line is fit to the data with $ t T_c = AR_s^{-b}$, yielding  the power-law exponent $b=0.86 \pm 0.015$, very close to $b \approx 0.9\sim 1.1$ widely observed in high-quality superconducting films\cite{PhysRevB.90.214515}.    
\label{fig:res_fit} } 
\end{center}
\end {figure}
%------------------------------------------------------------------------------------------------------------------------

%% ================================================================= Magnetic field ====================
 ~\\ \indent In Fig.~\ref{fig:dc_res}(c), we show plots of resistance versus perpendicular magnetic field $B$, measured at 2.0~K, for  films of different thickness. The resistance  values are normalized by the normal-state resistance obtained for each of these samples,  measured above the critical magnetic field.   
 The increase in resistance with increasing $B$ signifies breakdown of superconductivity by introduction of vortices.  For the thick films, \textit{i.e.} the 20~nm and 50~nm films, the transition is rather sharp, with a critical magnetic field around 1.3~T, which is very large compared to critical magnetic field of standard superconductors, such as aluminum, traditionally used in superconducting quantum circuits. This suggests suitability of usage of TiN in fabricating quantum circuits where large magnetic field is required for device operation.
 In a type-II superconductor, magnetic field penetrates the sample beyond lower critical field $H_{C1}$ and vortices are created. Movement of the vortices either because of electric field or because of thermal energy results into dissipation. In a disordered superconductor, vortices are pinned to defects  and the zero-resistance state is maintained over  a finite temperature range. Stronger pinning strength in thick films (in the 3D limit) allows larger magnetic field to penetrate the devices before dissipation sets in. In thin films near the 2D limit, however,  much smaller magnetic field sets in  dissipative electron transport~\cite{PhysRevB.100.174501}.

 \begin{center}
   \begin{table}[!ht]
     \caption{ Thin film characteristics  \label{tbl:resistivity} }
     \begin{tabular}{| p{2cm}  p{1.5cm}  p{1.5cm} p{1.5cm} p{1.5cm} |  }
       \hline
       \hline
       Thickness & RRR & $T_C$  & $R_S$  & $L_K$ \\ 
       (nm) & & (K)  & ($\Omega/\square$)  & (pH/$\square$) \\ \hline
       \hline
       3   & 1.10  & 3.0  &  522.0   & 239.0    \\ \hline
       5   & 1.17 & 3.4  &  254.4   & 103.0    \\ \hline
       10  & 1.17 & 3.9  &  93.5    & 33.4     \\ \hline
       20  & 1.23 & 4.1  &  41.2    & 14.0     \\ \hline
       50  & 1.25 & 4.3  &  13.3    & 4.2      \\ \hline
       100 & 1.33 & 4.5  &  5.7     & 1.7      \\ \hline
       \hline
     \end{tabular}
   \end{table}
 \end{center}

 ~\\ \indent 
Furthermore, we investigate the inter-dependence of $T_C$, $R_S$ and film thickness $t$.  A power-law $ t T_c = AR_s^{-b}$  has been  commonly observed in  superconductors with different materials and disorder~\cite{Faverzani_2020, PhysRevB.90.214515}.
With $A$ being a material-dependent quantity,  $b \approx 0.9\sim 1.1$ has been  observed to be an universal exponent for high-quality superconducting films~\cite{PhysRevB.90.214515}. A deviation points towards large granularity in the films. Fig.~\ref{fig:res_fit} shows a plot  of $tT_c$ versus $R_s$, extracted from measurements carried out with our TiN thin films (Fig.~\ref{fig:dc_res}).  While $b=0.67$ was recently reported~\cite{doi:10.1063/1.5053461} for ALD-grown TiN films,  we extract, from the fit, $b=0.86 \pm 0.015$. This value is very close to the universality, and suggests a very high  quality of our TiN thin films, even down to 3~nm thickness limit.

In Tbl.~\ref{tbl:resistivity}, we summarize different characteristic parameters of our TiN thin films.

\subsection*{Microwave characterization}

%-------------------------------------------------- resonator --------------------------------------------------
\begin {figure*}[!ht]
\begin{center}
    \includegraphics[width=.850\textwidth, keepaspectratio]{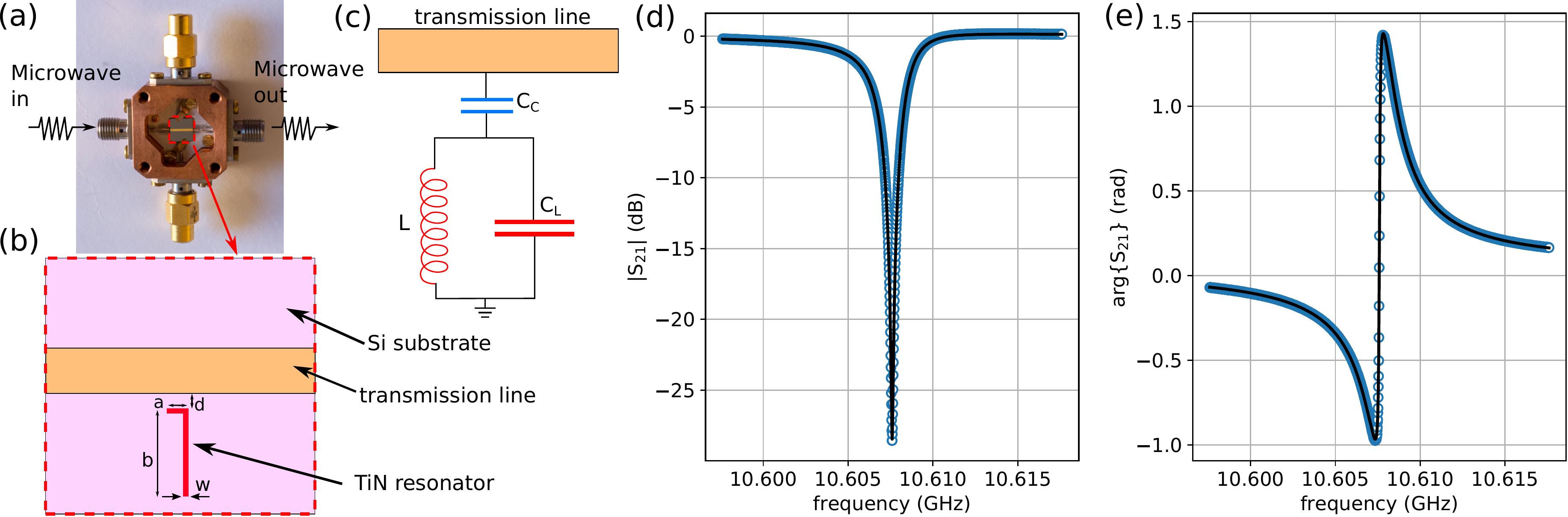}%{TiN_2D}%{184702_VNA_trace_standard_resonator2}%{fig_resonator_1}
    \caption[]{ (a)  Photograph of a custom-built copper sample holder, assembled with low-loss PCB and clamps to attach the sample chip, that allows microwave transmission measurements.  Unused connectors are terminated with a matched impedance.  (b) Schematic top view of a sample chip, showing an impedance matched microwave transmission line and a $\lambda/2$ TiN resonator (in red), capacitively coupled to the transmission line. (c)  Lumped model of the microwave resonator. The inductance is obtained from $L_K$ and the width $w$ and length $l=a+b$ of the resonator. While $C_L$ is the capacitance of the resonator obtained with the metallic ground plane at the bottom of the wafer, the coupling capacitance $C_c$ is determined by dimensions of the coupling arm $a$ and the separation from the transmission line $d$. Suitable coupling parameters are determined by FEM simulations using Sonnet$^{\circledR}$. (d-e) Transmission magnitude (d) and phase (e) of $S_{21}$ showing a microwave resonance   of 5~nm thick TiN resonator measured at 25~mK in  2D microstrip configuration. For this resonator, width  $w=500$~nm, $a=125~\mu$m,  $b=375~\mu$m, and  the separation $d=200~\mu$m.  In open circles, we plot $S_{21}$ magnitude (d) and phase (e) data points, while the thick lines are the fit to the data.  From the fits, we extract internal quality factor $Q_i \approx 1.16\times10^5$, and coupling quality factor $Q_c\approx5000$ measured with $n\sim5\times10^5$ circulating photons.
\label{fig:reson} } 
\end{center}
\end {figure*}
%------------------------------------------------------------------------------------------------------------------------

We begin microwave characterization with the study of 2D microstrip  $\lambda/2$ resonators of different film thickness and different aspect ratio. We have studied a total 3 resonators made with  3~nm TiN film, 6 resonators made with  5~nm TiN film, and 5 resonators made with  10~nm TiN films. All the microwave measurements described below have been  performed at the base temperature of the dilution refrigerator of 25~mK, unless otherwise specified. We obtain the coupling quality factor $Q_C$, internal quality factor $Q_i$ and mode frequency $f_R$ by fitting the data with standard fit procedures~\cite{doi:10.1063/1.4907935}. The average photon number $n$ circulating in the resonator is  estimated using $n = P_{in} Q_l^2 / (\hbar \pi^2)f_R^2 Q_C$, where $P_{in}$ is the attenuated input power, and the loaded quality factor $Q_l^{-1} = Q_i^{-1} + Q_C^{-1}$. In Fig.~\ref{fig:reson}(d-e), we show representative plots of $S_{21}$ data, and fit to the data,  obtained for a 5~nm thin film TiN resonator.

\subsection*{Kinetic inductance extraction}

%-------------------------------------------------- dispersion --------------------------------------------------
\begin {figure}[!ht]
\begin{center}
    \includegraphics[width=.45\textwidth, keepaspectratio]{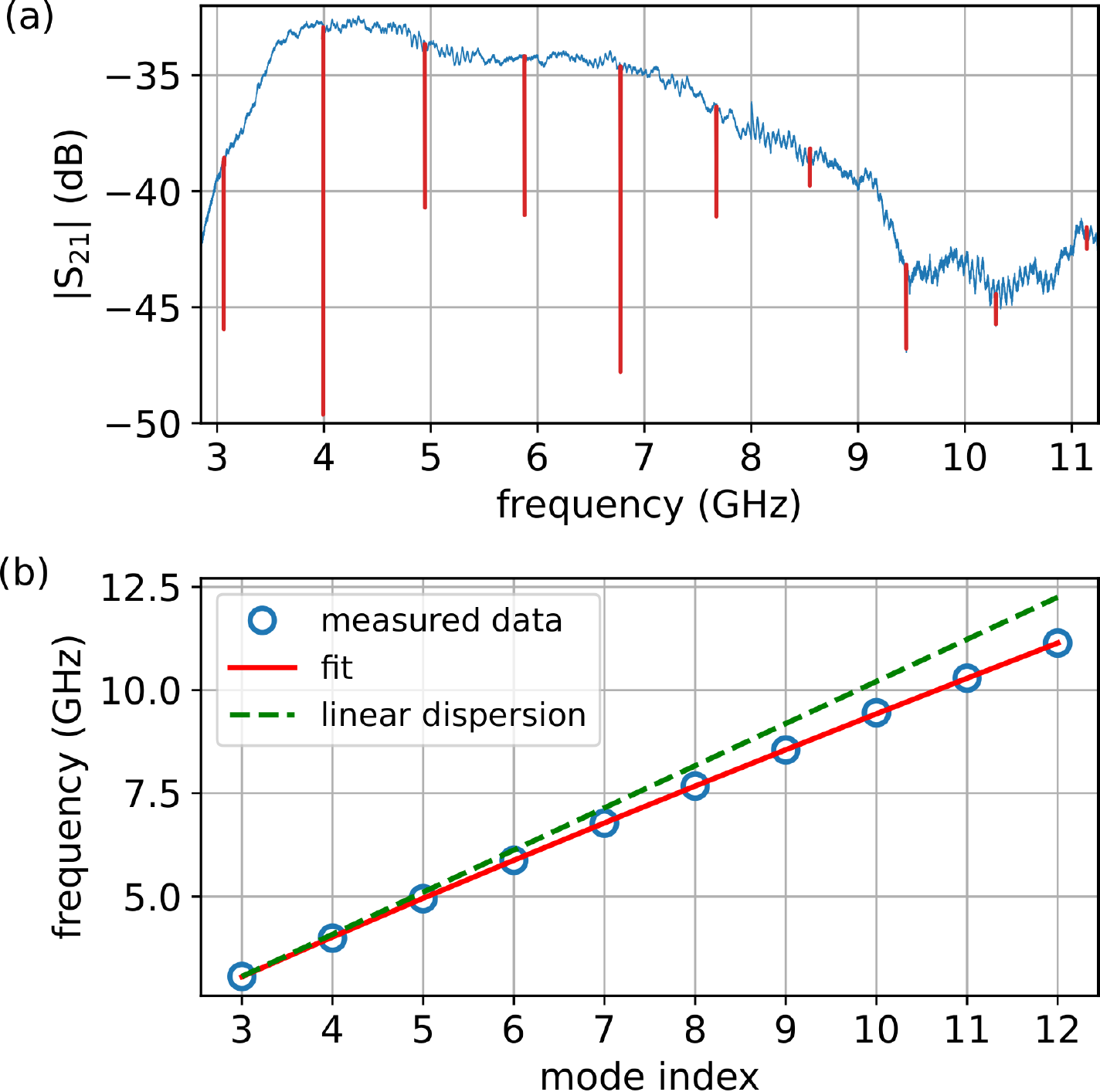}%{mode_dispersion}
    \caption[]{Mode dispersion in high kinetic impedance TiN microwave resonators. (a) Magnitude of microwave transmission parameter $S_{21}$(raw data) for a 3 nm thin film TiN resonator, fabricated in a microstrip configuration. 10 resonance modes were observed in the measurement bandwidth, and are plotted in red, for clarity.
      (b) Mode dispersion relation for the resonator, extracted from measurement shown in (a). The blue open circles are plots of mode frequency for different mode index. The green dashed line is the expected linear dispersion relation, obtained by fitting the data at low mode number. The red thick line is the fit to the data with the dispersion relation using long range Coulomb interaction model, that give $L_K$ as the only fit parameter.  
\label{fig:disp} } 
\end{center}
\end {figure}
%------------------------------------------------------------------------------------------------------------------------

~\\ \indent 
We now extract sheet kinetic inductance $L_K$ of the films from both  microwave measurements and low-frequency resistivity measurements.
We measure microwave resonators with length 6.5~mm and width of 2~$\mu$m, which, depending upon total inductance  determined by  $L_K$, exhibit  multiple modes in our measurement bandwidth. In Fig.~\ref{fig:disp}(a), we show plot of  $|S_{21}|$ obtained for such a resonator made with 3~nm TiN film, showing 10 modes.  In Fig.~\ref{fig:disp}(b), we  plot the mode frequencies, obtained from fit to the $S_{21}$ data, versus mode index, for all  10 modes. We observe that the data points deviate progressively from the expected linear dispersion at higher frequencies as the effective plasma frequency is approached asymptotically.  Furthermore, we fit the data using a modified version of long-range Coulomb interaction model~\cite{PhysRevB.98.094516}, developed taking into account the charge screening by the presence of a ground plane at the back of the device chip~\cite{tibo}, where total kinetic inductance  is  the only  free parameter which we then extract from the measured dispersion..

  Furthermore, we also perform a finite-element electromagnetic simulation using  Sonnet$^{\circledR}$, and  obtain resonant frequencies for our specific device geometry, and different values of sheet kinetic inductances $L_K$, which is an input parameter in the simulations. The experimentally obtained mode frequency is then used to fit into this simulated mode frequency versus $L_K$  relationship, and  estimate the kinetic inductance for out TiN films of a given thickness.    

\begin {figure}[!ht]
\begin{center}
    \includegraphics[width=0.5\textwidth, keepaspectratio]{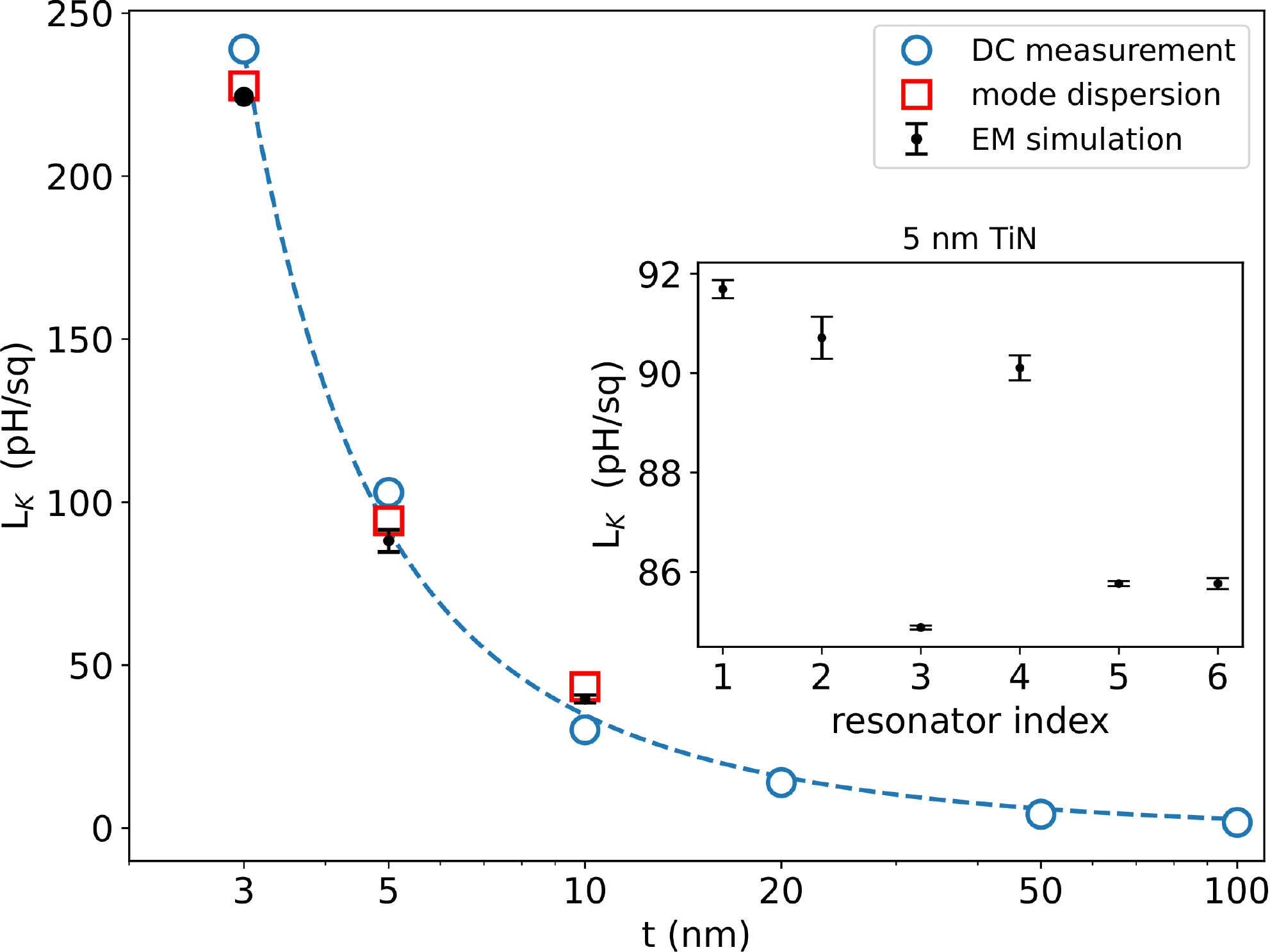}
        \caption[]{Kinetic inductance of TiN thin films. Plots of sheet kinetic inductance $L_K$ versus TiN film thickness $t$, extracted using three different methods. The blue dashed line is a guide to the eye. $L_K$  computed from superconducting transition temperature $T_C$ and normal state sheet resistance $R_s$ obtained just before superconducting transition,  from resistance versus temperature measurements of TiN films of different thickness as $L_K = \hbar R_s/1.76 \pi k_B T_C$ is plotted with blue open circles. $L_K$ obtained from resonance frequency versus mode number dispersion relation for $\lambda/2$ microwave resonators in a microstrip configuration fabricated from 3, 5, 10~nm TiN films are shown with red open box.  In black filled circles, we show plots of $L_K$ obtained from electromagnetic simulation using Sonnet$^{\circledR}$, where $L_K$ were tuned as a parameter of simulation to match mode frequency for a resonator of a given dimensions with experimentally obtained mode frequency. The errorbars represent spread in absolute value of the estimation from multiple resonators of different dimensions.  Inset: Plot of $L_K$ obtained using Sonnet simulations from 6 different microwave resonators, made with 5~nm TiN film, show a spread of less than $8\%$ in $L_K$. The errorbars are estimated using resonance linewidth as an error estimate for resonance frequency of the observed resonance. 
\label{fig:lk_all} } 
\end{center}
\end {figure}

~\\ \indent 
  Sheet kinetic inductance, being directly proportional to the superfluid density of the superconducting film, is also obtained from low-frequency measurements of the superconducting transitions. We extract $L_K$ of the films from the resistivity measurements [shown in Fig.~\ref{fig:dc_res}] using the relationship $L_K = \hbar R_S / \pi \Delta_0$~\cite{doi:10.1063/1.5053461,PhysRevApplied.5.044004}, where the zero temperature superconducting gap is obtained using the BCS relationship $\Delta_0 = 1.76 k_BT_C$. In Fig.~\ref{fig:lk_all} we combine $L_K$ obtained from all the above mentioned measurements, for different film thicknesses.
We observe excellent agreement in $L_K$  between the three different methods, which gives high confidence in the extracted values. We emphasize here that all the DC and microwave measurements presented until now and to be discussed later for  a given film thickness, are carried out from different parts of 200~mm diameter wafers. The agreement of $L_K$ over all such measurements indicates excellent  homogeneity of the stoichiometry and thickness, which is an asset of VLSI fabrication.
For the 100~nm film, we obtain very small value of $L_K=1.0$~pH/$\square$, at the limit where $L_K$ contributes negligibly to the total inductance. The $L_K$ increases with decreasing thickness, reaching a significant value of 239~pH/$\square$ (Tbl.~\ref{tbl:resistivity}) for the 3~nm TiN film, two orders of magnitude larger than the value for 100~nm film. This increase in $L_K$ with decrease in film thickness is a consequence of reduction of superfluid density in the superconducting state. Our observation of such an increase in the $L_K$ with decreasing thickness is in qualitative agreement with  TiN films deposited using other methods~\cite{doi:10.1063/1.5053461} and other disordered thin-film superconductors.

%-------------------------------------------------- Lk --------------------------------------------------
%------------------------------------------------------------------------------------------------------------------------

\subsection*{Investigation of microwave loss mechanisms}

 We now probe into the loss mechanisms, that determine the intrinsic performances of the TiN microwave resonators.   A very common source of loss in microwave resonators and other superconducting quantum devices are two-level systems (TLS)~\cite{M_ller_2019} which are present in the vicinity of the devices  and coupled to them. A series of detailed investigations showed that impurities residing in the metal-substrate interface is the major contributor to the TLS loss.   This can be attributed  to the order of magnitude larger participation of stored electric field in metal-substrate interfaces, as compared to metal-air or substrate-air interfaces~\cite{M_ller_2019}.
On the other hand, as the kinetic inductance fraction $\alpha= L_{kinetic} / L_{total}$ approaches unity, susceptibility to quasiparticles (QP), and in turn, induced microwave losses due to QP increases~\cite{Day2003,PhysRevLett.121.117001}. Non-equilibrium  quasiparticles in high kinetic inductance granular aluminum resonators have been found to be the dominant source of microwave loss~\cite{PhysRevLett.121.117001,doi:10.1063/1.5124967,Cardani2021}. We study the evolution of internal quality factors of multiple TiN resonators,  both in planar microstrip geometry and in rectangular 3D waveguide, with variation of microwave power and temperature to unravel the underlying loss mechanisms.  

%% ========================================= 3D resonator =========================================
\begin {figure*}[!t]
\begin{center}
    \includegraphics[width=.95\textwidth, keepaspectratio]{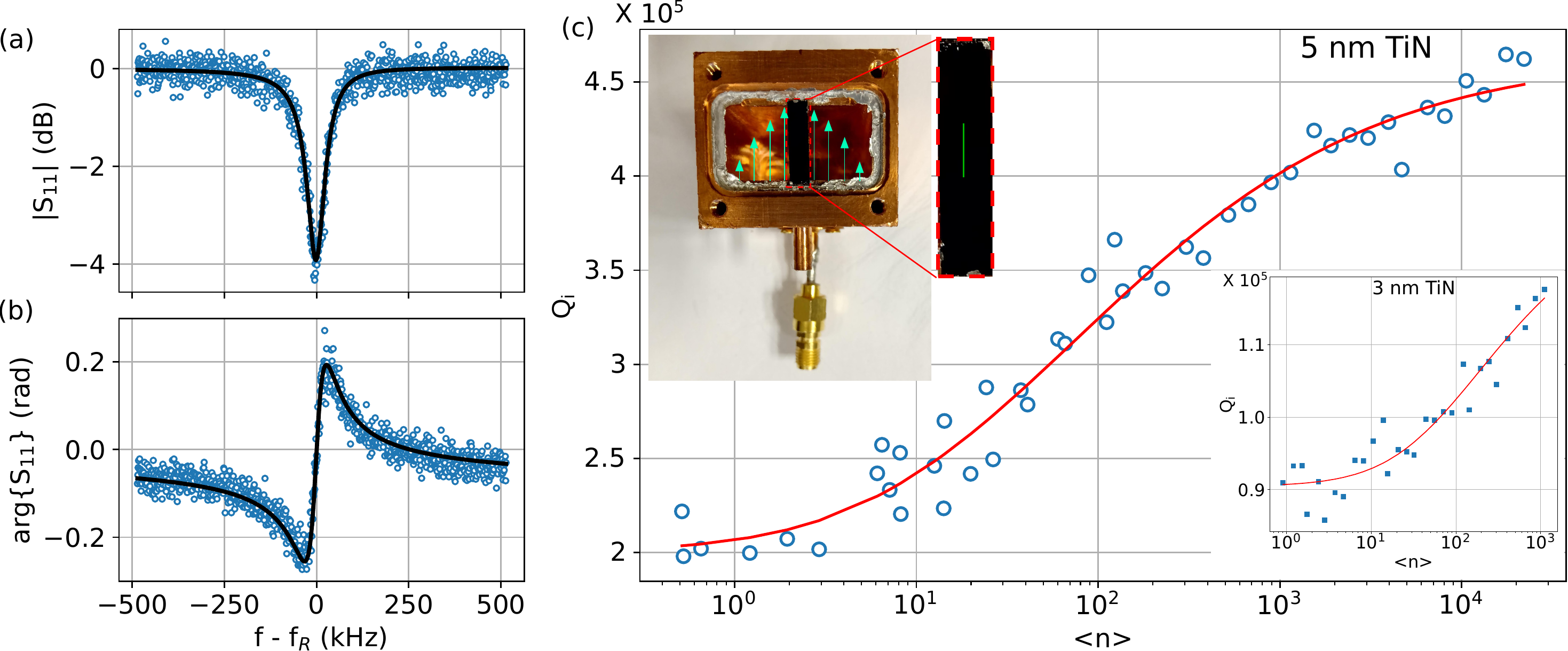}
    \caption[]{Photon number dependent microwave losses in TiN thin films resonators. Plots of (a) magnitude and (b) phase of $S_{11}$ parameter of a 3 nm thick film TiN resonator, with $f_R=5.563$~GHz measured using a 3D waveguide, shown in left-top inset of (c), with single circulating photon in the resonator. The blue open circles are data points while the black lines are the fit to the data points~\cite{doi:10.1063/1.4907935}. We extract $Q_i=0.9\times10^5$, $Q_c=4.1\times10^5$ and $<n>=1.3$ from the fits. 
        (c) Plots of $Q_i$ versus average number of photons $n$ measured for a 5~nm thick film TiN resonator. The blue open circles are the data points while the  red     lines are fit to the data points using  QP (Eqn.~\ref{eqn:qp}) induced loss model. Left-top inset: Optical image of the copper 3D waveguide used in the experiments. A $\lambda/2$ resonator is represented (not to scale) in the silicon chip, shown magnified in the right. The electric field magnitude of the fundamental TE$_{10}$ mode is represented by arrows. 
        Right-bottom inset: Plots of $Q_i$ versus $n$ measured for a  3~nm thick film TiN resonator (blue open circles), and fit to the data points, similar to that shown in the main panel.
\label{fig:3nm_3dres} } 
\end{center}
\end {figure*}
%% ===============================================================================================

We study multiple microstrip version of $\lambda/2$ resonators, embedded into  a 3D rectangular copper waveguide  similar to that in ref.~\cite{PhysRevLett.121.117001}. An  impedance-matched wave port is used to excite the resonator. With the help  of finite element simulations using the \textit{Ansys$^{\copyright}$ HFSS} software, we chose the  dimensions of the resonators,  in order to achieve a coupling quality factor $Q_c$ in the range $10^5 - 10^6$  and  the frequency in range of $4.0~-~6.0$~GHz, below the pass-band of the waveguide. We  couple the resonators with the evanescent wave of the waveguide in order to achieve $Q_c$ within an order of magnitude of $Q_i$ to have good confidence in the fit procedure~\cite{doi:10.1063/1.4907935}. 
In Fig.~\ref{fig:3nm_3dres}, we show representative plots of magnitude (Fig.~\ref{fig:3nm_3dres}(a)) and phase of (Fig.~\ref{fig:3nm_3dres}(b)) $S_{11}$, for a 3~nm TiN resonator.
The blue open circles are the data points, while the thick black  lines are the fit to the data. From the fit, we obtain $Q_i \sim 0.9 \times 10^5$, measured in the single photon limit. Internal quality factor of $\sim10^5$ in the single photon limit for such a thin film ($\sim$~3~nm) high kinetic inductance resonator is comparable to the best values  reported thus far~\cite{PhysRevLett.121.117001,PhysRevX.10.031032,doi:10.1063/1.5053461,doi:10.1063/1.5124967,Cardani2021}.

%% ========================================= Participation ratio =========================================
\begin {figure}[!t]
\begin{center}
    \includegraphics[width=0.5\textwidth, keepaspectratio]{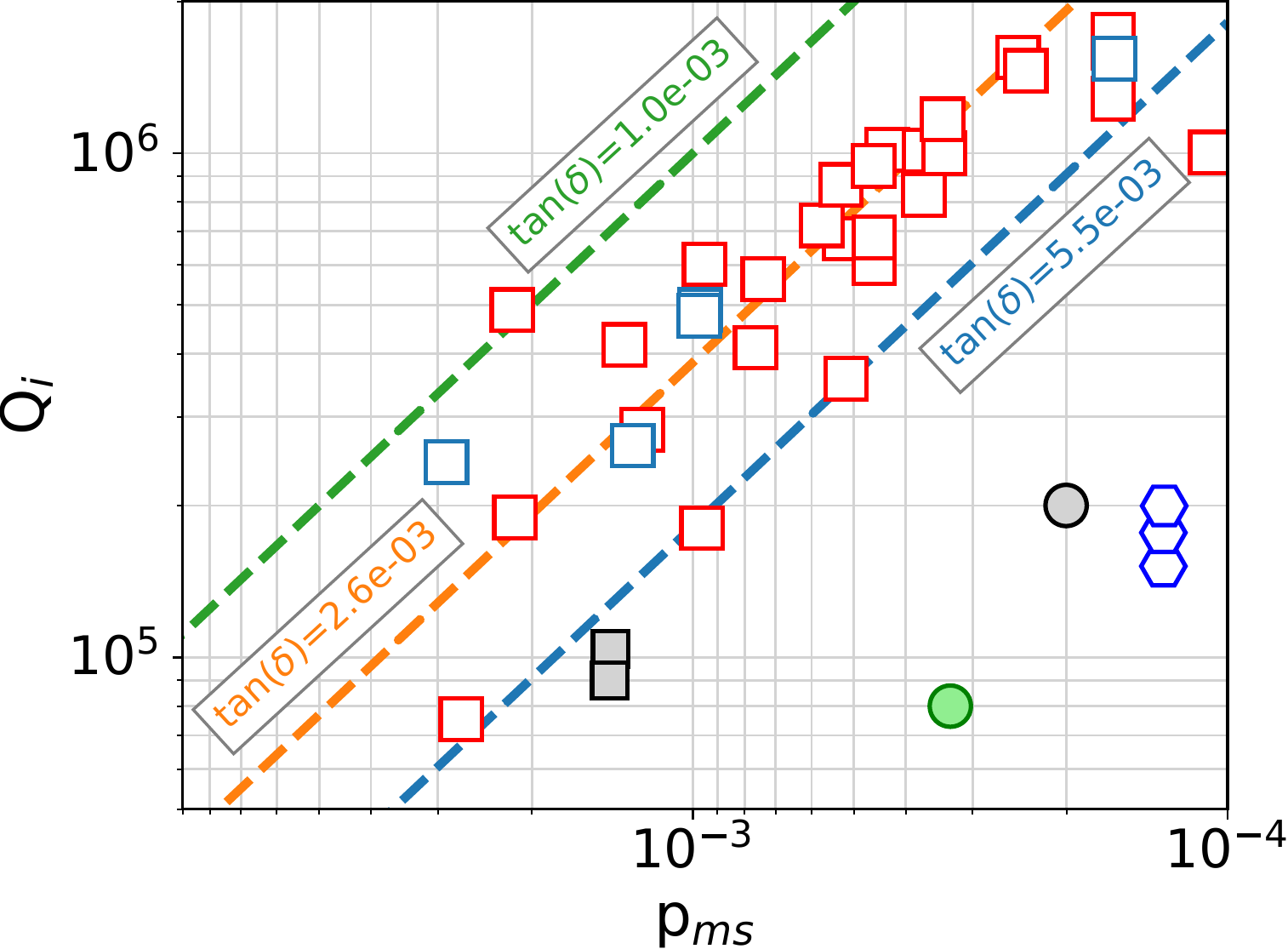}
    \caption[]{  Internal quality factor versus metal-substrate participation ratio.  The open squares represent characteristics of superconducting microwave resonators and  transmon qubits from literature [red squares from Ref.~\cite{Wang2015} and blue squares from Ref.~\cite{PhysRevLett.121.117001}].  Open pentagons represent high kinetic inductance GrAl microwave resonators~\cite{PhysRevLett.121.117001}, where $Q_i$ were  limited by quasiparticle loss.  Data points from microwave resonators  used in this work are represented by black (5~nm TiN) and green (3~nm TiN) filled squares and circles. The black filled squares correspond to $Q_i$ of resonator measured in 2D microstrip configuration, while black filled circle recpresent $Q_i$ of resonator measured using rectangular 3D waveguide.  The dashed lines represent $Q_i=[p_{ms} \tan{\delta}]^{-1}$ for three different values of dielectric loss tangent $\tan{\delta}$, in range with typically measured values for high quality Si or sapphire substrates~\cite{doi:10.1063/1.3637047,Wang2015,doi:10.1063/1.5006888,PhysRevLett.121.117001}. 
\label{fig:participation} } 
\end{center}
\end {figure}
%% =========================================================================================================

\begin{center}
  \begin{table}[!hb]
    \caption{Fit parameters for the data dispalyed in Fig~\ref{fig:3nm_3dres}(c)\label{tbl:qpfit} }
  \begin{tabular}{| p{2.5cm}  p{2.5cm}  p{2.5cm} |  }
    \hline
      \hline
      parameter & 3 nm TiN  & 5 nm TiN  \\ \hline
      \hline
    $Q_0$ & 90442 $\pm$997  & 200072 $\pm$ 6361 \\ \hline
    $\beta$ & 3.29$\times10^{-6}$  &  2.85$\times10^{-6}$   \\ \hline
    $\gamma$ & 0.011  &  0.063    \\ \hline
    \hline
  \end{tabular}
  \end{table}
\end{center}

We now present additional experiments to understand different loss mechanisms in the TiN resonators. First, we measure resonance traces for a wide range of input powers to extract  $Q_i$ as a function of photon number. In Fig.~\ref{fig:3nm_3dres}(c) we show, in blue open circles, plot of $Q_i$ versus average number of circulating photons $n$ in the resonator, measured for a 5~nm thick  TiN resonator. We observe  $Q_i \sim 2.0 \times 10^5$ in the single photon limit.  
We could measure the mode with up to $\sim~2\times10^4$ photons before the resonance mode bifurcates because of the intrinsic non-linearity of TiN. 
$Q_i$ increases from single photon limit by $\sim$~2 times to $\sim~4.5~\times 10^5$ and tends to saturate before bifurcation, over four orders of magnitude increase in the photon number. 
In the inset of Fig.~\ref{fig:3nm_3dres}(c), we show plots of $Q_i$ versus $n$, measured for  a  resonator made with 3~nm thick TiN.
We observe here a distinct behaviour for the $Q_i$ versus $n$  as compared to the 5~nm thick resonator.
Firstly, the bifurcation onsets  at $\sim 10^3$ photons in the resonator, almost an order of magnitude smaller than that in the 5~nm resonator. This can be attributed to larger Kerr nonlinearity arising because of larger $L_K$.
Moreover, for 3~nm TiN, we observe  $Q_i \sim 0.9 \times 10^5$ in the single photon limit, which indicates larger loss in the 3~nm resonators as compared to 5~nm TiN resonators. Over the three orders of magnitude increase in $n$,  $Q_i$ shows only a weak dependence on $n$ for the 3~nm thick resonator.  Such larger losses in 3~nm resonator as compared to 5~nm resonator have been consistently observed over multiple planar 2D microstrip resonators with different aspect ratios and frequencies, while the  substrate and the entire fabrication procedure are  same.

%============= participation disussion
 To gain more insight on the origin of loss mechanisms, in Fig.~\ref{fig:participation}, we combine $Q_i$ versus metal-substrate interface participation ratio $p_{ms}$ obtained  for our TiN resonators (filled circles and squares)  and also from literature survey of resonators (open squares and pentagons), and transmon qubits dominated by capacitive losses.   We obtain $p_{ms}$ for our resonators by computing the fraction of energy stored in a 3~nm thick region under the resonator, following the standard  methodology described in Ref.~\cite{doi:10.1063/1.3637047,Wang2015,doi:10.1063/1.5006888}. $Q_i$ in the single photon limit, determined  solely by dielectric loss in substrate scales as $Q_i=[p_{ms} \tan{\delta}]^{-1}$.   We can identify two distinct types of data points.  A cluster of data points from literature (open squares), for which  the dominant loss is from two-level systems at metal substrate interface, obey the empirical relation $Q_i=[p_{ms} \tan{\delta}]^{-1}$. We observe a contrasting behaviour for TiN resonators measured in 3D waveguide. In a 3D rectangular waveguide, the ground plane is provided by the metal body of the waveguide, which in turn significantly reduces metal-substrate participation ratio by diluting the electric field in the substrate. However, we observe that the measured $Q_i$ is limited to $\sim10^5$, one order of magnitude less that that predicted by $Q_i=[p_{ms} \tan{\delta}]^{-1}$ for standard high-quality Si substrates. 
Similar suppression of  $Q_i$ by order of magnitude compared to what is predicted by  $[p_{ms} \tan{\delta}]^{-1}$ were observed in high kinetic inductance GrAl resonators measured in similar 3D waveguides (open pentagons). Here the losses were found to be dominated by quasiparticle ~\cite{PhysRevLett.121.117001}. $Q_i$ in 2D microstrip TiN resonator (black filled square), with $p_{ms}$ orders of magnitude larger than that obtained in 3D waveguides, indeed match well with predicted value with $\tan\delta~\sim5.5\times10^{-3}$. We emphasis here that all of these resonators for  a given film thickness are fabricated from the same 200~mm wafer of TiN films on Si, following the same fabrication recipe.  For large $p_{ms}$ ($\geq 10^{-3}$, [black filled squares]) the dominant loss is naturally from TLS at metal-surface interface, as  any other  loss mechanisms are masked off. On the other hand, as the $p_{ms}$ is reduced for the TiN resonators measured in 3D waveguide [filled circles], larger suppression of $Q_i$ than  predicted by $Q_i=[p_{ms} \tan{\delta}]^{-1}$  clearly   suggests that loss mechanisms other than TLS become progressively dominant for high kinetic inductance TiN resonators.

%% ======================== QP loss fitting

We now analyze our data  with the model of photon number-dependent loss due to localized QP~\cite{PhysRevLett.121.117001}:
\begin{equation}
  \frac{1}{Q_i} = \frac{1}{Q_0} + \beta \bigg( \frac{1}{1+\frac{\gamma n}{1+ \frac{1}{2}\sqrt{1+4\gamma n}-1 } } -1 \bigg),
  \label{eqn:qp}
\end{equation}
with $Q_0$ being quality factor characterizing other loss mechanisms independent of $n$, and $\beta$ quantifies QP-photon coupling strength. The red thick lines in Fig.~\ref{fig:3nm_3dres}(c) are the fit to the data points with Eqn.~\ref{eqn:qp}.  
We obtain a good  agreement between the measured data  and QP induced loss model, which suggests that QP induced losses are probably the dominant loss mechanism in our resonators, similar to previous observations in  high kinetic inductance resonators~\cite{PhysRevLett.121.117001, Cardani2021, PhysRevApplied.11.054087}. 
QP can be generated in these devices because of photons, phonons or other high-energy particles, and often manifest as sudden frequency jumps of the resonators by `bursts' of 	quasiparticles. However, identifying  the exact origin of these QP is an active field of research, and beyond the scope of this work.
We present the fit results in Tbl.~\ref{tbl:qpfit}.

%-------------------------------------------------- Qi versus T -------------------------------------------------------
\begin {figure}[!t]
\begin{center}
    \includegraphics[width=0.45\textwidth, keepaspectratio]{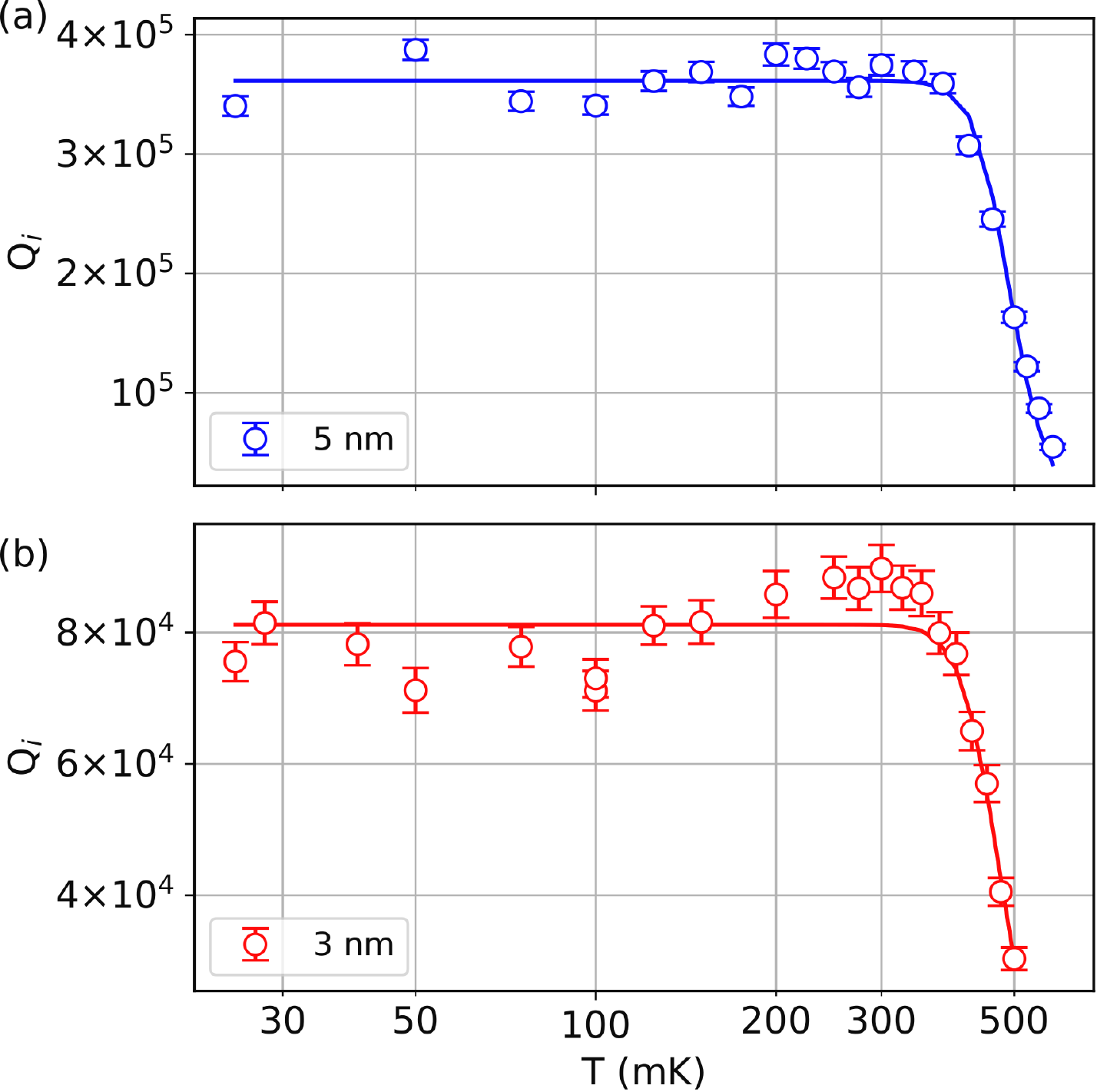}%{TiN3nm_3D_agpaint}
    \caption[]{Temperature dependence of losses in TiN microwave resonators. Plots of $Q_i$ versus $T$ measured for (a) 5~nm (b) and 3~nm  thick TiN resonator. The thick lines are fit to the  data using Eqn.~\ref{eqn:qi_temp}, which determine the loss of microwave resonators governed by increase in thermal quasiparticle density as $T$ increases.
\label{fig:Qi_temp} } 
\end{center}
\end {figure}
%------------------------------------------------------------------------------------------------------------------------

We now study dependence of $Q_i$ on  temperature $T$, by controlled heating of the mixing chamber, in a  subsequent cooldown. 
 In Fig.~\ref{fig:Qi_temp}, we show plots of $Q_i$ versus $T$, measured for 5 nm TiN (top panel) and 3 nm TiN (bottom panel). $Q_i$ is essentially temperature independent until about 400~mK, and then decreases monotonously, but sharply with  further increasing $T$.
As temperature increases, the thermal equilibrium quasiparticle density $n_{qp}$  increases~\cite{doi:10.1063/1.3638063,Gao2008}, as given by $n_{qp}(T) = D(E_F) \sqrt{2\pi k_BT\Delta} \exp(\Delta_0/k_BT)$, 
where $ D(E_F)$ is the density of states at the Fermi energy, and $\Delta_0$ is the  zero temperature superconducting gap. The increasing $n_{qp}$ results in additional loss in microwave resonators, governed by:
\begin{equation}
  \frac{1}{Qi} = 2\alpha  \sqrt{\frac{k_B T}{\pi h f_R} } \exp\bigg(-\frac{\Delta_0}{k_BT}\bigg) + \frac{1}{Q_a}
  \label{eqn:qi_temp}
\end{equation}
with $\alpha= L_{kinetic} / L_{total}$  being the kinetic inductance fraction, and $\frac{1}{Q_a}$ is loss because of other loss mechanisms, which, in our case, is dominated by loss due to non equilibrium localized QP, as discussed previously.  The thick lines in Fig.~\ref{fig:Qi_temp} show the fit to the data using Eqn.~\ref{eqn:qi_temp}, with only one free parameter $\Delta_0$, and $\alpha=1$. The estimated value     of $\Delta_0$ obtained from  the fits  agrees well, within $10\%$ uncertainty~\cite{PhysRevB.93.100503, doi:10.1063/1.5053461}, with the BCS prediction of $1.76 k_B T_C$.    
We note here that, observed saturation in $Q_i$  for $T\leq400$~mK is contrasting to TLS predictions, where saturation of TLS because of provided thermal energy results in an increase in $Q_i$. Hence the measured temperature dependance of $Q_i$ is in agreement with the previous discussion suggesting that the microwave loss in the  single photon limit in TiN resonators in 3D waveguide is not predominantly TLS limited. Our results and  interpretations inspire further investigation of the microwave loss mechanism in thin-film high kinetic inductance TiN resonators.

%% =======================================================================================================================================================================
\subsection*{Thin film stability}

%-------------------------------------------------- aging --------------------------------------------------
\begin {figure*}[!ht]
\begin{center}
\includegraphics[width=0.95\textwidth, keepaspectratio]{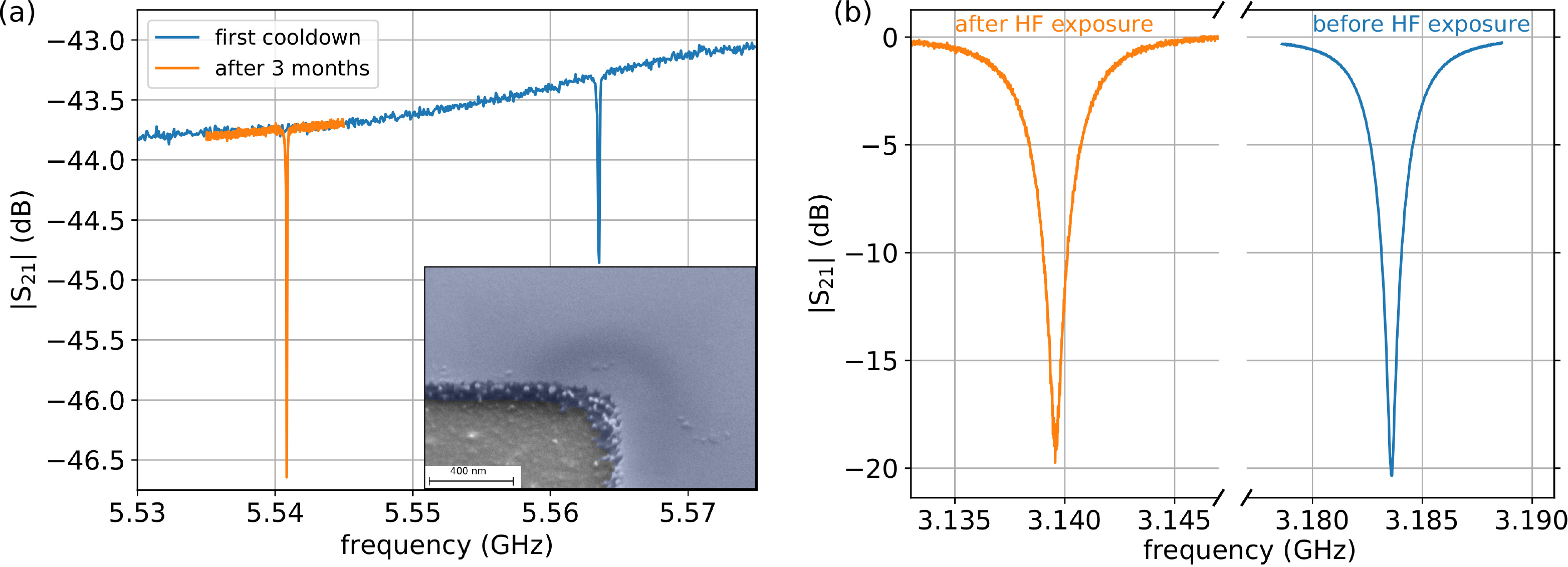}%{TiN3nm_aging}
\caption[]{    Stability of the TiN thin film resonators. (a) Plots of microwave transmission parameter $|S_{21}|$ for a 3~nm thick TiN resonator, measured with a gap of three months. Less that $1.0\%$ reduction in the resonance frequency indicates only negligible aging affect in the resonator.
(b) Plot of $|S_{21}|$ versus frequency for a 5~nm thick TiN resonator before and after exposure with HF. After the first measurement, the device was exposed to vapour HF of 4 steps of 300~s each, with an etch rate of 48~nm/min for thermal silicon oxide. Very small change in resonance frequency($\sim1.5\%$) and $Q_i$ suggest excellent  robustness of the films under HF treatment.   False color high resolution SEM image of a device after exposure to HF is shown in inset of (a).    }
\label{fig:aging}
\end{center}
\end {figure*}
%------------------------------------------------------------------------------------------------------------------------

A very important aspect for future integration and use in large scale circuits is device stability. At present,  superconducting quantum devices, made with conventional Al-AlO$_x$-Al Josephson junctions suffer from a major drawback, which is aging of the Josephson     junction. With time, characteristics of the materials alter,  resulting in degradation of the device quality. These aging effects are  often attributed to oxidation of the superconducting building block. Aging was also observed in the thinnest ALD TiN films reported in \cite{doi:10.1063/1.5053461}.
We investigate this aspect in our TiN films. We carry out multiple measurements of superconducting transitions in TiN thin films with a time gap of more that two months, and do not observe, within the measurement uncertainty, any change in the $T_C$ or the sheet resistance. Furthermore, we  study shift in resonance frequency or quality factors in TiN microwave resonators over a period of nearly three months, and do not observe any significant aging in our devices. In Fig.~\ref{fig:aging}(a) we show plots of $|S_{21}|$  measured in one of our resonators made with 3~nm thin TiN film initially, and   measured  after three months. We observe a decrease in resonant frequency by only $\sim$20~MHz, which is a less than 1~$\%$ change, and is often observed in different cooldowns, while any significant oxidation would have resulted in an increase in $L_K$ and an appreciable change in $f_R$. We also do not observe any significant change in the $Q_i$ of the device, extracted from the resonances,  over this time period. The devices were stored in ambient atmosphere inside a clean-room for the entire period of time.
We note here that the first measurement were carried out after a couple of weeks of the film deposition, and we cannot rule out that a fast aging happens faster than this time scale. Nevertheless our experiments demonstrate long term stability which is the key feature sought for.

A large variety of device architecture used in modern quantum technology  often require exposure of the devices to HF for removal of oxides~\cite{PhysRevApplied.6.014013,Melville2020}. While it has been reported that the quality of relatively thicker devices improve, because of removal of native oxide from device surface, after exposure to HF for a very short amount of time~\cite{Melville2020},  we investigate the robustness of very thin film microwave resonators under HF exposure for significantly longer time.  We expose the thin film resonators with both liquid and vapour HF for sufficient long time, often equivalent to removal of 1~$\mu$m thermal silicon  oxide, and observe only negligible alteration of the device quality, both in terms of resonant frequency and internal quality factors, suggesting extreme robustness of the thin films.  In Fig.~\ref{fig:aging}(b), we show representative plot of normalized $|S_{21}|$ of a 5~nm thick TiN resonator before and after exposure to HF.  We also  examine  the device surface with high-resolution SEM, and observe that the surface quality is maintained after such HF exposures (inset of Fig.~\ref{fig:aging}(a)).  The stability of the device under such extreme condition opens up the possibility to integrate such devices with SOI wafers, commonly used in different device architectures. 
The very high quality and robustness of the grown TiN films make them a very suitable candidate to fabricate building blocks of superconducting quantum devices.
%% =======================================================================================================================================================================

\subsection*{Impedance of the resonators}

 The total kinetic inductance of a superinductor of length $l$ and width $w$ can be estimated as $L_{kinetic} = \frac{l}{w} L_K =  \frac{\hbar}{1.76 \pi k_B} \frac{l}{w}\frac{R_N}{T_C}$, whereas,  in the limit of $l \gg w \gg t$, $t$ being the thickness, the geometric inductance can be approximated to  $L_{geo} \approx \frac{\mu_0}{2 \pi}l \ln(\frac{2l}{w})$~\cite{PhysRevApplied.11.054087, tinkham}.   For our 3~nm thick TiN resonator (Fig.~\ref{fig:3nm_3dres}), we obtain the kinetic inductance of the resonator $L_{kin} = 290$~nH. This value  is comparable to the kinetic inductance of typical~$\sim100$~nH of high-impedance superinductors used in high-coherence  fluxonium qubits~\cite{Pop2014,PhysRevLett.122.010504, PhysRevApplied.14.064038, Gruenhaupt2019}.  We also obtain $\alpha=0.98$ and $\alpha=0.99$ for 5~nm and 3~nm thick TiN resonators, respectively. Hence the constraints $\alpha=1$ in the fit in Fig.~\ref{fig:Qi_temp} are well justified.  
Such large kinetic inductance, or equivalently $\alpha \sim 1$, is an approach towards obtaining high characteristic impedance.   We estimate  mode impedance $Z_c = \sqrt{\mathcal{L}/c}$, where $\mathcal{L}$ and $c$ are inductance and capacitance per unit length of the resonator, of $3.2$~k$\Omega$ for our 3~nm TiN resonator, measured in the 3D waveguide.  We measure impedance as high as 4.2~k$\Omega$ for a 3~nm thick TiN resonator with mode frequency of 8.7~GHz in a 2D geometry.   This mode impedance can be further modulated by modifying  length and width of the  resonator. While reduction of width increases the mode impedance, the $Q_i$ were often observed to suffer from different loss mechanisms~\cite{PhysRevApplied.11.044014}.

\section{Conclusions}

In summary, we studied the microwave properties of superconducting TiN thin films fabricated with a VLSI  platform. We showed that the films remain superconducting down to at least 3~nm, with a critical temperature still exceeding 3~K. When reducing the film thickness, the kinetic inductance increases up to 239~pH/$\square$ for a 3~nm thick film. In microwave resonators, we demonstrate very large total inductance of several hundreds of nH and characteristic impedance $Z \approx 4.2~$k$\Omega$ together with state-of-the-art losses in the single photon regime, i.e. internal quality factors $Q_i \approx 10^5$. We show evidence that the remaining losses can be attributed to non-equilibrium quasiparticles. Mitigation strategies, such as an improved shielding or phonon traps might help to reduce the losses even further in the future.  
Our TiN showed negligible degradation due to aging, contrary to what is often observed in very thin films, and can withstand extended exposition to HF. All these demonstrations open up the possibility to develop industrial scale fabrication of superconducting microwave circuits. The compatibility of TiN with large magnetic field will also allow to integrate it into hybrid circuits using semiconductor spins and superconducting circuits.

\section{Acknowledgment}
This work was supported by the French National Research Agency (ANR) in the framework of the Graphmon project (ANR-19-CE47-0007) and the QNEMS project from LETI Carnot Institute.
JR acknowledges F. Balestro, V. Bouchiat, E. Eyraud and W. Wernsdorfer for help with the cryogenic system. KRA acknowledge help of F. Faroughi in EM simulations and with 3D waveguide. We acknowledge the help of P. Lachkar for DC resistivity measurements. We acknowledge the work of J. Jarreau and L. Del-Rey for the fabrication of the 3D waveguide. We acknowledge support from the Nanofab team of Institut N\'eel, and from  Guillaume Rodriguez of CEA Leti.

\section*{Appendix A: Methods}

%-------------------------------------------------- Thickness measurement  -----------------------------------------
\begin {figure}[!ht]
\begin{center}
    \includegraphics[width=0.5\textwidth, keepaspectratio]{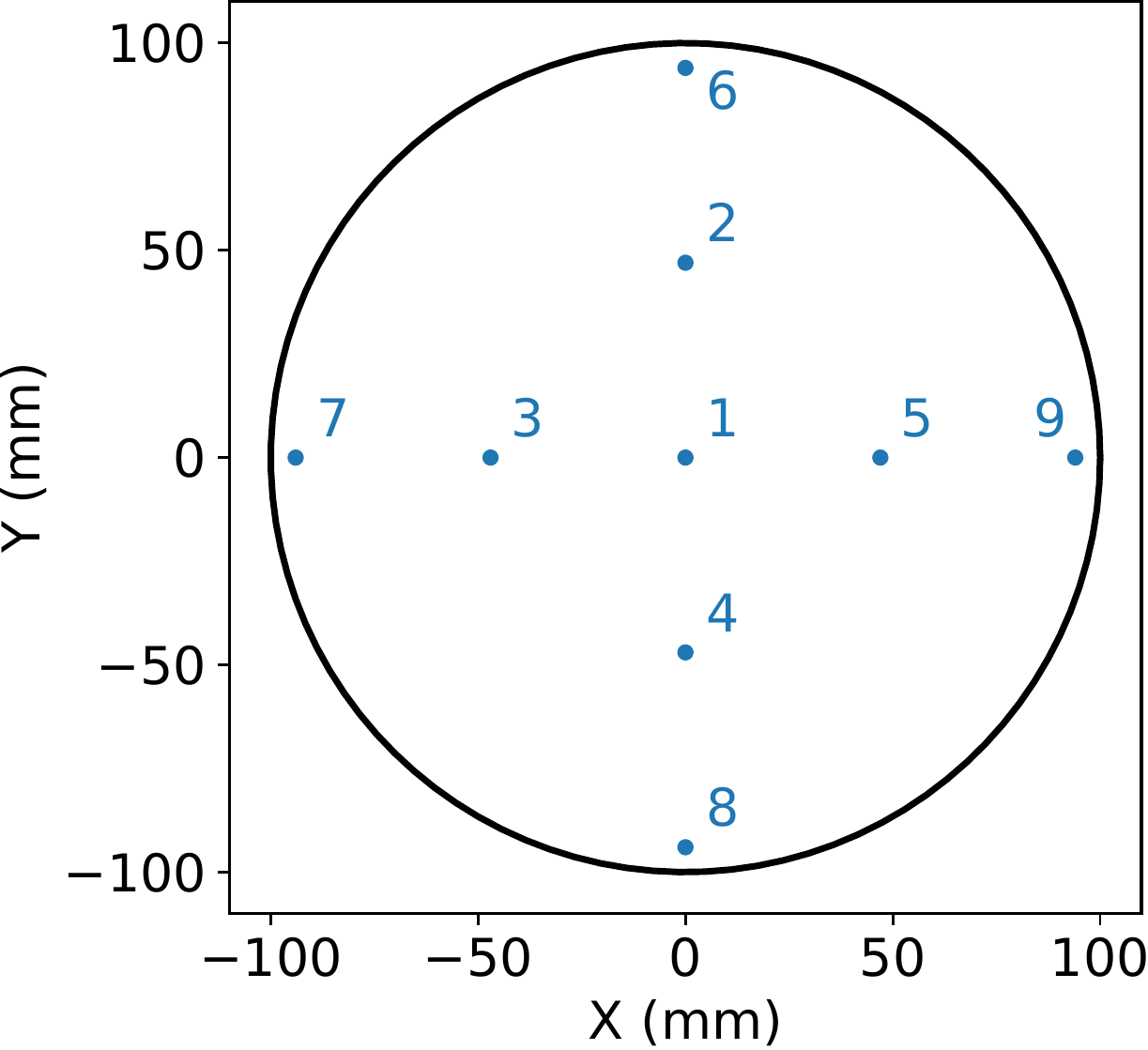}%{supfig1_measure_ellip_10nm_v2} 
    \caption[]{ Measurement of thickness of TiN film by ellipsometry.  The black circle represents outline of the silicon wafer of 200~mm diameter, with 10~nm TiN. The  blue dots represent the positions where the film thicknesses were measured by ellipsometry.  The measured values yield an average thickness of 10.6~nm, and a spread of 1.2~nm. The values are given in Tbl.~\ref{tbl:elips}.
\label{fig:ellipso} } 
\end{center}
\end {figure}
%------------------------------------------------------------------------------------------------------------------------

TiN thin films of six different thicknesses were deposited with a VLSI, CMOS-compatible physical vapour deposition (PVD) method at 350$^\circ$C on high resistive, 725~$\mu$m thick  intrinsic high-resistive silicon (100)  wafer of 200~mm diameters. The thicknesses and film uniformity were characterized  using ellipsometry and sheet resistance measurements using a probe-station. Both measurements were carried out at 9 different positions spread over the wafers [Fig~\ref{fig:ellipso}], and suggested excellent homogeneity of the films. For a film of thickness targeted 10~nm, measured values from elipsometry yield an average thickness of 10.6~nm, and a spread of only 1.2~nm over the 200~mm diameter wafer. The values are given in Tbl.~\ref{tbl:elips}.
 Resistivity measurements from 300~K down to 1.9~K, and magnetic field 0-8~T  were then performed  in a Physical Property Measurement Systems (PPMS), by a standard low-frequency a.c. four-probe method, with bias current kept sufficiently low to avoid Joule heating effect.  Ti-Au ohmic current and voltage contact pads were deposited on the samples by electron beam evaporation. 
 Microwave resonators were patterned by negative electron-beam lithography followed by SF$_6$ based dry-etching of the TiN~\cite{doi:10.1063/1.4729623}.
Microwave  $\lambda/2$   resonators, in a 2D microstrip configuration, were capacitively coupled to low-loss impedance matched transmission lines. The coupling was optimized with aid of EM simulations.  Transmission line, and ground plane  at the backside of the wafers were deposited using electron-beam evaporation. The device chips were wire-bonded to a custom-built high-frequency copper-made sample holder.  Separately, microstrip $\lambda/2$ resonators were  measured using a rectangular copper 3D waveguide, where the copper waveguide determines the ground, and  provides a low-loss, clean microwave environment free of lossy components in close proximity to the resonator~\cite{doi:10.1063/1.4935541}. Microwave resonators were measured in a dilution refrigerator, with a base temperature of 25~mK.

\begin{center}
  \begin{table}[!h]
    \caption{ Measurement of film thickness using ellipsometry. X and Y are the coordinates of the position where the measurements were taken.  \label{tbl:elips} }
    \begin{tabular}{| p{2cm} p{2cm}  p{2cm}  p{2cm}  |  }
      \hline
      \hline
      Sl. No. & X & Y & thicnkess   \\ 
      & (mm) & (mm) & (nm) \\ \hline  
      1 & 0   &  0  &  10.96    \\ \hline
      2 & 0   &  47 & 11.28   \\ \hline
      3 & -47 & 0   &  11.2    \\ \hline
      4 & 0   & -47 & 11.2    \\ \hline
      5 & 47  & 0   & 11.17     \\ \hline
      6 & 0   &  94  & 10.13       \\ \hline
      7 & -94  & 0  & 10.08       \\ \hline
      8 & 0  & -94  & 10.11       \\ \hline
      9 & 94  & 0  & 10.12       \\ \hline
      \hline
    \end{tabular}
  \end{table}
\end{center}

\clearpage
\section*{Appendix B: Additional plots of resistivity measurements}

%-------------------------------------------------- DC resistivity --------------------------------------------------
\begin {figure}[!h]
\begin{center}
    \includegraphics[width=0.5\textwidth, keepaspectratio]{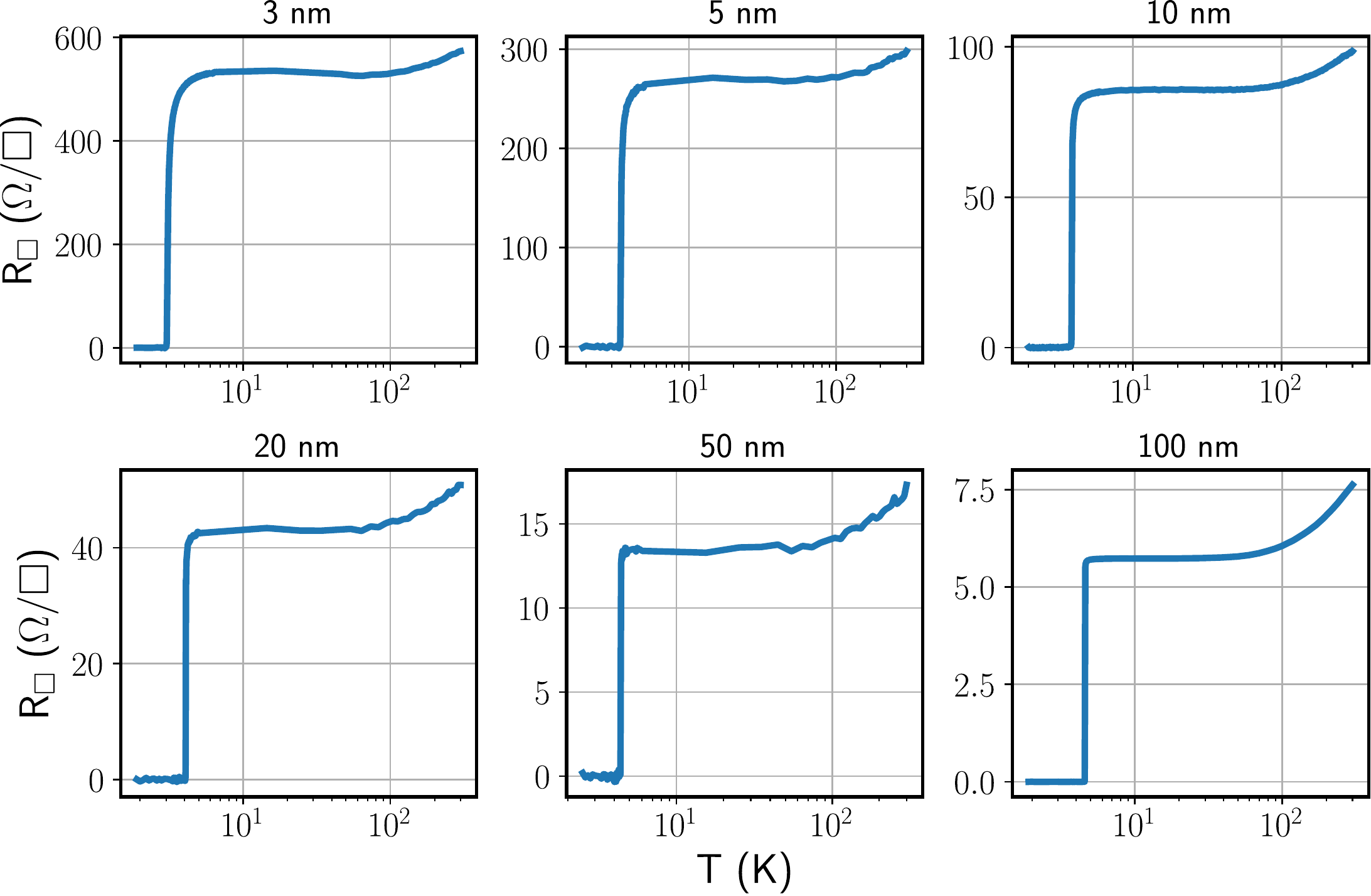}
    \caption[]{ Temperature dependance of sheet resistance for TiN films of different  thickness, from room temperature down to 1.9~K. Plots of the data zoomed in near the superconducting transitions are shown in Fig.~\ref{fig:dc_res}. 
\label{fig:dc_res_full} } 
\end{center}
\end {figure}
%------------------------------------------------------------------------------------------------------------------------

\section*{Appendix C: Impedance of the microwave resonators}

The mode impedance of a microwave resonator is given by $Z_c = \sqrt{\mathcal{L}/c}$, where $\mathcal{L}$ and $c$ are inductance and capacitance per unit length of the resonator. $\mathcal{L}$ for a microwave resonator of length $l$ and width $w$ can be approximated from $L_K$, in the limit of kinetic inductance fraction $\alpha=1$ as  $\mathcal{L} = L_K/w$~\cite{PhysRevApplied.11.044014}. Equating the phase velocity $v_p= \dfrac{2\pi f_R}{k} =  \dfrac{1}{\sqrt{\mathcal{L}c} }$, we can estimate the mode impedance from $Z_c = 2 f_r  L_K ( \frac{ l}{w})$. In Tbl.~\ref{tbl:inductance}, we summarize mode impedance of multiple $\lambda/2$ resonators of different thickness.%, and also compare mode inductance, reported in literature for different materials.. 

\begin{center}
  \begin{table}[!h]
    \caption{ Mode impedance of TiN resonators. \label{tbl:inductance}  }
    \begin{tabular}{| p{1.25cm} p{1.20cm} p{1.0cm} p{1.0cm} p{1.10cm}    p{1.25cm}  |  }
      \hline
      \hline
      Sl. No. &  $f_R$ &  t  & l  & w  &   $Z_c$     \\ 
     % &   &       \\  \hline
      &  (GHz) &  (nm) & ($\mu$m)  & ($\mu$m) & (k$\Omega$)    \\ \hline
      1 &    8.69  & 3 & 300 & 0.3 & 4.15     \\  %\hline
      2 &   5.57    &  3 & 600 & 0.5 & 3.19     \\ %\hline

      3 &    10.6  & 5 & 500 & 0.5 & 2.18    \\  %\hline
      4 &    3.93  & 5 & 2000 & 1.0 & 1.62     \\  %\hline
 
      \hline
    \end{tabular}
  \end{table}
\end{center}

%\begin{thebibliography}{68}%
\bibliography{tin}

\begin{thebibliography}{68}%
\makeatletter
\providecommand \@ifxundefined [1]{%
 \@ifx{#1\undefined}
}%
\providecommand \@ifnum [1]{%
 \ifnum #1\expandafter \@firstoftwo
 \else \expandafter \@secondoftwo
 \fi
}%
\providecommand \@ifx [1]{%
 \ifx #1\expandafter \@firstoftwo
 \else \expandafter \@secondoftwo
 \fi
}%
\providecommand \natexlab [1]{#1}%
\providecommand \enquote  [1]{``#1''}%
\providecommand \bibnamefont  [1]{#1}%
\providecommand \bibfnamefont [1]{#1}%
\providecommand \citenamefont [1]{#1}%
\providecommand \href@noop [0]{\@secondoftwo}%
\providecommand \href [0]{\begingroup \@sanitize@url \@href}%
\providecommand \@href[1]{\@@startlink{#1}\@@href}%
\providecommand \@@href[1]{\endgroup#1\@@endlink}%
\providecommand \@sanitize@url [0]{\catcode `\\12\catcode `\$12\catcode
  `\&12\catcode `\#12\catcode `\^12\catcode `\_12\catcode `\%12\relax}%
\providecommand \@@startlink[1]{}%
\providecommand \@@endlink[0]{}%
\providecommand \url  [0]{\begingroup\@sanitize@url \@url }%
\providecommand \@url [1]{\endgroup\@href {#1}{\urlprefix }}%
\providecommand \urlprefix  [0]{URL }%
\providecommand \Eprint [0]{\href }%
\providecommand \doibase [0]{https://doi.org/}%
\providecommand \selectlanguage [0]{\@gobble}%
\providecommand \bibinfo  [0]{\@secondoftwo}%
\providecommand \bibfield  [0]{\@secondoftwo}%
\providecommand \translation [1]{[#1]}%
\providecommand \BibitemOpen [0]{}%
\providecommand \bibitemStop [0]{}%
\providecommand \bibitemNoStop [0]{.\EOS\space}%
\providecommand \EOS [0]{\spacefactor3000\relax}%
\providecommand \BibitemShut  [1]{\csname bibitem#1\endcsname}%
\let\auto@bib@innerbib\@empty
%</preamble>
\bibitem [{\citenamefont {Arute}\ \emph {et~al.}(2019)\citenamefont {Arute},
  \citenamefont {Arya}, \citenamefont {Babbush}, \citenamefont {Bacon},
  \citenamefont {Bardin}, \citenamefont {Barends}, \citenamefont {Biswas},
  \citenamefont {Boixo}, \citenamefont {Brandao}, \citenamefont {Buell},
  \citenamefont {Burkett}, \citenamefont {Chen}, \citenamefont {Chen},
  \citenamefont {Chiaro}, \citenamefont {Collins}, \citenamefont {Courtney},
  \citenamefont {Dunsworth}, \citenamefont {Farhi}, \citenamefont {Foxen},
  \citenamefont {Fowler}, \citenamefont {Gidney}, \citenamefont {Giustina},
  \citenamefont {Graff}, \citenamefont {Guerin}, \citenamefont {Habegger},
  \citenamefont {Harrigan}, \citenamefont {Hartmann}, \citenamefont {Ho},
  \citenamefont {Hoffmann}, \citenamefont {Huang}, \citenamefont {Humble},
  \citenamefont {Isakov}, \citenamefont {Jeffrey}, \citenamefont {Jiang},
  \citenamefont {Kafri}, \citenamefont {Kechedzhi}, \citenamefont {Kelly},
  \citenamefont {Klimov}, \citenamefont {Knysh}, \citenamefont {Korotkov},
  \citenamefont {Kostritsa}, \citenamefont {Landhuis}, \citenamefont
  {Lindmark}, \citenamefont {Lucero}, \citenamefont {Lyakh}, \citenamefont
  {Mandr{\`a}}, \citenamefont {McClean}, \citenamefont {McEwen}, \citenamefont
  {Megrant}, \citenamefont {Mi}, \citenamefont {Michielsen}, \citenamefont
  {Mohseni}, \citenamefont {Mutus}, \citenamefont {Naaman}, \citenamefont
  {Neeley}, \citenamefont {Neill}, \citenamefont {Niu}, \citenamefont {Ostby},
  \citenamefont {Petukhov}, \citenamefont {Platt}, \citenamefont {Quintana},
  \citenamefont {Rieffel}, \citenamefont {Roushan}, \citenamefont {Rubin},
  \citenamefont {Sank}, \citenamefont {Satzinger}, \citenamefont {Smelyanskiy},
  \citenamefont {Sung}, \citenamefont {Trevithick}, \citenamefont
  {Vainsencher}, \citenamefont {Villalonga}, \citenamefont {White},
  \citenamefont {Yao}, \citenamefont {Yeh}, \citenamefont {Zalcman},
  \citenamefont {Neven},\ and\ \citenamefont {Martinis}}]{Arute2019}%
  \BibitemOpen
  \bibfield  {author} {\bibinfo {author} {\bibfnamefont {F.}~\bibnamefont
  {Arute}}, \bibinfo {author} {\bibfnamefont {K.}~\bibnamefont {Arya}},
  \bibinfo {author} {\bibfnamefont {R.}~\bibnamefont {Babbush}}, \bibinfo
  {author} {\bibfnamefont {D.}~\bibnamefont {Bacon}}, \bibinfo {author}
  {\bibfnamefont {J.~C.}\ \bibnamefont {Bardin}}, \bibinfo {author}
  {\bibfnamefont {R.}~\bibnamefont {Barends}}, \bibinfo {author} {\bibfnamefont
  {R.}~\bibnamefont {Biswas}}, \bibinfo {author} {\bibfnamefont
  {S.}~\bibnamefont {Boixo}}, \bibinfo {author} {\bibfnamefont {F.~G. S.~L.}\
  \bibnamefont {Brandao}}, \bibinfo {author} {\bibfnamefont {D.~A.}\
  \bibnamefont {Buell}}, \bibinfo {author} {\bibfnamefont {B.}~\bibnamefont
  {Burkett}}, \bibinfo {author} {\bibfnamefont {Y.}~\bibnamefont {Chen}},
  \bibinfo {author} {\bibfnamefont {Z.}~\bibnamefont {Chen}}, \bibinfo {author}
  {\bibfnamefont {B.}~\bibnamefont {Chiaro}}, \bibinfo {author} {\bibfnamefont
  {R.}~\bibnamefont {Collins}}, \bibinfo {author} {\bibfnamefont
  {W.}~\bibnamefont {Courtney}}, \bibinfo {author} {\bibfnamefont
  {A.}~\bibnamefont {Dunsworth}}, \bibinfo {author} {\bibfnamefont
  {E.}~\bibnamefont {Farhi}}, \bibinfo {author} {\bibfnamefont
  {B.}~\bibnamefont {Foxen}}, \bibinfo {author} {\bibfnamefont
  {A.}~\bibnamefont {Fowler}}, \bibinfo {author} {\bibfnamefont
  {C.}~\bibnamefont {Gidney}}, \bibinfo {author} {\bibfnamefont
  {M.}~\bibnamefont {Giustina}}, \bibinfo {author} {\bibfnamefont
  {R.}~\bibnamefont {Graff}}, \bibinfo {author} {\bibfnamefont
  {K.}~\bibnamefont {Guerin}}, \bibinfo {author} {\bibfnamefont
  {S.}~\bibnamefont {Habegger}}, \bibinfo {author} {\bibfnamefont {M.~P.}\
  \bibnamefont {Harrigan}}, \bibinfo {author} {\bibfnamefont {M.~J.}\
  \bibnamefont {Hartmann}}, \bibinfo {author} {\bibfnamefont {A.}~\bibnamefont
  {Ho}}, \bibinfo {author} {\bibfnamefont {M.}~\bibnamefont {Hoffmann}},
  \bibinfo {author} {\bibfnamefont {T.}~\bibnamefont {Huang}}, \bibinfo
  {author} {\bibfnamefont {T.~S.}\ \bibnamefont {Humble}}, \bibinfo {author}
  {\bibfnamefont {S.~V.}\ \bibnamefont {Isakov}}, \bibinfo {author}
  {\bibfnamefont {E.}~\bibnamefont {Jeffrey}}, \bibinfo {author} {\bibfnamefont
  {Z.}~\bibnamefont {Jiang}}, \bibinfo {author} {\bibfnamefont
  {D.}~\bibnamefont {Kafri}}, \bibinfo {author} {\bibfnamefont
  {K.}~\bibnamefont {Kechedzhi}}, \bibinfo {author} {\bibfnamefont
  {J.}~\bibnamefont {Kelly}}, \bibinfo {author} {\bibfnamefont {P.~V.}\
  \bibnamefont {Klimov}}, \bibinfo {author} {\bibfnamefont {S.}~\bibnamefont
  {Knysh}}, \bibinfo {author} {\bibfnamefont {A.}~\bibnamefont {Korotkov}},
  \bibinfo {author} {\bibfnamefont {F.}~\bibnamefont {Kostritsa}}, \bibinfo
  {author} {\bibfnamefont {D.}~\bibnamefont {Landhuis}}, \bibinfo {author}
  {\bibfnamefont {M.}~\bibnamefont {Lindmark}}, \bibinfo {author}
  {\bibfnamefont {E.}~\bibnamefont {Lucero}}, \bibinfo {author} {\bibfnamefont
  {D.}~\bibnamefont {Lyakh}}, \bibinfo {author} {\bibfnamefont
  {S.}~\bibnamefont {Mandr{\`a}}}, \bibinfo {author} {\bibfnamefont {J.~R.}\
  \bibnamefont {McClean}}, \bibinfo {author} {\bibfnamefont {M.}~\bibnamefont
  {McEwen}}, \bibinfo {author} {\bibfnamefont {A.}~\bibnamefont {Megrant}},
  \bibinfo {author} {\bibfnamefont {X.}~\bibnamefont {Mi}}, \bibinfo {author}
  {\bibfnamefont {K.}~\bibnamefont {Michielsen}}, \bibinfo {author}
  {\bibfnamefont {M.}~\bibnamefont {Mohseni}}, \bibinfo {author} {\bibfnamefont
  {J.}~\bibnamefont {Mutus}}, \bibinfo {author} {\bibfnamefont
  {O.}~\bibnamefont {Naaman}}, \bibinfo {author} {\bibfnamefont
  {M.}~\bibnamefont {Neeley}}, \bibinfo {author} {\bibfnamefont
  {C.}~\bibnamefont {Neill}}, \bibinfo {author} {\bibfnamefont {M.~Y.}\
  \bibnamefont {Niu}}, \bibinfo {author} {\bibfnamefont {E.}~\bibnamefont
  {Ostby}}, \bibinfo {author} {\bibfnamefont {A.}~\bibnamefont {Petukhov}},
  \bibinfo {author} {\bibfnamefont {J.~C.}\ \bibnamefont {Platt}}, \bibinfo
  {author} {\bibfnamefont {C.}~\bibnamefont {Quintana}}, \bibinfo {author}
  {\bibfnamefont {E.~G.}\ \bibnamefont {Rieffel}}, \bibinfo {author}
  {\bibfnamefont {P.}~\bibnamefont {Roushan}}, \bibinfo {author} {\bibfnamefont
  {N.~C.}\ \bibnamefont {Rubin}}, \bibinfo {author} {\bibfnamefont
  {D.}~\bibnamefont {Sank}}, \bibinfo {author} {\bibfnamefont {K.~J.}\
  \bibnamefont {Satzinger}}, \bibinfo {author} {\bibfnamefont {V.}~\bibnamefont
  {Smelyanskiy}}, \bibinfo {author} {\bibfnamefont {K.~J.}\ \bibnamefont
  {Sung}}, \bibinfo {author} {\bibfnamefont {M.~D.}\ \bibnamefont
  {Trevithick}}, \bibinfo {author} {\bibfnamefont {A.}~\bibnamefont
  {Vainsencher}}, \bibinfo {author} {\bibfnamefont {B.}~\bibnamefont
  {Villalonga}}, \bibinfo {author} {\bibfnamefont {T.}~\bibnamefont {White}},
  \bibinfo {author} {\bibfnamefont {Z.~J.}\ \bibnamefont {Yao}}, \bibinfo
  {author} {\bibfnamefont {P.}~\bibnamefont {Yeh}}, \bibinfo {author}
  {\bibfnamefont {A.}~\bibnamefont {Zalcman}}, \bibinfo {author} {\bibfnamefont
  {H.}~\bibnamefont {Neven}},\ and\ \bibinfo {author} {\bibfnamefont {J.~M.}\
  \bibnamefont {Martinis}},\ }\bibfield  {title} {\bibinfo {title} {Quantum
  supremacy using a programmable superconducting processor},\ }\href
  {https://doi.org/10.1038/s41586-019-1666-5} {\bibfield  {journal} {\bibinfo
  {journal} {Nature}\ }\textbf {\bibinfo {volume} {574}},\ \bibinfo {pages}
  {505} (\bibinfo {year} {2019})}\BibitemShut {NoStop}%
\bibitem [{\citenamefont {Wu}\ \emph {et~al.}(2021)\citenamefont {Wu},
  \citenamefont {Bao}, \citenamefont {Cao}, \citenamefont {Chen}, \citenamefont
  {Chen}, \citenamefont {Chen}, \citenamefont {Chung}, \citenamefont {Deng},
  \citenamefont {Du}, \citenamefont {Fan}, \citenamefont {Gong}, \citenamefont
  {Guo}, \citenamefont {Guo}, \citenamefont {Guo}, \citenamefont {Han},
  \citenamefont {Hong}, \citenamefont {Huang}, \citenamefont {Huo},
  \citenamefont {Li}, \citenamefont {Li}, \citenamefont {Li}, \citenamefont
  {Li}, \citenamefont {Liang}, \citenamefont {Lin}, \citenamefont {Lin},
  \citenamefont {Qian}, \citenamefont {Qiao}, \citenamefont {Rong},
  \citenamefont {Su}, \citenamefont {Sun}, \citenamefont {Wang}, \citenamefont
  {Wang}, \citenamefont {Wu}, \citenamefont {Xu}, \citenamefont {Yan},
  \citenamefont {Yang}, \citenamefont {Yang}, \citenamefont {Ye}, \citenamefont
  {Yin}, \citenamefont {Ying}, \citenamefont {Yu}, \citenamefont {Zha},
  \citenamefont {Zhang}, \citenamefont {Zhang}, \citenamefont {Zhang},
  \citenamefont {Zhang}, \citenamefont {Zhao}, \citenamefont {Zhao},
  \citenamefont {Zhou}, \citenamefont {Zhu}, \citenamefont {Lu}, \citenamefont
  {Peng}, \citenamefont {Zhu},\ and\ \citenamefont {Pan}}]{wu2021strong}%
  \BibitemOpen
  \bibfield  {author} {\bibinfo {author} {\bibfnamefont {Y.}~\bibnamefont
  {Wu}}, \bibinfo {author} {\bibfnamefont {W.-S.}\ \bibnamefont {Bao}},
  \bibinfo {author} {\bibfnamefont {S.}~\bibnamefont {Cao}}, \bibinfo {author}
  {\bibfnamefont {F.}~\bibnamefont {Chen}}, \bibinfo {author} {\bibfnamefont
  {M.-C.}\ \bibnamefont {Chen}}, \bibinfo {author} {\bibfnamefont
  {X.}~\bibnamefont {Chen}}, \bibinfo {author} {\bibfnamefont {T.-H.}\
  \bibnamefont {Chung}}, \bibinfo {author} {\bibfnamefont {H.}~\bibnamefont
  {Deng}}, \bibinfo {author} {\bibfnamefont {Y.}~\bibnamefont {Du}}, \bibinfo
  {author} {\bibfnamefont {D.}~\bibnamefont {Fan}}, \bibinfo {author}
  {\bibfnamefont {M.}~\bibnamefont {Gong}}, \bibinfo {author} {\bibfnamefont
  {C.}~\bibnamefont {Guo}}, \bibinfo {author} {\bibfnamefont {C.}~\bibnamefont
  {Guo}}, \bibinfo {author} {\bibfnamefont {S.}~\bibnamefont {Guo}}, \bibinfo
  {author} {\bibfnamefont {L.}~\bibnamefont {Han}}, \bibinfo {author}
  {\bibfnamefont {L.}~\bibnamefont {Hong}}, \bibinfo {author} {\bibfnamefont
  {H.-L.}\ \bibnamefont {Huang}}, \bibinfo {author} {\bibfnamefont {Y.-H.}\
  \bibnamefont {Huo}}, \bibinfo {author} {\bibfnamefont {L.}~\bibnamefont
  {Li}}, \bibinfo {author} {\bibfnamefont {N.}~\bibnamefont {Li}}, \bibinfo
  {author} {\bibfnamefont {S.}~\bibnamefont {Li}}, \bibinfo {author}
  {\bibfnamefont {Y.}~\bibnamefont {Li}}, \bibinfo {author} {\bibfnamefont
  {F.}~\bibnamefont {Liang}}, \bibinfo {author} {\bibfnamefont
  {C.}~\bibnamefont {Lin}}, \bibinfo {author} {\bibfnamefont {J.}~\bibnamefont
  {Lin}}, \bibinfo {author} {\bibfnamefont {H.}~\bibnamefont {Qian}}, \bibinfo
  {author} {\bibfnamefont {D.}~\bibnamefont {Qiao}}, \bibinfo {author}
  {\bibfnamefont {H.}~\bibnamefont {Rong}}, \bibinfo {author} {\bibfnamefont
  {H.}~\bibnamefont {Su}}, \bibinfo {author} {\bibfnamefont {L.}~\bibnamefont
  {Sun}}, \bibinfo {author} {\bibfnamefont {L.}~\bibnamefont {Wang}}, \bibinfo
  {author} {\bibfnamefont {S.}~\bibnamefont {Wang}}, \bibinfo {author}
  {\bibfnamefont {D.}~\bibnamefont {Wu}}, \bibinfo {author} {\bibfnamefont
  {Y.}~\bibnamefont {Xu}}, \bibinfo {author} {\bibfnamefont {K.}~\bibnamefont
  {Yan}}, \bibinfo {author} {\bibfnamefont {W.}~\bibnamefont {Yang}}, \bibinfo
  {author} {\bibfnamefont {Y.}~\bibnamefont {Yang}}, \bibinfo {author}
  {\bibfnamefont {Y.}~\bibnamefont {Ye}}, \bibinfo {author} {\bibfnamefont
  {J.}~\bibnamefont {Yin}}, \bibinfo {author} {\bibfnamefont {C.}~\bibnamefont
  {Ying}}, \bibinfo {author} {\bibfnamefont {J.}~\bibnamefont {Yu}}, \bibinfo
  {author} {\bibfnamefont {C.}~\bibnamefont {Zha}}, \bibinfo {author}
  {\bibfnamefont {C.}~\bibnamefont {Zhang}}, \bibinfo {author} {\bibfnamefont
  {H.}~\bibnamefont {Zhang}}, \bibinfo {author} {\bibfnamefont
  {K.}~\bibnamefont {Zhang}}, \bibinfo {author} {\bibfnamefont
  {Y.}~\bibnamefont {Zhang}}, \bibinfo {author} {\bibfnamefont
  {H.}~\bibnamefont {Zhao}}, \bibinfo {author} {\bibfnamefont {Y.}~\bibnamefont
  {Zhao}}, \bibinfo {author} {\bibfnamefont {L.}~\bibnamefont {Zhou}}, \bibinfo
  {author} {\bibfnamefont {Q.}~\bibnamefont {Zhu}}, \bibinfo {author}
  {\bibfnamefont {C.-Y.}\ \bibnamefont {Lu}}, \bibinfo {author} {\bibfnamefont
  {C.-Z.}\ \bibnamefont {Peng}}, \bibinfo {author} {\bibfnamefont
  {X.}~\bibnamefont {Zhu}},\ and\ \bibinfo {author} {\bibfnamefont {J.-W.}\
  \bibnamefont {Pan}},\ }\href@noop {} {\bibinfo {title} {Strong quantum
  computational advantage using a superconducting quantum processor}} (\bibinfo
  {year} {2021}),\ \Eprint {https://arxiv.org/abs/2106.14734} {arXiv:2106.14734
  [quant-ph]} \BibitemShut {NoStop}%
\bibitem [{\citenamefont {Forn-D\'{\i}az}\ \emph {et~al.}(2019)\citenamefont
  {Forn-D\'{\i}az}, \citenamefont {Lamata}, \citenamefont {Rico}, \citenamefont
  {Kono},\ and\ \citenamefont {Solano}}]{FornDiaz2019}%
  \BibitemOpen
  \bibfield  {author} {\bibinfo {author} {\bibfnamefont {P.}~\bibnamefont
  {Forn-D\'{\i}az}}, \bibinfo {author} {\bibfnamefont {L.}~\bibnamefont
  {Lamata}}, \bibinfo {author} {\bibfnamefont {E.}~\bibnamefont {Rico}},
  \bibinfo {author} {\bibfnamefont {J.}~\bibnamefont {Kono}},\ and\ \bibinfo
  {author} {\bibfnamefont {E.}~\bibnamefont {Solano}},\ }\bibfield  {title}
  {\bibinfo {title} {Ultrastrong coupling regimes of light-matter
  interaction},\ }\href {https://doi.org/10.1103/RevModPhys.91.025005}
  {\bibfield  {journal} {\bibinfo  {journal} {Rev. Mod. Phys.}\ }\textbf
  {\bibinfo {volume} {91}},\ \bibinfo {pages} {025005} (\bibinfo {year}
  {2019})}\BibitemShut {NoStop}%
\bibitem [{\citenamefont {Frisk~Kockum}\ \emph {et~al.}(2019)\citenamefont
  {Frisk~Kockum}, \citenamefont {Miranowicz}, \citenamefont {De~Liberato},
  \citenamefont {Savasta},\ and\ \citenamefont {Nori}}]{FriskKockum2019}%
  \BibitemOpen
  \bibfield  {author} {\bibinfo {author} {\bibfnamefont {A.}~\bibnamefont
  {Frisk~Kockum}}, \bibinfo {author} {\bibfnamefont {A.}~\bibnamefont
  {Miranowicz}}, \bibinfo {author} {\bibfnamefont {S.}~\bibnamefont
  {De~Liberato}}, \bibinfo {author} {\bibfnamefont {S.}~\bibnamefont
  {Savasta}},\ and\ \bibinfo {author} {\bibfnamefont {F.}~\bibnamefont
  {Nori}},\ }\bibfield  {title} {\bibinfo {title} {Ultrastrong coupling between
  light and matter},\ }\href {https://doi.org/10.1038/s42254-018-0006-2}
  {\bibfield  {journal} {\bibinfo  {journal} {Nature Reviews Physics}\ }\textbf
  {\bibinfo {volume} {1}},\ \bibinfo {pages} {19} (\bibinfo {year}
  {2019})}\BibitemShut {NoStop}%
\bibitem [{\citenamefont {L{\'e}ger}\ \emph {et~al.}(2019)\citenamefont
  {L{\'e}ger}, \citenamefont {Puertas-Mart\'inez}, \citenamefont {Bharadwaj},
  \citenamefont {Dassonneville}, \citenamefont {Delaforce}, \citenamefont
  {Foroughi}, \citenamefont {Milchakov}, \citenamefont {Planat}, \citenamefont
  {Buisson}, \citenamefont {Naud}, \citenamefont {Hasch-Guichard},
  \citenamefont {Florens}, \citenamefont {Snyman},\ and\ \citenamefont
  {Roch}}]{Leger2019}%
  \BibitemOpen
  \bibfield  {author} {\bibinfo {author} {\bibfnamefont {S.}~\bibnamefont
  {L{\'e}ger}}, \bibinfo {author} {\bibfnamefont {J.}~\bibnamefont
  {Puertas-Mart\'inez}}, \bibinfo {author} {\bibfnamefont {K.}~\bibnamefont
  {Bharadwaj}}, \bibinfo {author} {\bibfnamefont {R.}~\bibnamefont
  {Dassonneville}}, \bibinfo {author} {\bibfnamefont {J.}~\bibnamefont
  {Delaforce}}, \bibinfo {author} {\bibfnamefont {F.}~\bibnamefont {Foroughi}},
  \bibinfo {author} {\bibfnamefont {V.}~\bibnamefont {Milchakov}}, \bibinfo
  {author} {\bibfnamefont {L.}~\bibnamefont {Planat}}, \bibinfo {author}
  {\bibfnamefont {O.}~\bibnamefont {Buisson}}, \bibinfo {author} {\bibfnamefont
  {C.}~\bibnamefont {Naud}}, \bibinfo {author} {\bibfnamefont {W.}~\bibnamefont
  {Hasch-Guichard}}, \bibinfo {author} {\bibfnamefont {S.}~\bibnamefont
  {Florens}}, \bibinfo {author} {\bibfnamefont {I.}~\bibnamefont {Snyman}},\
  and\ \bibinfo {author} {\bibfnamefont {N.}~\bibnamefont {Roch}},\ }\bibfield
  {title} {\bibinfo {title} {Observation of quantum many-body effects due to
  zero point fluctuations in superconducting circuits},\ }\href
  {https://doi.org/10.1038/s41467-019-13199-x} {\bibfield  {journal} {\bibinfo
  {journal} {Nature Communications}\ }\textbf {\bibinfo {volume} {10}},\
  \bibinfo {pages} {5259} (\bibinfo {year} {2019})}\BibitemShut {NoStop}%
\bibitem [{\citenamefont {Kuzmin}\ \emph {et~al.}(2019)\citenamefont {Kuzmin},
  \citenamefont {Mehta}, \citenamefont {Grabon}, \citenamefont {Mencia},\ and\
  \citenamefont {Manucharyan}}]{Kuzmin2019}%
  \BibitemOpen
  \bibfield  {author} {\bibinfo {author} {\bibfnamefont {R.}~\bibnamefont
  {Kuzmin}}, \bibinfo {author} {\bibfnamefont {N.}~\bibnamefont {Mehta}},
  \bibinfo {author} {\bibfnamefont {N.}~\bibnamefont {Grabon}}, \bibinfo
  {author} {\bibfnamefont {R.}~\bibnamefont {Mencia}},\ and\ \bibinfo {author}
  {\bibfnamefont {V.~E.}\ \bibnamefont {Manucharyan}},\ }\bibfield  {title}
  {\bibinfo {title} {Superstrong coupling in circuit quantum electrodynamics},\
  }\href {https://doi.org/10.1038/s41534-019-0134-2} {\bibfield  {journal}
  {\bibinfo  {journal} {npj Quantum Information}\ }\textbf {\bibinfo {volume}
  {5}},\ \bibinfo {pages} {20} (\bibinfo {year} {2019})}\BibitemShut {NoStop}%
\bibitem [{\citenamefont {Roushan}\ \emph {et~al.}(2017)\citenamefont
  {Roushan}, \citenamefont {Neill}, \citenamefont {Tangpanitanon},
  \citenamefont {Bastidas}, \citenamefont {Megrant}, \citenamefont {Barends},
  \citenamefont {Chen}, \citenamefont {Chen}, \citenamefont {Chiaro},
  \citenamefont {Dunsworth}, \citenamefont {Fowler}, \citenamefont {Foxen},
  \citenamefont {Giustina}, \citenamefont {Jeffrey}, \citenamefont {Kelly},
  \citenamefont {Lucero}, \citenamefont {Mutus}, \citenamefont {Neeley},
  \citenamefont {Quintana}, \citenamefont {Sank}, \citenamefont {Vainsencher},
  \citenamefont {Wenner}, \citenamefont {White}, \citenamefont {Neven},
  \citenamefont {Angelakis},\ and\ \citenamefont {Martinis}}]{Roushan2017}%
  \BibitemOpen
  \bibfield  {author} {\bibinfo {author} {\bibfnamefont {P.}~\bibnamefont
  {Roushan}}, \bibinfo {author} {\bibfnamefont {C.}~\bibnamefont {Neill}},
  \bibinfo {author} {\bibfnamefont {J.}~\bibnamefont {Tangpanitanon}}, \bibinfo
  {author} {\bibfnamefont {V.~M.}\ \bibnamefont {Bastidas}}, \bibinfo {author}
  {\bibfnamefont {A.}~\bibnamefont {Megrant}}, \bibinfo {author} {\bibfnamefont
  {R.}~\bibnamefont {Barends}}, \bibinfo {author} {\bibfnamefont
  {Y.}~\bibnamefont {Chen}}, \bibinfo {author} {\bibfnamefont {Z.}~\bibnamefont
  {Chen}}, \bibinfo {author} {\bibfnamefont {B.}~\bibnamefont {Chiaro}},
  \bibinfo {author} {\bibfnamefont {A.}~\bibnamefont {Dunsworth}}, \bibinfo
  {author} {\bibfnamefont {A.}~\bibnamefont {Fowler}}, \bibinfo {author}
  {\bibfnamefont {B.}~\bibnamefont {Foxen}}, \bibinfo {author} {\bibfnamefont
  {M.}~\bibnamefont {Giustina}}, \bibinfo {author} {\bibfnamefont
  {E.}~\bibnamefont {Jeffrey}}, \bibinfo {author} {\bibfnamefont
  {J.}~\bibnamefont {Kelly}}, \bibinfo {author} {\bibfnamefont
  {E.}~\bibnamefont {Lucero}}, \bibinfo {author} {\bibfnamefont
  {J.}~\bibnamefont {Mutus}}, \bibinfo {author} {\bibfnamefont
  {M.}~\bibnamefont {Neeley}}, \bibinfo {author} {\bibfnamefont
  {C.}~\bibnamefont {Quintana}}, \bibinfo {author} {\bibfnamefont
  {D.}~\bibnamefont {Sank}}, \bibinfo {author} {\bibfnamefont {A.}~\bibnamefont
  {Vainsencher}}, \bibinfo {author} {\bibfnamefont {J.}~\bibnamefont {Wenner}},
  \bibinfo {author} {\bibfnamefont {T.}~\bibnamefont {White}}, \bibinfo
  {author} {\bibfnamefont {H.}~\bibnamefont {Neven}}, \bibinfo {author}
  {\bibfnamefont {D.~G.}\ \bibnamefont {Angelakis}},\ and\ \bibinfo {author}
  {\bibfnamefont {J.}~\bibnamefont {Martinis}},\ }\bibfield  {title} {\bibinfo
  {title} {Spectroscopic signatures of localization with interacting photons in
  superconducting qubits},\ }\href {https://doi.org/10.1126/science.aao1401}
  {\bibfield  {journal} {\bibinfo  {journal} {Science}\ }\textbf {\bibinfo
  {volume} {358}},\ \bibinfo {pages} {1175} (\bibinfo {year}
  {2017})}\BibitemShut {NoStop}%
\bibitem [{\citenamefont {Ma}\ \emph {et~al.}(2019)\citenamefont {Ma},
  \citenamefont {Saxberg}, \citenamefont {Owens}, \citenamefont {Leung},
  \citenamefont {Lu}, \citenamefont {Simon},\ and\ \citenamefont
  {Schuster}}]{Ma2019}%
  \BibitemOpen
  \bibfield  {author} {\bibinfo {author} {\bibfnamefont {R.}~\bibnamefont
  {Ma}}, \bibinfo {author} {\bibfnamefont {B.}~\bibnamefont {Saxberg}},
  \bibinfo {author} {\bibfnamefont {C.}~\bibnamefont {Owens}}, \bibinfo
  {author} {\bibfnamefont {N.}~\bibnamefont {Leung}}, \bibinfo {author}
  {\bibfnamefont {Y.}~\bibnamefont {Lu}}, \bibinfo {author} {\bibfnamefont
  {J.}~\bibnamefont {Simon}},\ and\ \bibinfo {author} {\bibfnamefont {D.~I.}\
  \bibnamefont {Schuster}},\ }\bibfield  {title} {\bibinfo {title} {A
  dissipatively stabilized mott insulator of photons},\ }\href
  {https://doi.org/10.1038/s41586-019-0897-9} {\bibfield  {journal} {\bibinfo
  {journal} {Nature}\ }\textbf {\bibinfo {volume} {566}},\ \bibinfo {pages}
  {51} (\bibinfo {year} {2019})}\BibitemShut {NoStop}%
\bibitem [{\citenamefont {Carusotto}\ \emph {et~al.}(2020)\citenamefont
  {Carusotto}, \citenamefont {Houck}, \citenamefont {Koll{\'a}r}, \citenamefont
  {Roushan}, \citenamefont {Schuster},\ and\ \citenamefont
  {Simon}}]{Carusotto2020}%
  \BibitemOpen
  \bibfield  {author} {\bibinfo {author} {\bibfnamefont {I.}~\bibnamefont
  {Carusotto}}, \bibinfo {author} {\bibfnamefont {A.~A.}\ \bibnamefont
  {Houck}}, \bibinfo {author} {\bibfnamefont {A.~J.}\ \bibnamefont
  {Koll{\'a}r}}, \bibinfo {author} {\bibfnamefont {P.}~\bibnamefont {Roushan}},
  \bibinfo {author} {\bibfnamefont {D.~I.}\ \bibnamefont {Schuster}},\ and\
  \bibinfo {author} {\bibfnamefont {J.}~\bibnamefont {Simon}},\ }\bibfield
  {title} {\bibinfo {title} {Photonic materials in circuit quantum
  electrodynamics},\ }\href {https://doi.org/10.1038/s41567-020-0815-y}
  {\bibfield  {journal} {\bibinfo  {journal} {Nature Physics}\ }\textbf
  {\bibinfo {volume} {16}},\ \bibinfo {pages} {268} (\bibinfo {year}
  {2020})}\BibitemShut {NoStop}%
\bibitem [{\citenamefont {Gyenis}\ \emph
  {et~al.}(2021{\natexlab{a}})\citenamefont {Gyenis}, \citenamefont {Mundada},
  \citenamefont {Di~Paolo}, \citenamefont {Hazard}, \citenamefont {You},
  \citenamefont {Schuster}, \citenamefont {Koch}, \citenamefont {Blais},\ and\
  \citenamefont {Houck}}]{PRXQuantum.2.010339}%
  \BibitemOpen
  \bibfield  {author} {\bibinfo {author} {\bibfnamefont {A.}~\bibnamefont
  {Gyenis}}, \bibinfo {author} {\bibfnamefont {P.~S.}\ \bibnamefont {Mundada}},
  \bibinfo {author} {\bibfnamefont {A.}~\bibnamefont {Di~Paolo}}, \bibinfo
  {author} {\bibfnamefont {T.~M.}\ \bibnamefont {Hazard}}, \bibinfo {author}
  {\bibfnamefont {X.}~\bibnamefont {You}}, \bibinfo {author} {\bibfnamefont
  {D.~I.}\ \bibnamefont {Schuster}}, \bibinfo {author} {\bibfnamefont
  {J.}~\bibnamefont {Koch}}, \bibinfo {author} {\bibfnamefont {A.}~\bibnamefont
  {Blais}},\ and\ \bibinfo {author} {\bibfnamefont {A.~A.}\ \bibnamefont
  {Houck}},\ }\bibfield  {title} {\bibinfo {title} {Experimental realization of
  a protected superconducting circuit derived from the $0$--$\ensuremath{\pi}$
  qubit},\ }\href {https://doi.org/10.1103/PRXQuantum.2.010339} {\bibfield
  {journal} {\bibinfo  {journal} {PRX Quantum}\ }\textbf {\bibinfo {volume}
  {2}},\ \bibinfo {pages} {010339} (\bibinfo {year}
  {2021}{\natexlab{a}})}\BibitemShut {NoStop}%
\bibitem [{\citenamefont {Gyenis}\ \emph
  {et~al.}(2021{\natexlab{b}})\citenamefont {Gyenis}, \citenamefont {Paolo},
  \citenamefont {Koch}, \citenamefont {Blais}, \citenamefont {Houck},\ and\
  \citenamefont {Schuster}}]{gyenis2021moving}%
  \BibitemOpen
  \bibfield  {author} {\bibinfo {author} {\bibfnamefont {A.}~\bibnamefont
  {Gyenis}}, \bibinfo {author} {\bibfnamefont {A.~D.}\ \bibnamefont {Paolo}},
  \bibinfo {author} {\bibfnamefont {J.}~\bibnamefont {Koch}}, \bibinfo {author}
  {\bibfnamefont {A.}~\bibnamefont {Blais}}, \bibinfo {author} {\bibfnamefont
  {A.~A.}\ \bibnamefont {Houck}},\ and\ \bibinfo {author} {\bibfnamefont
  {D.~I.}\ \bibnamefont {Schuster}},\ }\href@noop {} {\bibinfo {title} {Moving
  beyond the transmon: Noise-protected superconducting quantum circuits}}
  (\bibinfo {year} {2021}{\natexlab{b}}),\ \Eprint
  {https://arxiv.org/abs/2106.10296} {arXiv:2106.10296 [quant-ph]} \BibitemShut
  {NoStop}%
\bibitem [{\citenamefont {Manucharyan}(2012)}]{Manucharyanthesis}%
  \BibitemOpen
  \bibfield  {author} {\bibinfo {author} {\bibfnamefont {V.~E.}\ \bibnamefont
  {Manucharyan}},\ }\emph {\bibinfo {title} {Superinductance}},\ \href@noop {}
  {Ph.D. thesis},\ \bibinfo  {school} {Yale University} (\bibinfo {year}
  {2012})\BibitemShut {NoStop}%
\bibitem [{\citenamefont {Pop}\ \emph {et~al.}(2014)\citenamefont {Pop},
  \citenamefont {Geerlings}, \citenamefont {Catelani}, \citenamefont
  {Schoelkopf}, \citenamefont {Glazman},\ and\ \citenamefont
  {Devoret}}]{Pop2014}%
  \BibitemOpen
  \bibfield  {author} {\bibinfo {author} {\bibfnamefont {I.~M.}\ \bibnamefont
  {Pop}}, \bibinfo {author} {\bibfnamefont {K.}~\bibnamefont {Geerlings}},
  \bibinfo {author} {\bibfnamefont {G.}~\bibnamefont {Catelani}}, \bibinfo
  {author} {\bibfnamefont {R.~J.}\ \bibnamefont {Schoelkopf}}, \bibinfo
  {author} {\bibfnamefont {L.~I.}\ \bibnamefont {Glazman}},\ and\ \bibinfo
  {author} {\bibfnamefont {M.~H.}\ \bibnamefont {Devoret}},\ }\bibfield
  {title} {\bibinfo {title} {Coherent suppression of electromagnetic
  dissipation due to superconducting quasiparticles},\ }\href
  {https://doi.org/10.1038/nature13017} {\bibfield  {journal} {\bibinfo
  {journal} {Nature}\ }\textbf {\bibinfo {volume} {508}},\ \bibinfo {pages}
  {369} (\bibinfo {year} {2014})}\BibitemShut {NoStop}%
\bibitem [{\citenamefont {Manucharyan}\ \emph {et~al.}(2009)\citenamefont
  {Manucharyan}, \citenamefont {Koch}, \citenamefont {Glazman},\ and\
  \citenamefont {Devoret}}]{Manucharyan113}%
  \BibitemOpen
  \bibfield  {author} {\bibinfo {author} {\bibfnamefont {V.~E.}\ \bibnamefont
  {Manucharyan}}, \bibinfo {author} {\bibfnamefont {J.}~\bibnamefont {Koch}},
  \bibinfo {author} {\bibfnamefont {L.~I.}\ \bibnamefont {Glazman}},\ and\
  \bibinfo {author} {\bibfnamefont {M.~H.}\ \bibnamefont {Devoret}},\
  }\bibfield  {title} {\bibinfo {title} {Fluxonium: Single cooper-pair circuit
  free of charge offsets},\ }\href {https://doi.org/10.1126/science.1175552}
  {\bibfield  {journal} {\bibinfo  {journal} {Science}\ }\textbf {\bibinfo
  {volume} {326}},\ \bibinfo {pages} {113} (\bibinfo {year}
  {2009})}\BibitemShut {NoStop}%
\bibitem [{\citenamefont {Brooks}\ \emph {et~al.}(2013)\citenamefont {Brooks},
  \citenamefont {Kitaev},\ and\ \citenamefont {Preskill}}]{Brooks2013}%
  \BibitemOpen
  \bibfield  {author} {\bibinfo {author} {\bibfnamefont {P.}~\bibnamefont
  {Brooks}}, \bibinfo {author} {\bibfnamefont {A.}~\bibnamefont {Kitaev}},\
  and\ \bibinfo {author} {\bibfnamefont {J.}~\bibnamefont {Preskill}},\
  }\bibfield  {title} {\bibinfo {title} {Protected gates for superconducting
  qubits},\ }\href {https://doi.org/10.1103/PhysRevA.87.052306} {\bibfield
  {journal} {\bibinfo  {journal} {Phys. Rev. A}\ }\textbf {\bibinfo {volume}
  {87}},\ \bibinfo {pages} {052306} (\bibinfo {year} {2013})}\BibitemShut
  {NoStop}%
\bibitem [{\citenamefont {Stockklauser}\ \emph {et~al.}(2017)\citenamefont
  {Stockklauser}, \citenamefont {Scarlino}, \citenamefont {Koski},
  \citenamefont {Gasparinetti}, \citenamefont {Andersen}, \citenamefont
  {Reichl}, \citenamefont {Wegscheider}, \citenamefont {Ihn}, \citenamefont
  {Ensslin},\ and\ \citenamefont {Wallraff}}]{Stockklauser2017}%
  \BibitemOpen
  \bibfield  {author} {\bibinfo {author} {\bibfnamefont {A.}~\bibnamefont
  {Stockklauser}}, \bibinfo {author} {\bibfnamefont {P.}~\bibnamefont
  {Scarlino}}, \bibinfo {author} {\bibfnamefont {J.~V.}\ \bibnamefont {Koski}},
  \bibinfo {author} {\bibfnamefont {S.}~\bibnamefont {Gasparinetti}}, \bibinfo
  {author} {\bibfnamefont {C.~K.}\ \bibnamefont {Andersen}}, \bibinfo {author}
  {\bibfnamefont {C.}~\bibnamefont {Reichl}}, \bibinfo {author} {\bibfnamefont
  {W.}~\bibnamefont {Wegscheider}}, \bibinfo {author} {\bibfnamefont
  {T.}~\bibnamefont {Ihn}}, \bibinfo {author} {\bibfnamefont {K.}~\bibnamefont
  {Ensslin}},\ and\ \bibinfo {author} {\bibfnamefont {A.}~\bibnamefont
  {Wallraff}},\ }\bibfield  {title} {\bibinfo {title} {Strong coupling cavity
  qed with gate-defined double quantum dots enabled by a high impedance
  resonator},\ }\href {https://doi.org/10.1103/PhysRevX.7.011030} {\bibfield
  {journal} {\bibinfo  {journal} {Phys. Rev. X}\ }\textbf {\bibinfo {volume}
  {7}},\ \bibinfo {pages} {011030} (\bibinfo {year} {2017})}\BibitemShut
  {NoStop}%
\bibitem [{\citenamefont {Puertas~Mart{\'i}nez}\ \emph
  {et~al.}(2019)\citenamefont {Puertas~Mart{\'i}nez}, \citenamefont
  {L{\'e}ger}, \citenamefont {Gheeraert}, \citenamefont {Dassonneville},
  \citenamefont {Planat}, \citenamefont {Foroughi}, \citenamefont {Krupko},
  \citenamefont {Buisson}, \citenamefont {Naud}, \citenamefont
  {Hasch-Guichard}, \citenamefont {Florens}, \citenamefont {Snyman},\ and\
  \citenamefont {Roch}}]{PuertasMartinez2019}%
  \BibitemOpen
  \bibfield  {author} {\bibinfo {author} {\bibfnamefont {J.}~\bibnamefont
  {Puertas~Mart{\'i}nez}}, \bibinfo {author} {\bibfnamefont {S.}~\bibnamefont
  {L{\'e}ger}}, \bibinfo {author} {\bibfnamefont {N.}~\bibnamefont
  {Gheeraert}}, \bibinfo {author} {\bibfnamefont {R.}~\bibnamefont
  {Dassonneville}}, \bibinfo {author} {\bibfnamefont {L.}~\bibnamefont
  {Planat}}, \bibinfo {author} {\bibfnamefont {F.}~\bibnamefont {Foroughi}},
  \bibinfo {author} {\bibfnamefont {Y.}~\bibnamefont {Krupko}}, \bibinfo
  {author} {\bibfnamefont {O.}~\bibnamefont {Buisson}}, \bibinfo {author}
  {\bibfnamefont {C.}~\bibnamefont {Naud}}, \bibinfo {author} {\bibfnamefont
  {W.}~\bibnamefont {Hasch-Guichard}}, \bibinfo {author} {\bibfnamefont
  {S.}~\bibnamefont {Florens}}, \bibinfo {author} {\bibfnamefont
  {I.}~\bibnamefont {Snyman}},\ and\ \bibinfo {author} {\bibfnamefont
  {N.}~\bibnamefont {Roch}},\ }\bibfield  {title} {\bibinfo {title} {A tunable
  josephson platform to explore many-body quantum optics in circuit-qed},\
  }\href {https://doi.org/10.1038/s41534-018-0104-0} {\bibfield  {journal}
  {\bibinfo  {journal} {npj Quantum Information}\ }\textbf {\bibinfo {volume}
  {5}},\ \bibinfo {pages} {19} (\bibinfo {year} {2019})}\BibitemShut {NoStop}%
\bibitem [{\citenamefont {Bell}\ \emph {et~al.}(2012)\citenamefont {Bell},
  \citenamefont {Sadovskyy}, \citenamefont {Ioffe}, \citenamefont {Kitaev},\
  and\ \citenamefont {Gershenson}}]{Bell2012}%
  \BibitemOpen
  \bibfield  {author} {\bibinfo {author} {\bibfnamefont {M.~T.}\ \bibnamefont
  {Bell}}, \bibinfo {author} {\bibfnamefont {I.~A.}\ \bibnamefont {Sadovskyy}},
  \bibinfo {author} {\bibfnamefont {L.~B.}\ \bibnamefont {Ioffe}}, \bibinfo
  {author} {\bibfnamefont {A.~Y.}\ \bibnamefont {Kitaev}},\ and\ \bibinfo
  {author} {\bibfnamefont {M.~E.}\ \bibnamefont {Gershenson}},\ }\bibfield
  {title} {\bibinfo {title} {Quantum superinductor with tunable nonlinearity},\
  }\href {https://doi.org/10.1103/PhysRevLett.109.137003} {\bibfield  {journal}
  {\bibinfo  {journal} {Phys. Rev. Lett.}\ }\textbf {\bibinfo {volume} {109}},\
  \bibinfo {pages} {137003} (\bibinfo {year} {2012})}\BibitemShut {NoStop}%
\bibitem [{\citenamefont {Masluk}\ \emph {et~al.}(2012)\citenamefont {Masluk},
  \citenamefont {Pop}, \citenamefont {Kamal}, \citenamefont {Minev},\ and\
  \citenamefont {Devoret}}]{Masluk2012}%
  \BibitemOpen
  \bibfield  {author} {\bibinfo {author} {\bibfnamefont {N.~A.}\ \bibnamefont
  {Masluk}}, \bibinfo {author} {\bibfnamefont {I.~M.}\ \bibnamefont {Pop}},
  \bibinfo {author} {\bibfnamefont {A.}~\bibnamefont {Kamal}}, \bibinfo
  {author} {\bibfnamefont {Z.~K.}\ \bibnamefont {Minev}},\ and\ \bibinfo
  {author} {\bibfnamefont {M.~H.}\ \bibnamefont {Devoret}},\ }\bibfield
  {title} {\bibinfo {title} {Microwave characterization of josephson junction
  arrays: Implementing a low loss superinductance},\ }\href
  {https://doi.org/10.1103/PhysRevLett.109.137002} {\bibfield  {journal}
  {\bibinfo  {journal} {Phys. Rev. Lett.}\ }\textbf {\bibinfo {volume} {109}},\
  \bibinfo {pages} {137002} (\bibinfo {year} {2012})}\BibitemShut {NoStop}%
\bibitem [{\citenamefont {Krupko}\ \emph {et~al.}(2018)\citenamefont {Krupko},
  \citenamefont {Nguyen}, \citenamefont {Wei\ss{}l}, \citenamefont {Dumur},
  \citenamefont {Puertas}, \citenamefont {Dassonneville}, \citenamefont {Naud},
  \citenamefont {Hekking}, \citenamefont {Basko}, \citenamefont {Buisson},
  \citenamefont {Roch},\ and\ \citenamefont
  {Hasch-Guichard}}]{PhysRevB.98.094516}%
  \BibitemOpen
  \bibfield  {author} {\bibinfo {author} {\bibfnamefont {Y.}~\bibnamefont
  {Krupko}}, \bibinfo {author} {\bibfnamefont {V.~D.}\ \bibnamefont {Nguyen}},
  \bibinfo {author} {\bibfnamefont {T.}~\bibnamefont {Wei\ss{}l}}, \bibinfo
  {author} {\bibfnamefont {E.}~\bibnamefont {Dumur}}, \bibinfo {author}
  {\bibfnamefont {J.}~\bibnamefont {Puertas}}, \bibinfo {author} {\bibfnamefont
  {R.}~\bibnamefont {Dassonneville}}, \bibinfo {author} {\bibfnamefont
  {C.}~\bibnamefont {Naud}}, \bibinfo {author} {\bibfnamefont {F.~W.~J.}\
  \bibnamefont {Hekking}}, \bibinfo {author} {\bibfnamefont {D.~M.}\
  \bibnamefont {Basko}}, \bibinfo {author} {\bibfnamefont {O.}~\bibnamefont
  {Buisson}}, \bibinfo {author} {\bibfnamefont {N.}~\bibnamefont {Roch}},\ and\
  \bibinfo {author} {\bibfnamefont {W.}~\bibnamefont {Hasch-Guichard}},\
  }\bibfield  {title} {\bibinfo {title} {Kerr nonlinearity in a superconducting
  josephson metamaterial},\ }\href {https://doi.org/10.1103/PhysRevB.98.094516}
  {\bibfield  {journal} {\bibinfo  {journal} {Phys. Rev. B}\ }\textbf {\bibinfo
  {volume} {98}},\ \bibinfo {pages} {094516} (\bibinfo {year}
  {2018})}\BibitemShut {NoStop}%
\bibitem [{\citenamefont {Gr{\"u}nhaupt}\ \emph {et~al.}(2019)\citenamefont
  {Gr{\"u}nhaupt}, \citenamefont {Spiecker}, \citenamefont {Gusenkova},
  \citenamefont {Maleeva}, \citenamefont {Skacel}, \citenamefont {Takmakov},
  \citenamefont {Valenti}, \citenamefont {Winkel}, \citenamefont {Rotzinger},
  \citenamefont {Wernsdorfer}, \citenamefont {Ustinov},\ and\ \citenamefont
  {Pop}}]{Gruenhaupt2019}%
  \BibitemOpen
  \bibfield  {author} {\bibinfo {author} {\bibfnamefont {L.}~\bibnamefont
  {Gr{\"u}nhaupt}}, \bibinfo {author} {\bibfnamefont {M.}~\bibnamefont
  {Spiecker}}, \bibinfo {author} {\bibfnamefont {D.}~\bibnamefont {Gusenkova}},
  \bibinfo {author} {\bibfnamefont {N.}~\bibnamefont {Maleeva}}, \bibinfo
  {author} {\bibfnamefont {S.~T.}\ \bibnamefont {Skacel}}, \bibinfo {author}
  {\bibfnamefont {I.}~\bibnamefont {Takmakov}}, \bibinfo {author}
  {\bibfnamefont {F.}~\bibnamefont {Valenti}}, \bibinfo {author} {\bibfnamefont
  {P.}~\bibnamefont {Winkel}}, \bibinfo {author} {\bibfnamefont
  {H.}~\bibnamefont {Rotzinger}}, \bibinfo {author} {\bibfnamefont
  {W.}~\bibnamefont {Wernsdorfer}}, \bibinfo {author} {\bibfnamefont {A.~V.}\
  \bibnamefont {Ustinov}},\ and\ \bibinfo {author} {\bibfnamefont {I.~M.}\
  \bibnamefont {Pop}},\ }\bibfield  {title} {\bibinfo {title} {Granular
  aluminium as a superconducting material for high-impedance quantum
  circuits},\ }\href {https://doi.org/10.1038/s41563-019-0350-3} {\bibfield
  {journal} {\bibinfo  {journal} {Nature Materials}\ }\textbf {\bibinfo
  {volume} {18}},\ \bibinfo {pages} {816} (\bibinfo {year} {2019})}\BibitemShut
  {NoStop}%
\bibitem [{\citenamefont {Place}\ \emph {et~al.}(2021)\citenamefont {Place},
  \citenamefont {Rodgers}, \citenamefont {Mundada}, \citenamefont {Smitham},
  \citenamefont {Fitzpatrick}, \citenamefont {Leng}, \citenamefont {Premkumar},
  \citenamefont {Bryon}, \citenamefont {Vrajitoarea}, \citenamefont {Sussman},
  \citenamefont {Cheng}, \citenamefont {Madhavan}, \citenamefont {Babla},
  \citenamefont {Le}, \citenamefont {Gang}, \citenamefont {Jäck},
  \citenamefont {Gyenis}, \citenamefont {Yao}, \citenamefont {Cava},
  \citenamefont {de~Leon},\ and\ \citenamefont {Houck}}]{Place2021}%
  \BibitemOpen
  \bibfield  {author} {\bibinfo {author} {\bibfnamefont {A.~P.~M.}\
  \bibnamefont {Place}}, \bibinfo {author} {\bibfnamefont {L.~V.~H.}\
  \bibnamefont {Rodgers}}, \bibinfo {author} {\bibfnamefont {P.}~\bibnamefont
  {Mundada}}, \bibinfo {author} {\bibfnamefont {B.~M.}\ \bibnamefont
  {Smitham}}, \bibinfo {author} {\bibfnamefont {M.}~\bibnamefont
  {Fitzpatrick}}, \bibinfo {author} {\bibfnamefont {Z.}~\bibnamefont {Leng}},
  \bibinfo {author} {\bibfnamefont {A.}~\bibnamefont {Premkumar}}, \bibinfo
  {author} {\bibfnamefont {J.}~\bibnamefont {Bryon}}, \bibinfo {author}
  {\bibfnamefont {A.}~\bibnamefont {Vrajitoarea}}, \bibinfo {author}
  {\bibfnamefont {S.}~\bibnamefont {Sussman}}, \bibinfo {author} {\bibfnamefont
  {G.}~\bibnamefont {Cheng}}, \bibinfo {author} {\bibfnamefont
  {T.}~\bibnamefont {Madhavan}}, \bibinfo {author} {\bibfnamefont {H.~K.}\
  \bibnamefont {Babla}}, \bibinfo {author} {\bibfnamefont {X.~H.}\ \bibnamefont
  {Le}}, \bibinfo {author} {\bibfnamefont {Y.}~\bibnamefont {Gang}}, \bibinfo
  {author} {\bibfnamefont {B.}~\bibnamefont {Jäck}}, \bibinfo {author}
  {\bibfnamefont {A.}~\bibnamefont {Gyenis}}, \bibinfo {author} {\bibfnamefont
  {N.}~\bibnamefont {Yao}}, \bibinfo {author} {\bibfnamefont {R.~J.}\
  \bibnamefont {Cava}}, \bibinfo {author} {\bibfnamefont {N.~P.}\ \bibnamefont
  {de~Leon}},\ and\ \bibinfo {author} {\bibfnamefont {A.~A.}\ \bibnamefont
  {Houck}},\ }\bibfield  {title} {\bibinfo {title} {New material platform for
  superconducting transmon qubits with coherence times exceeding 0.3
  milliseconds},\ }\href {https://doi.org/10.1038/s41467-021-22030-5}
  {\bibfield  {journal} {\bibinfo  {journal} {Nature Communications}\ }\textbf
  {\bibinfo {volume} {12}},\ \bibinfo {pages} {1779} (\bibinfo {year}
  {2021})}\BibitemShut {NoStop}%
\bibitem [{\citenamefont {Wang}\ \emph {et~al.}(2021)\citenamefont {Wang},
  \citenamefont {Li}, \citenamefont {Xu}, \citenamefont {Li}, \citenamefont
  {Wang}, \citenamefont {Yang}, \citenamefont {Mi}, \citenamefont {Liang},
  \citenamefont {Su}, \citenamefont {Yang}, \citenamefont {Wang}, \citenamefont
  {Wang}, \citenamefont {Li}, \citenamefont {Chen}, \citenamefont {Li},
  \citenamefont {Linghu}, \citenamefont {Han}, \citenamefont {Zhang},
  \citenamefont {Feng}, \citenamefont {Song}, \citenamefont {Ma}, \citenamefont
  {Zhang}, \citenamefont {Wang}, \citenamefont {Zhao}, \citenamefont {Liu},
  \citenamefont {Xue}, \citenamefont {Jin},\ and\ \citenamefont
  {Yu}}]{wang2021transmon}%
  \BibitemOpen
  \bibfield  {author} {\bibinfo {author} {\bibfnamefont {C.}~\bibnamefont
  {Wang}}, \bibinfo {author} {\bibfnamefont {X.}~\bibnamefont {Li}}, \bibinfo
  {author} {\bibfnamefont {H.}~\bibnamefont {Xu}}, \bibinfo {author}
  {\bibfnamefont {Z.}~\bibnamefont {Li}}, \bibinfo {author} {\bibfnamefont
  {J.}~\bibnamefont {Wang}}, \bibinfo {author} {\bibfnamefont {Z.}~\bibnamefont
  {Yang}}, \bibinfo {author} {\bibfnamefont {Z.}~\bibnamefont {Mi}}, \bibinfo
  {author} {\bibfnamefont {X.}~\bibnamefont {Liang}}, \bibinfo {author}
  {\bibfnamefont {T.}~\bibnamefont {Su}}, \bibinfo {author} {\bibfnamefont
  {C.}~\bibnamefont {Yang}}, \bibinfo {author} {\bibfnamefont {G.}~\bibnamefont
  {Wang}}, \bibinfo {author} {\bibfnamefont {W.}~\bibnamefont {Wang}}, \bibinfo
  {author} {\bibfnamefont {Y.}~\bibnamefont {Li}}, \bibinfo {author}
  {\bibfnamefont {M.}~\bibnamefont {Chen}}, \bibinfo {author} {\bibfnamefont
  {C.}~\bibnamefont {Li}}, \bibinfo {author} {\bibfnamefont {K.}~\bibnamefont
  {Linghu}}, \bibinfo {author} {\bibfnamefont {J.}~\bibnamefont {Han}},
  \bibinfo {author} {\bibfnamefont {Y.}~\bibnamefont {Zhang}}, \bibinfo
  {author} {\bibfnamefont {Y.}~\bibnamefont {Feng}}, \bibinfo {author}
  {\bibfnamefont {Y.}~\bibnamefont {Song}}, \bibinfo {author} {\bibfnamefont
  {T.}~\bibnamefont {Ma}}, \bibinfo {author} {\bibfnamefont {J.}~\bibnamefont
  {Zhang}}, \bibinfo {author} {\bibfnamefont {R.}~\bibnamefont {Wang}},
  \bibinfo {author} {\bibfnamefont {P.}~\bibnamefont {Zhao}}, \bibinfo {author}
  {\bibfnamefont {W.}~\bibnamefont {Liu}}, \bibinfo {author} {\bibfnamefont
  {G.}~\bibnamefont {Xue}}, \bibinfo {author} {\bibfnamefont {Y.}~\bibnamefont
  {Jin}},\ and\ \bibinfo {author} {\bibfnamefont {H.}~\bibnamefont {Yu}},\
  }\href@noop {} {\bibinfo {title} {Transmon qubit with relaxation time
  exceeding 0.5 milliseconds}} (\bibinfo {year} {2021}),\ \Eprint
  {https://arxiv.org/abs/2105.09890} {arXiv:2105.09890 [quant-ph]} \BibitemShut
  {NoStop}%
\bibitem [{\citenamefont {{Makise}}\ \emph {et~al.}(2015)\citenamefont
  {{Makise}}, \citenamefont {{Sun}}, \citenamefont {{Terai}},\ and\
  \citenamefont {{Wang}}}]{6933905}%
  \BibitemOpen
  \bibfield  {author} {\bibinfo {author} {\bibfnamefont {K.}~\bibnamefont
  {{Makise}}}, \bibinfo {author} {\bibfnamefont {R.}~\bibnamefont {{Sun}}},
  \bibinfo {author} {\bibfnamefont {H.}~\bibnamefont {{Terai}}},\ and\ \bibinfo
  {author} {\bibfnamefont {Z.}~\bibnamefont {{Wang}}},\ }\bibfield  {title}
  {\bibinfo {title} {Fabrication and characterization of epitaxial tin-based
  josephson junctions for superconducting circuit applications},\ }\href@noop
  {} {\bibfield  {journal} {\bibinfo  {journal} {IEEE Transactions on Applied
  Superconductivity}\ }\textbf {\bibinfo {volume} {25}},\ \bibinfo {pages} {1}
  (\bibinfo {year} {2015})}\BibitemShut {NoStop}%
\bibitem [{\citenamefont {Vissers}\ \emph {et~al.}(2010)\citenamefont
  {Vissers}, \citenamefont {Gao}, \citenamefont {Wisbey}, \citenamefont {Hite},
  \citenamefont {Tsuei}, \citenamefont {Corcoles}, \citenamefont {Steffen},\
  and\ \citenamefont {Pappas}}]{doi:10.1063/1.3517252}%
  \BibitemOpen
  \bibfield  {author} {\bibinfo {author} {\bibfnamefont {M.~R.}\ \bibnamefont
  {Vissers}}, \bibinfo {author} {\bibfnamefont {J.}~\bibnamefont {Gao}},
  \bibinfo {author} {\bibfnamefont {D.~S.}\ \bibnamefont {Wisbey}}, \bibinfo
  {author} {\bibfnamefont {D.~A.}\ \bibnamefont {Hite}}, \bibinfo {author}
  {\bibfnamefont {C.~C.}\ \bibnamefont {Tsuei}}, \bibinfo {author}
  {\bibfnamefont {A.~D.}\ \bibnamefont {Corcoles}}, \bibinfo {author}
  {\bibfnamefont {M.}~\bibnamefont {Steffen}},\ and\ \bibinfo {author}
  {\bibfnamefont {D.~P.}\ \bibnamefont {Pappas}},\ }\bibfield  {title}
  {\bibinfo {title} {Low loss superconducting titanium nitride coplanar
  waveguide resonators},\ }\href {https://doi.org/10.1063/1.3517252} {\bibfield
   {journal} {\bibinfo  {journal} {Applied Physics Letters}\ }\textbf {\bibinfo
  {volume} {97}},\ \bibinfo {pages} {232509} (\bibinfo {year} {2010})},\
  \Eprint {https://arxiv.org/abs/https://doi.org/10.1063/1.3517252}
  {https://doi.org/10.1063/1.3517252} \BibitemShut {NoStop}%
\bibitem [{\citenamefont {Leduc}\ \emph {et~al.}(2010)\citenamefont {Leduc},
  \citenamefont {Bumble}, \citenamefont {Day}, \citenamefont {Eom},
  \citenamefont {Gao}, \citenamefont {Golwala}, \citenamefont {Mazin},
  \citenamefont {McHugh}, \citenamefont {Merrill}, \citenamefont {Moore},
  \citenamefont {Noroozian}, \citenamefont {Turner},\ and\ \citenamefont
  {Zmuidzinas}}]{doi:10.1063/1.3480420}%
  \BibitemOpen
  \bibfield  {author} {\bibinfo {author} {\bibfnamefont {H.~G.}\ \bibnamefont
  {Leduc}}, \bibinfo {author} {\bibfnamefont {B.}~\bibnamefont {Bumble}},
  \bibinfo {author} {\bibfnamefont {P.~K.}\ \bibnamefont {Day}}, \bibinfo
  {author} {\bibfnamefont {B.~H.}\ \bibnamefont {Eom}}, \bibinfo {author}
  {\bibfnamefont {J.}~\bibnamefont {Gao}}, \bibinfo {author} {\bibfnamefont
  {S.}~\bibnamefont {Golwala}}, \bibinfo {author} {\bibfnamefont {B.~A.}\
  \bibnamefont {Mazin}}, \bibinfo {author} {\bibfnamefont {S.}~\bibnamefont
  {McHugh}}, \bibinfo {author} {\bibfnamefont {A.}~\bibnamefont {Merrill}},
  \bibinfo {author} {\bibfnamefont {D.~C.}\ \bibnamefont {Moore}}, \bibinfo
  {author} {\bibfnamefont {O.}~\bibnamefont {Noroozian}}, \bibinfo {author}
  {\bibfnamefont {A.~D.}\ \bibnamefont {Turner}},\ and\ \bibinfo {author}
  {\bibfnamefont {J.}~\bibnamefont {Zmuidzinas}},\ }\bibfield  {title}
  {\bibinfo {title} {Titanium nitride films for ultrasensitive microresonator
  detectors},\ }\href {https://doi.org/10.1063/1.3480420} {\bibfield  {journal}
  {\bibinfo  {journal} {Applied Physics Letters}\ }\textbf {\bibinfo {volume}
  {97}},\ \bibinfo {pages} {102509} (\bibinfo {year} {2010})},\ \Eprint
  {https://arxiv.org/abs/https://doi.org/10.1063/1.3480420}
  {https://doi.org/10.1063/1.3480420} \BibitemShut {NoStop}%
\bibitem [{\citenamefont {Shearrow}\ \emph {et~al.}(2018)\citenamefont
  {Shearrow}, \citenamefont {Koolstra}, \citenamefont {Whiteley}, \citenamefont
  {Earnest}, \citenamefont {Barry}, \citenamefont {Heremans}, \citenamefont
  {Awschalom}, \citenamefont {Shirokoff},\ and\ \citenamefont
  {Schuster}}]{doi:10.1063/1.5053461}%
  \BibitemOpen
  \bibfield  {author} {\bibinfo {author} {\bibfnamefont {A.}~\bibnamefont
  {Shearrow}}, \bibinfo {author} {\bibfnamefont {G.}~\bibnamefont {Koolstra}},
  \bibinfo {author} {\bibfnamefont {S.~J.}\ \bibnamefont {Whiteley}}, \bibinfo
  {author} {\bibfnamefont {N.}~\bibnamefont {Earnest}}, \bibinfo {author}
  {\bibfnamefont {P.~S.}\ \bibnamefont {Barry}}, \bibinfo {author}
  {\bibfnamefont {F.~J.}\ \bibnamefont {Heremans}}, \bibinfo {author}
  {\bibfnamefont {D.~D.}\ \bibnamefont {Awschalom}}, \bibinfo {author}
  {\bibfnamefont {E.}~\bibnamefont {Shirokoff}},\ and\ \bibinfo {author}
  {\bibfnamefont {D.~I.}\ \bibnamefont {Schuster}},\ }\bibfield  {title}
  {\bibinfo {title} {Atomic layer deposition of titanium nitride for quantum
  circuits},\ }\href {https://doi.org/10.1063/1.5053461} {\bibfield  {journal}
  {\bibinfo  {journal} {Applied Physics Letters}\ }\textbf {\bibinfo {volume}
  {113}},\ \bibinfo {pages} {212601} (\bibinfo {year} {2018})},\ \Eprint
  {https://arxiv.org/abs/https://doi.org/10.1063/1.5053461}
  {https://doi.org/10.1063/1.5053461} \BibitemShut {NoStop}%
\bibitem [{\citenamefont {Samkharadze}\ \emph {et~al.}(2016)\citenamefont
  {Samkharadze}, \citenamefont {Bruno}, \citenamefont {Scarlino}, \citenamefont
  {Zheng}, \citenamefont {DiVincenzo}, \citenamefont {DiCarlo},\ and\
  \citenamefont {Vandersypen}}]{PhysRevApplied.5.044004}%
  \BibitemOpen
  \bibfield  {author} {\bibinfo {author} {\bibfnamefont {N.}~\bibnamefont
  {Samkharadze}}, \bibinfo {author} {\bibfnamefont {A.}~\bibnamefont {Bruno}},
  \bibinfo {author} {\bibfnamefont {P.}~\bibnamefont {Scarlino}}, \bibinfo
  {author} {\bibfnamefont {G.}~\bibnamefont {Zheng}}, \bibinfo {author}
  {\bibfnamefont {D.~P.}\ \bibnamefont {DiVincenzo}}, \bibinfo {author}
  {\bibfnamefont {L.}~\bibnamefont {DiCarlo}},\ and\ \bibinfo {author}
  {\bibfnamefont {L.~M.~K.}\ \bibnamefont {Vandersypen}},\ }\bibfield  {title}
  {\bibinfo {title} {High-kinetic-inductance superconducting nanowire
  resonators for circuit qed in a magnetic field},\ }\href
  {https://doi.org/10.1103/PhysRevApplied.5.044004} {\bibfield  {journal}
  {\bibinfo  {journal} {Phys. Rev. Applied}\ }\textbf {\bibinfo {volume} {5}},\
  \bibinfo {pages} {044004} (\bibinfo {year} {2016})}\BibitemShut {NoStop}%
\bibitem [{\citenamefont {Niepce}\ \emph {et~al.}(2019)\citenamefont {Niepce},
  \citenamefont {Burnett},\ and\ \citenamefont
  {Bylander}}]{PhysRevApplied.11.044014}%
  \BibitemOpen
  \bibfield  {author} {\bibinfo {author} {\bibfnamefont {D.}~\bibnamefont
  {Niepce}}, \bibinfo {author} {\bibfnamefont {J.}~\bibnamefont {Burnett}},\
  and\ \bibinfo {author} {\bibfnamefont {J.}~\bibnamefont {Bylander}},\
  }\bibfield  {title} {\bibinfo {title} {High kinetic inductance
  $\mathrm{Nb}\mathrm{N}$ nanowire superinductors},\ }\href
  {https://doi.org/10.1103/PhysRevApplied.11.044014} {\bibfield  {journal}
  {\bibinfo  {journal} {Phys. Rev. Applied}\ }\textbf {\bibinfo {volume}
  {11}},\ \bibinfo {pages} {044014} (\bibinfo {year} {2019})}\BibitemShut
  {NoStop}%
\bibitem [{\citenamefont {Yu}\ \emph {et~al.}(2021)\citenamefont {Yu},
  \citenamefont {Zihlmann}, \citenamefont {Troncoso Fern\'andez-Bada},
  \citenamefont {Thomassin}, \citenamefont {Gustavo}, \citenamefont {Dumur},\
  and\ \citenamefont {Maurand}}]{Yu2021}%
  \BibitemOpen
  \bibfield  {author} {\bibinfo {author} {\bibfnamefont {C.~X.}\ \bibnamefont
  {Yu}}, \bibinfo {author} {\bibfnamefont {S.}~\bibnamefont {Zihlmann}},
  \bibinfo {author} {\bibfnamefont {G.}~\bibnamefont {Troncoso
  Fern\'andez-Bada}}, \bibinfo {author} {\bibfnamefont {J.-L.}\ \bibnamefont
  {Thomassin}}, \bibinfo {author} {\bibfnamefont {F.}~\bibnamefont {Gustavo}},
  \bibinfo {author} {\bibfnamefont {E.}~\bibnamefont {Dumur}},\ and\ \bibinfo
  {author} {\bibfnamefont {R.}~\bibnamefont {Maurand}},\ }\bibfield  {title}
  {\bibinfo {title} {Magnetic field resilient high kinetic inductance
  superconducting niobium nitride coplanar waveguide resonators},\ }\href
  {https://doi.org/10.1063/5.0039945} {\bibfield  {journal} {\bibinfo
  {journal} {Applied Physics Letters}\ }\textbf {\bibinfo {volume} {118}},\
  \bibinfo {pages} {054001} (\bibinfo {year} {2021})},\ \Eprint
  {https://arxiv.org/abs/https://doi.org/10.1063/5.0039945}
  {https://doi.org/10.1063/5.0039945} \BibitemShut {NoStop}%
\bibitem [{\citenamefont {Gr\"unhaupt}\ \emph {et~al.}(2018)\citenamefont
  {Gr\"unhaupt}, \citenamefont {Maleeva}, \citenamefont {Skacel}, \citenamefont
  {Calvo}, \citenamefont {Levy-Bertrand}, \citenamefont {Ustinov},
  \citenamefont {Rotzinger}, \citenamefont {Monfardini}, \citenamefont
  {Catelani},\ and\ \citenamefont {Pop}}]{PhysRevLett.121.117001}%
  \BibitemOpen
  \bibfield  {author} {\bibinfo {author} {\bibfnamefont {L.}~\bibnamefont
  {Gr\"unhaupt}}, \bibinfo {author} {\bibfnamefont {N.}~\bibnamefont
  {Maleeva}}, \bibinfo {author} {\bibfnamefont {S.~T.}\ \bibnamefont {Skacel}},
  \bibinfo {author} {\bibfnamefont {M.}~\bibnamefont {Calvo}}, \bibinfo
  {author} {\bibfnamefont {F.}~\bibnamefont {Levy-Bertrand}}, \bibinfo {author}
  {\bibfnamefont {A.~V.}\ \bibnamefont {Ustinov}}, \bibinfo {author}
  {\bibfnamefont {H.}~\bibnamefont {Rotzinger}}, \bibinfo {author}
  {\bibfnamefont {A.}~\bibnamefont {Monfardini}}, \bibinfo {author}
  {\bibfnamefont {G.}~\bibnamefont {Catelani}},\ and\ \bibinfo {author}
  {\bibfnamefont {I.~M.}\ \bibnamefont {Pop}},\ }\bibfield  {title} {\bibinfo
  {title} {Loss mechanisms and quasiparticle dynamics in superconducting
  microwave resonators made of thin-film granular aluminum},\ }\href
  {https://doi.org/10.1103/PhysRevLett.121.117001} {\bibfield  {journal}
  {\bibinfo  {journal} {Phys. Rev. Lett.}\ }\textbf {\bibinfo {volume} {121}},\
  \bibinfo {pages} {117001} (\bibinfo {year} {2018})}\BibitemShut {NoStop}%
\bibitem [{\citenamefont {Zhang}\ \emph {et~al.}(2019)\citenamefont {Zhang},
  \citenamefont {Kalashnikov}, \citenamefont {Lu}, \citenamefont {Kamenov},
  \citenamefont {DiNapoli},\ and\ \citenamefont
  {Gershenson}}]{PhysRevApplied.11.011003}%
  \BibitemOpen
  \bibfield  {author} {\bibinfo {author} {\bibfnamefont {W.}~\bibnamefont
  {Zhang}}, \bibinfo {author} {\bibfnamefont {K.}~\bibnamefont {Kalashnikov}},
  \bibinfo {author} {\bibfnamefont {W.-S.}\ \bibnamefont {Lu}}, \bibinfo
  {author} {\bibfnamefont {P.}~\bibnamefont {Kamenov}}, \bibinfo {author}
  {\bibfnamefont {T.}~\bibnamefont {DiNapoli}},\ and\ \bibinfo {author}
  {\bibfnamefont {M.}~\bibnamefont {Gershenson}},\ }\bibfield  {title}
  {\bibinfo {title} {Microresonators fabricated from high-kinetic-inductance
  aluminum films},\ }\href {https://doi.org/10.1103/PhysRevApplied.11.011003}
  {\bibfield  {journal} {\bibinfo  {journal} {Phys. Rev. Applied}\ }\textbf
  {\bibinfo {volume} {11}},\ \bibinfo {pages} {011003} (\bibinfo {year}
  {2019})}\BibitemShut {NoStop}%
\bibitem [{\citenamefont {Astafiev}\ \emph {et~al.}(2012)\citenamefont
  {Astafiev}, \citenamefont {Ioffe}, \citenamefont {Kafanov}, \citenamefont
  {Pashkin}, \citenamefont {Arutyunov}, \citenamefont {Shahar}, \citenamefont
  {Cohen},\ and\ \citenamefont {Tsai}}]{Astafiev2012}%
  \BibitemOpen
  \bibfield  {author} {\bibinfo {author} {\bibfnamefont {O.~V.}\ \bibnamefont
  {Astafiev}}, \bibinfo {author} {\bibfnamefont {L.~B.}\ \bibnamefont {Ioffe}},
  \bibinfo {author} {\bibfnamefont {S.}~\bibnamefont {Kafanov}}, \bibinfo
  {author} {\bibfnamefont {Y.~A.}\ \bibnamefont {Pashkin}}, \bibinfo {author}
  {\bibfnamefont {K.~Y.}\ \bibnamefont {Arutyunov}}, \bibinfo {author}
  {\bibfnamefont {D.}~\bibnamefont {Shahar}}, \bibinfo {author} {\bibfnamefont
  {O.}~\bibnamefont {Cohen}},\ and\ \bibinfo {author} {\bibfnamefont {J.~S.}\
  \bibnamefont {Tsai}},\ }\bibfield  {title} {\bibinfo {title} {Coherent
  quantum phase slip},\ }\href {https://doi.org/10.1038/nature10930} {\bibfield
   {journal} {\bibinfo  {journal} {Nature}\ }\textbf {\bibinfo {volume}
  {484}},\ \bibinfo {pages} {355} (\bibinfo {year} {2012})}\BibitemShut
  {NoStop}%
\bibitem [{\citenamefont {Dupr{\'{e}}}\ \emph {et~al.}(2017)\citenamefont
  {Dupr{\'{e}}}, \citenamefont {Beno{\^{\i}}t}, \citenamefont {Calvo},
  \citenamefont {Catalano}, \citenamefont {Goupy}, \citenamefont {Hoarau},
  \citenamefont {Klein}, \citenamefont {Calvez}, \citenamefont
  {Sac{\'{e}}p{\'{e}}}, \citenamefont {Monfardini},\ and\ \citenamefont
  {Levy-Bertrand}}]{Dupre2017}%
  \BibitemOpen
  \bibfield  {author} {\bibinfo {author} {\bibfnamefont {O.}~\bibnamefont
  {Dupr{\'{e}}}}, \bibinfo {author} {\bibfnamefont {A.}~\bibnamefont
  {Beno{\^{\i}}t}}, \bibinfo {author} {\bibfnamefont {M.}~\bibnamefont
  {Calvo}}, \bibinfo {author} {\bibfnamefont {A.}~\bibnamefont {Catalano}},
  \bibinfo {author} {\bibfnamefont {J.}~\bibnamefont {Goupy}}, \bibinfo
  {author} {\bibfnamefont {C.}~\bibnamefont {Hoarau}}, \bibinfo {author}
  {\bibfnamefont {T.}~\bibnamefont {Klein}}, \bibinfo {author} {\bibfnamefont
  {K.~L.}\ \bibnamefont {Calvez}}, \bibinfo {author} {\bibfnamefont
  {B.}~\bibnamefont {Sac{\'{e}}p{\'{e}}}}, \bibinfo {author} {\bibfnamefont
  {A.}~\bibnamefont {Monfardini}},\ and\ \bibinfo {author} {\bibfnamefont
  {F.}~\bibnamefont {Levy-Bertrand}},\ }\bibfield  {title} {\bibinfo {title}
  {Tunable sub-gap radiation detection with superconducting resonators},\
  }\href {https://doi.org/10.1088/1361-6668/aa5b14} {\bibfield  {journal}
  {\bibinfo  {journal} {Superconductor Science and Technology}\ }\textbf
  {\bibinfo {volume} {30}},\ \bibinfo {pages} {045007} (\bibinfo {year}
  {2017})}\BibitemShut {NoStop}%
\bibitem [{\citenamefont {Bonnet}\ \emph {et~al.}(2021)\citenamefont {Bonnet},
  \citenamefont {Chiodi}, \citenamefont {Flanigan}, \citenamefont {Delagrange},
  \citenamefont {Brochu}, \citenamefont {D{\'{e}}barre},\ and\ \citenamefont
  {le~Sueur}}]{bonnet2021strongly}%
  \BibitemOpen
  \bibfield  {author} {\bibinfo {author} {\bibfnamefont {P.}~\bibnamefont
  {Bonnet}}, \bibinfo {author} {\bibfnamefont {F.}~\bibnamefont {Chiodi}},
  \bibinfo {author} {\bibfnamefont {D.}~\bibnamefont {Flanigan}}, \bibinfo
  {author} {\bibfnamefont {R.}~\bibnamefont {Delagrange}}, \bibinfo {author}
  {\bibfnamefont {N.}~\bibnamefont {Brochu}}, \bibinfo {author} {\bibfnamefont
  {D.}~\bibnamefont {D{\'{e}}barre}},\ and\ \bibinfo {author} {\bibfnamefont
  {H.}~\bibnamefont {le~Sueur}},\ }\href@noop {} {\bibinfo {title} {Strongly
  non-linear superconducting silicon resonators}} (\bibinfo {year} {2021}),\
  \Eprint {https://arxiv.org/abs/2101.11125} {arXiv:2101.11125
  [cond-mat.supr-con]} \BibitemShut {NoStop}%
\bibitem [{\citenamefont {Basset}\ \emph {et~al.}(2019)\citenamefont {Basset},
  \citenamefont {Watfa}, \citenamefont {Aiello}, \citenamefont {F{\'{e}}chant},
  \citenamefont {Morvan}, \citenamefont {Est{\'{e}}ve}, \citenamefont
  {Gabelli}, \citenamefont {Aprili}, \citenamefont {Weil}, \citenamefont
  {Kasumov}, \citenamefont {Bouchiat},\ and\ \citenamefont
  {Deblock}}]{Basset2019}%
  \BibitemOpen
  \bibfield  {author} {\bibinfo {author} {\bibfnamefont {J.}~\bibnamefont
  {Basset}}, \bibinfo {author} {\bibfnamefont {D.}~\bibnamefont {Watfa}},
  \bibinfo {author} {\bibfnamefont {G.}~\bibnamefont {Aiello}}, \bibinfo
  {author} {\bibfnamefont {M.}~\bibnamefont {F{\'{e}}chant}}, \bibinfo {author}
  {\bibfnamefont {A.}~\bibnamefont {Morvan}}, \bibinfo {author} {\bibfnamefont
  {J.}~\bibnamefont {Est{\'{e}}ve}}, \bibinfo {author} {\bibfnamefont
  {J.}~\bibnamefont {Gabelli}}, \bibinfo {author} {\bibfnamefont
  {M.}~\bibnamefont {Aprili}}, \bibinfo {author} {\bibfnamefont
  {R.}~\bibnamefont {Weil}}, \bibinfo {author} {\bibfnamefont {A.}~\bibnamefont
  {Kasumov}}, \bibinfo {author} {\bibfnamefont {H.}~\bibnamefont {Bouchiat}},\
  and\ \bibinfo {author} {\bibfnamefont {R.}~\bibnamefont {Deblock}},\
  }\bibfield  {title} {\bibinfo {title} {High kinetic inductance microwave
  resonators made by he-beam assisted deposition of tungsten nanowires},\
  }\href {https://doi.org/10.1063/1.5080925} {\bibfield  {journal} {\bibinfo
  {journal} {Applied Physics Letters}\ }\textbf {\bibinfo {volume} {114}},\
  \bibinfo {pages} {102601} (\bibinfo {year} {2019})},\ \Eprint
  {https://arxiv.org/abs/https://doi.org/10.1063/1.5080925}
  {https://doi.org/10.1063/1.5080925} \BibitemShut {NoStop}%
\bibitem [{\citenamefont {Driessen}\ \emph {et~al.}(2012)\citenamefont
  {Driessen}, \citenamefont {Coumou}, \citenamefont {Tromp}, \citenamefont
  {de~Visser},\ and\ \citenamefont {Klapwijk}}]{PhysRevLett.109.107003}%
  \BibitemOpen
  \bibfield  {author} {\bibinfo {author} {\bibfnamefont {E.~F.~C.}\
  \bibnamefont {Driessen}}, \bibinfo {author} {\bibfnamefont {P.~C. J.~J.}\
  \bibnamefont {Coumou}}, \bibinfo {author} {\bibfnamefont {R.~R.}\
  \bibnamefont {Tromp}}, \bibinfo {author} {\bibfnamefont {P.~J.}\ \bibnamefont
  {de~Visser}},\ and\ \bibinfo {author} {\bibfnamefont {T.~M.}\ \bibnamefont
  {Klapwijk}},\ }\bibfield  {title} {\bibinfo {title} {Strongly disordered tin
  and nbtin $s$-wave superconductors probed by microwave electrodynamics},\
  }\href {https://doi.org/10.1103/PhysRevLett.109.107003} {\bibfield  {journal}
  {\bibinfo  {journal} {Phys. Rev. Lett.}\ }\textbf {\bibinfo {volume} {109}},\
  \bibinfo {pages} {107003} (\bibinfo {year} {2012})}\BibitemShut {NoStop}%
\bibitem [{\citenamefont {Coumou}\ \emph {et~al.}(2013)\citenamefont {Coumou},
  \citenamefont {Driessen}, \citenamefont {Bueno}, \citenamefont {Chapelier},\
  and\ \citenamefont {Klapwijk}}]{PhysRevB.88.180505}%
  \BibitemOpen
  \bibfield  {author} {\bibinfo {author} {\bibfnamefont {P.~C. J.~J.}\
  \bibnamefont {Coumou}}, \bibinfo {author} {\bibfnamefont {E.~F.~C.}\
  \bibnamefont {Driessen}}, \bibinfo {author} {\bibfnamefont {J.}~\bibnamefont
  {Bueno}}, \bibinfo {author} {\bibfnamefont {C.}~\bibnamefont {Chapelier}},\
  and\ \bibinfo {author} {\bibfnamefont {T.~M.}\ \bibnamefont {Klapwijk}},\
  }\bibfield  {title} {\bibinfo {title} {Electrodynamic response and local
  tunneling spectroscopy of strongly disordered superconducting tin films},\
  }\href {https://doi.org/10.1103/PhysRevB.88.180505} {\bibfield  {journal}
  {\bibinfo  {journal} {Phys. Rev. B}\ }\textbf {\bibinfo {volume} {88}},\
  \bibinfo {pages} {180505} (\bibinfo {year} {2013})}\BibitemShut {NoStop}%
\bibitem [{\citenamefont {Sandberg}\ \emph {et~al.}(2012)\citenamefont
  {Sandberg}, \citenamefont {Vissers}, \citenamefont {Kline}, \citenamefont
  {Weides}, \citenamefont {Gao}, \citenamefont {Wisbey},\ and\ \citenamefont
  {Pappas}}]{doi:10.1063/1.4729623}%
  \BibitemOpen
  \bibfield  {author} {\bibinfo {author} {\bibfnamefont {M.}~\bibnamefont
  {Sandberg}}, \bibinfo {author} {\bibfnamefont {M.~R.}\ \bibnamefont
  {Vissers}}, \bibinfo {author} {\bibfnamefont {J.~S.}\ \bibnamefont {Kline}},
  \bibinfo {author} {\bibfnamefont {M.}~\bibnamefont {Weides}}, \bibinfo
  {author} {\bibfnamefont {J.}~\bibnamefont {Gao}}, \bibinfo {author}
  {\bibfnamefont {D.~S.}\ \bibnamefont {Wisbey}},\ and\ \bibinfo {author}
  {\bibfnamefont {D.~P.}\ \bibnamefont {Pappas}},\ }\bibfield  {title}
  {\bibinfo {title} {Etch induced microwave losses in titanium nitride
  superconducting resonators},\ }\href {https://doi.org/10.1063/1.4729623}
  {\bibfield  {journal} {\bibinfo  {journal} {Applied Physics Letters}\
  }\textbf {\bibinfo {volume} {100}},\ \bibinfo {pages} {262605} (\bibinfo
  {year} {2012})},\ \Eprint
  {https://arxiv.org/abs/https://doi.org/10.1063/1.4729623}
  {https://doi.org/10.1063/1.4729623} \BibitemShut {NoStop}%
\bibitem [{\citenamefont {Hazard}\ \emph {et~al.}(2019)\citenamefont {Hazard},
  \citenamefont {Gyenis}, \citenamefont {Di~Paolo}, \citenamefont {Asfaw},
  \citenamefont {Lyon}, \citenamefont {Blais},\ and\ \citenamefont
  {Houck}}]{PhysRevLett.122.010504}%
  \BibitemOpen
  \bibfield  {author} {\bibinfo {author} {\bibfnamefont {T.~M.}\ \bibnamefont
  {Hazard}}, \bibinfo {author} {\bibfnamefont {A.}~\bibnamefont {Gyenis}},
  \bibinfo {author} {\bibfnamefont {A.}~\bibnamefont {Di~Paolo}}, \bibinfo
  {author} {\bibfnamefont {A.~T.}\ \bibnamefont {Asfaw}}, \bibinfo {author}
  {\bibfnamefont {S.~A.}\ \bibnamefont {Lyon}}, \bibinfo {author}
  {\bibfnamefont {A.}~\bibnamefont {Blais}},\ and\ \bibinfo {author}
  {\bibfnamefont {A.~A.}\ \bibnamefont {Houck}},\ }\bibfield  {title} {\bibinfo
  {title} {Nanowire superinductance fluxonium qubit},\ }\href
  {https://doi.org/10.1103/PhysRevLett.122.010504} {\bibfield  {journal}
  {\bibinfo  {journal} {Phys. Rev. Lett.}\ }\textbf {\bibinfo {volume} {122}},\
  \bibinfo {pages} {010504} (\bibinfo {year} {2019})}\BibitemShut {NoStop}%
\bibitem [{\citenamefont {Ohya}\ \emph {et~al.}(2013)\citenamefont {Ohya},
  \citenamefont {Chiaro}, \citenamefont {Megrant}, \citenamefont {Neill},
  \citenamefont {Barends}, \citenamefont {Chen}, \citenamefont {Kelly},
  \citenamefont {Low}, \citenamefont {Mutus}, \citenamefont {O'Malley},
  \citenamefont {Roushan}, \citenamefont {Sank}, \citenamefont {Vainsencher},
  \citenamefont {Wenner}, \citenamefont {White}, \citenamefont {Yin},
  \citenamefont {Schultz}, \citenamefont {Palmstr{\o}m}, \citenamefont {Mazin},
  \citenamefont {Cleland},\ and\ \citenamefont {Martinis}}]{Ohya_2013}%
  \BibitemOpen
  \bibfield  {author} {\bibinfo {author} {\bibfnamefont {S.}~\bibnamefont
  {Ohya}}, \bibinfo {author} {\bibfnamefont {B.}~\bibnamefont {Chiaro}},
  \bibinfo {author} {\bibfnamefont {A.}~\bibnamefont {Megrant}}, \bibinfo
  {author} {\bibfnamefont {C.}~\bibnamefont {Neill}}, \bibinfo {author}
  {\bibfnamefont {R.}~\bibnamefont {Barends}}, \bibinfo {author} {\bibfnamefont
  {Y.}~\bibnamefont {Chen}}, \bibinfo {author} {\bibfnamefont {J.}~\bibnamefont
  {Kelly}}, \bibinfo {author} {\bibfnamefont {D.}~\bibnamefont {Low}}, \bibinfo
  {author} {\bibfnamefont {J.}~\bibnamefont {Mutus}}, \bibinfo {author}
  {\bibfnamefont {P.~J.~J.}\ \bibnamefont {O'Malley}}, \bibinfo {author}
  {\bibfnamefont {P.}~\bibnamefont {Roushan}}, \bibinfo {author} {\bibfnamefont
  {D.}~\bibnamefont {Sank}}, \bibinfo {author} {\bibfnamefont {A.}~\bibnamefont
  {Vainsencher}}, \bibinfo {author} {\bibfnamefont {J.}~\bibnamefont {Wenner}},
  \bibinfo {author} {\bibfnamefont {T.~C.}\ \bibnamefont {White}}, \bibinfo
  {author} {\bibfnamefont {Y.}~\bibnamefont {Yin}}, \bibinfo {author}
  {\bibfnamefont {B.~D.}\ \bibnamefont {Schultz}}, \bibinfo {author}
  {\bibfnamefont {C.~J.}\ \bibnamefont {Palmstr{\o}m}}, \bibinfo {author}
  {\bibfnamefont {B.~A.}\ \bibnamefont {Mazin}}, \bibinfo {author}
  {\bibfnamefont {A.~N.}\ \bibnamefont {Cleland}},\ and\ \bibinfo {author}
  {\bibfnamefont {J.~M.}\ \bibnamefont {Martinis}},\ }\bibfield  {title}
  {\bibinfo {title} {Room temperature deposition of sputtered {TiN} films for
  superconducting coplanar waveguide resonators},\ }\href
  {https://doi.org/10.1088/0953-2048/27/1/015009} {\bibfield  {journal}
  {\bibinfo  {journal} {Superconductor Science and Technology}\ }\textbf
  {\bibinfo {volume} {27}},\ \bibinfo {pages} {015009} (\bibinfo {year}
  {2013})}\BibitemShut {NoStop}%
\bibitem [{\citenamefont {Verhaverbeke}\ and\ \citenamefont
  {Parker}(1997)}]{verhaverbeke_parker_1997}%
  \BibitemOpen
  \bibfield  {author} {\bibinfo {author} {\bibfnamefont {S.}~\bibnamefont
  {Verhaverbeke}}\ and\ \bibinfo {author} {\bibfnamefont {J.~W.}\ \bibnamefont
  {Parker}},\ }\bibfield  {title} {\bibinfo {title} {A model for the etching of
  ti and tin in sc-1 solutions},\ }\href {https://doi.org/10.1557/PROC-477-447}
  {\bibfield  {journal} {\bibinfo  {journal} {MRS Proceedings}\ }\textbf
  {\bibinfo {volume} {477}},\ \bibinfo {pages} {447} (\bibinfo {year}
  {1997})}\BibitemShut {NoStop}%
\bibitem [{\citenamefont {Murray}(2021)}]{murray2021material}%
  \BibitemOpen
  \bibfield  {author} {\bibinfo {author} {\bibfnamefont {C.~E.}\ \bibnamefont
  {Murray}},\ }\href@noop {} {\bibinfo {title} {Material matters in
  superconducting qubits}} (\bibinfo {year} {2021}),\ \Eprint
  {https://arxiv.org/abs/2106.05919} {arXiv:2106.05919 [quant-ph]} \BibitemShut
  {NoStop}%
\bibitem [{\citenamefont {Saveskul}\ \emph {et~al.}(2019)\citenamefont
  {Saveskul}, \citenamefont {Titova}, \citenamefont {Baeva}, \citenamefont
  {Semenov}, \citenamefont {Lubenchenko}, \citenamefont {Saha}, \citenamefont
  {Reddy}, \citenamefont {Bogdanov}, \citenamefont {Marinero}, \citenamefont
  {Shalaev}, \citenamefont {Boltasseva}, \citenamefont {Khrapai}, \citenamefont
  {Kardakova},\ and\ \citenamefont {Goltsman}}]{PhysRevApplied.12.054001}%
  \BibitemOpen
  \bibfield  {author} {\bibinfo {author} {\bibfnamefont {N.}~\bibnamefont
  {Saveskul}}, \bibinfo {author} {\bibfnamefont {N.}~\bibnamefont {Titova}},
  \bibinfo {author} {\bibfnamefont {E.}~\bibnamefont {Baeva}}, \bibinfo
  {author} {\bibfnamefont {A.}~\bibnamefont {Semenov}}, \bibinfo {author}
  {\bibfnamefont {A.}~\bibnamefont {Lubenchenko}}, \bibinfo {author}
  {\bibfnamefont {S.}~\bibnamefont {Saha}}, \bibinfo {author} {\bibfnamefont
  {H.}~\bibnamefont {Reddy}}, \bibinfo {author} {\bibfnamefont
  {S.}~\bibnamefont {Bogdanov}}, \bibinfo {author} {\bibfnamefont
  {E.}~\bibnamefont {Marinero}}, \bibinfo {author} {\bibfnamefont
  {V.}~\bibnamefont {Shalaev}}, \bibinfo {author} {\bibfnamefont
  {A.}~\bibnamefont {Boltasseva}}, \bibinfo {author} {\bibfnamefont
  {V.}~\bibnamefont {Khrapai}}, \bibinfo {author} {\bibfnamefont
  {A.}~\bibnamefont {Kardakova}},\ and\ \bibinfo {author} {\bibfnamefont
  {G.}~\bibnamefont {Goltsman}},\ }\bibfield  {title} {\bibinfo {title}
  {Superconductivity behavior in epitaxial tin films points to surface magnetic
  disorder},\ }\href {https://doi.org/10.1103/PhysRevApplied.12.054001}
  {\bibfield  {journal} {\bibinfo  {journal} {Phys. Rev. Applied}\ }\textbf
  {\bibinfo {volume} {12}},\ \bibinfo {pages} {054001} (\bibinfo {year}
  {2019})}\BibitemShut {NoStop}%
\bibitem [{\citenamefont {Torgovkin}\ \emph {et~al.}(2018)\citenamefont
  {Torgovkin}, \citenamefont {Chaudhuri}, \citenamefont {Ruhtinas},
  \citenamefont {Lahtinen}, \citenamefont {Sajavaara},\ and\ \citenamefont
  {Maasilta}}]{Torgovkin2018}%
  \BibitemOpen
  \bibfield  {author} {\bibinfo {author} {\bibfnamefont {A.}~\bibnamefont
  {Torgovkin}}, \bibinfo {author} {\bibfnamefont {S.}~\bibnamefont
  {Chaudhuri}}, \bibinfo {author} {\bibfnamefont {A.}~\bibnamefont {Ruhtinas}},
  \bibinfo {author} {\bibfnamefont {M.}~\bibnamefont {Lahtinen}}, \bibinfo
  {author} {\bibfnamefont {T.}~\bibnamefont {Sajavaara}},\ and\ \bibinfo
  {author} {\bibfnamefont {I.~J.}\ \bibnamefont {Maasilta}},\ }\bibfield
  {title} {\bibinfo {title} {High quality superconducting titanium nitride thin
  film growth using infrared pulsed laser deposition},\ }\href
  {https://doi.org/10.1088/1361-6668/aab7d6} {\bibfield  {journal} {\bibinfo
  {journal} {Superconductor Science and Technology}\ }\textbf {\bibinfo
  {volume} {31}},\ \bibinfo {pages} {055017} (\bibinfo {year}
  {2018})}\BibitemShut {NoStop}%
\bibitem [{\citenamefont {Sac{\'e}p{\'e}}\ \emph {et~al.}(2008)\citenamefont
  {Sac{\'e}p{\'e}}, \citenamefont {Chapelier}, \citenamefont {Baturina},
  \citenamefont {Vinokur}, \citenamefont {Baklanov},\ and\ \citenamefont
  {Sanquer}}]{Sacepe:2008jx}%
  \BibitemOpen
  \bibfield  {author} {\bibinfo {author} {\bibfnamefont {B.}~\bibnamefont
  {Sac{\'e}p{\'e}}}, \bibinfo {author} {\bibfnamefont {C.}~\bibnamefont
  {Chapelier}}, \bibinfo {author} {\bibfnamefont {T.~I.}\ \bibnamefont
  {Baturina}}, \bibinfo {author} {\bibfnamefont {V.~M.}\ \bibnamefont
  {Vinokur}}, \bibinfo {author} {\bibfnamefont {M.~R.}\ \bibnamefont
  {Baklanov}},\ and\ \bibinfo {author} {\bibfnamefont {M.}~\bibnamefont
  {Sanquer}},\ }\bibfield  {title} {{\selectlanguage {English}\bibinfo {title}
  {{Disorder-Induced Inhomogeneities of the Superconducting State Close to the
  Superconductor-Insulator Transition}}},\ }\href
  {https://doi.org/10.1103/PhysRevLett.101.157006} {\bibfield  {journal}
  {\bibinfo  {journal} {Physical Review Letters}\ }\textbf {\bibinfo {volume}
  {101}},\ \bibinfo {pages} {1765} (\bibinfo {year} {2008})}\BibitemShut
  {NoStop}%
\bibitem [{\citenamefont {Ivry}\ \emph {et~al.}(2014)\citenamefont {Ivry},
  \citenamefont {Kim}, \citenamefont {Dane}, \citenamefont {De~Fazio},
  \citenamefont {McCaughan}, \citenamefont {Sunter}, \citenamefont {Zhao},\
  and\ \citenamefont {Berggren}}]{PhysRevB.90.214515}%
  \BibitemOpen
  \bibfield  {author} {\bibinfo {author} {\bibfnamefont {Y.}~\bibnamefont
  {Ivry}}, \bibinfo {author} {\bibfnamefont {C.-S.}\ \bibnamefont {Kim}},
  \bibinfo {author} {\bibfnamefont {A.~E.}\ \bibnamefont {Dane}}, \bibinfo
  {author} {\bibfnamefont {D.}~\bibnamefont {De~Fazio}}, \bibinfo {author}
  {\bibfnamefont {A.~N.}\ \bibnamefont {McCaughan}}, \bibinfo {author}
  {\bibfnamefont {K.~A.}\ \bibnamefont {Sunter}}, \bibinfo {author}
  {\bibfnamefont {Q.}~\bibnamefont {Zhao}},\ and\ \bibinfo {author}
  {\bibfnamefont {K.~K.}\ \bibnamefont {Berggren}},\ }\bibfield  {title}
  {\bibinfo {title} {Universal scaling of the critical temperature for thin
  films near the superconducting-to-insulating transition},\ }\href
  {https://doi.org/10.1103/PhysRevB.90.214515} {\bibfield  {journal} {\bibinfo
  {journal} {Phys. Rev. B}\ }\textbf {\bibinfo {volume} {90}},\ \bibinfo
  {pages} {214515} (\bibinfo {year} {2014})}\BibitemShut {NoStop}%
\bibitem [{\citenamefont {Kundu}\ \emph {et~al.}(2019)\citenamefont {Kundu},
  \citenamefont {Amin}, \citenamefont {Jesudasan}, \citenamefont
  {Raychaudhuri}, \citenamefont {Mukerjee},\ and\ \citenamefont
  {Bid}}]{PhysRevB.100.174501}%
  \BibitemOpen
  \bibfield  {author} {\bibinfo {author} {\bibfnamefont {H.~K.}\ \bibnamefont
  {Kundu}}, \bibinfo {author} {\bibfnamefont {K.~R.}\ \bibnamefont {Amin}},
  \bibinfo {author} {\bibfnamefont {J.}~\bibnamefont {Jesudasan}}, \bibinfo
  {author} {\bibfnamefont {P.}~\bibnamefont {Raychaudhuri}}, \bibinfo {author}
  {\bibfnamefont {S.}~\bibnamefont {Mukerjee}},\ and\ \bibinfo {author}
  {\bibfnamefont {A.}~\bibnamefont {Bid}},\ }\bibfield  {title} {\bibinfo
  {title} {Effect of dimensionality on the vortex dynamics in a type-ii
  superconductor},\ }\href {https://doi.org/10.1103/PhysRevB.100.174501}
  {\bibfield  {journal} {\bibinfo  {journal} {Phys. Rev. B}\ }\textbf {\bibinfo
  {volume} {100}},\ \bibinfo {pages} {174501} (\bibinfo {year}
  {2019})}\BibitemShut {NoStop}%
\bibitem [{\citenamefont {Faverzani}\ \emph {et~al.}(2020)\citenamefont
  {Faverzani}, \citenamefont {Ferri}, \citenamefont {Giachero}, \citenamefont
  {Giordano}, \citenamefont {Margesin}, \citenamefont {Mezzena}, \citenamefont
  {Nucciotti},\ and\ \citenamefont {Puiu}}]{Faverzani_2020}%
  \BibitemOpen
  \bibfield  {author} {\bibinfo {author} {\bibfnamefont {M.}~\bibnamefont
  {Faverzani}}, \bibinfo {author} {\bibfnamefont {E.}~\bibnamefont {Ferri}},
  \bibinfo {author} {\bibfnamefont {A.}~\bibnamefont {Giachero}}, \bibinfo
  {author} {\bibfnamefont {C.}~\bibnamefont {Giordano}}, \bibinfo {author}
  {\bibfnamefont {B.}~\bibnamefont {Margesin}}, \bibinfo {author}
  {\bibfnamefont {R.}~\bibnamefont {Mezzena}}, \bibinfo {author} {\bibfnamefont
  {A.}~\bibnamefont {Nucciotti}},\ and\ \bibinfo {author} {\bibfnamefont
  {A.}~\bibnamefont {Puiu}},\ }\bibfield  {title} {\bibinfo {title}
  {Characterization of the low temperature behavior of thin titanium/titanium
  nitride multilayer films},\ }\href {https://doi.org/10.1088/1361-6668/ab7435}
  {\bibfield  {journal} {\bibinfo  {journal} {Superconductor Science and
  Technology}\ }\textbf {\bibinfo {volume} {33}},\ \bibinfo {pages} {045009}
  (\bibinfo {year} {2020})}\BibitemShut {NoStop}%
\bibitem [{\citenamefont {Probst}\ \emph {et~al.}(2015)\citenamefont {Probst},
  \citenamefont {Song}, \citenamefont {Bushev}, \citenamefont {Ustinov},\ and\
  \citenamefont {Weides}}]{doi:10.1063/1.4907935}%
  \BibitemOpen
  \bibfield  {author} {\bibinfo {author} {\bibfnamefont {S.}~\bibnamefont
  {Probst}}, \bibinfo {author} {\bibfnamefont {F.~B.}\ \bibnamefont {Song}},
  \bibinfo {author} {\bibfnamefont {P.~A.}\ \bibnamefont {Bushev}}, \bibinfo
  {author} {\bibfnamefont {A.~V.}\ \bibnamefont {Ustinov}},\ and\ \bibinfo
  {author} {\bibfnamefont {M.}~\bibnamefont {Weides}},\ }\bibfield  {title}
  {\bibinfo {title} {Efficient and robust analysis of complex scattering data
  under noise in microwave resonators},\ }\href
  {https://doi.org/10.1063/1.4907935} {\bibfield  {journal} {\bibinfo
  {journal} {Review of Scientific Instruments}\ }\textbf {\bibinfo {volume}
  {86}},\ \bibinfo {pages} {024706} (\bibinfo {year} {2015})},\ \Eprint
  {https://arxiv.org/abs/https://doi.org/10.1063/1.4907935}
  {https://doi.org/10.1063/1.4907935} \BibitemShut {NoStop}%
\bibitem [{\citenamefont {Charpentier}\ and\ \citenamefont {et. al.}()}]{tibo}%
  \BibitemOpen
  \bibfield  {author} {\bibinfo {author} {\bibfnamefont {T.}~\bibnamefont
  {Charpentier}}\ and\ \bibinfo {author} {\bibnamefont {et. al.}},\ }\bibinfo
  {title} {Manuscript under preparation}\BibitemShut {NoStop}%
\bibitem [{\citenamefont {M{\"u}ller}\ \emph {et~al.}(2019)\citenamefont
  {M{\"u}ller}, \citenamefont {Cole},\ and\ \citenamefont
  {Lisenfeld}}]{M_ller_2019}%
  \BibitemOpen
\bibfield  {title} {  }\bibfield  {author} {\bibinfo {author} {\bibfnamefont
  {C.}~\bibnamefont {M{\"u}ller}}, \bibinfo {author} {\bibfnamefont {J.~H.}\
  \bibnamefont {Cole}},\ and\ \bibinfo {author} {\bibfnamefont
  {J.}~\bibnamefont {Lisenfeld}},\ }\bibfield  {title} {\bibinfo {title}
  {Towards understanding two-level-systems in amorphous solids: insights from
  quantum circuits},\ }\href {https://doi.org/10.1088/1361-6633/ab3a7e}
  {\bibfield  {journal} {\bibinfo  {journal} {Reports on Progress in Physics}\
  }\textbf {\bibinfo {volume} {82}},\ \bibinfo {pages} {124501} (\bibinfo
  {year} {2019})}\BibitemShut {NoStop}%
\bibitem [{\citenamefont {Day}\ \emph {et~al.}(2003)\citenamefont {Day},
  \citenamefont {LeDuc}, \citenamefont {Mazin}, \citenamefont {Vayonakis},\
  and\ \citenamefont {Zmuidzinas}}]{Day2003}%
  \BibitemOpen
  \bibfield  {author} {\bibinfo {author} {\bibfnamefont {P.~K.}\ \bibnamefont
  {Day}}, \bibinfo {author} {\bibfnamefont {H.~G.}\ \bibnamefont {LeDuc}},
  \bibinfo {author} {\bibfnamefont {B.~A.}\ \bibnamefont {Mazin}}, \bibinfo
  {author} {\bibfnamefont {A.}~\bibnamefont {Vayonakis}},\ and\ \bibinfo
  {author} {\bibfnamefont {J.}~\bibnamefont {Zmuidzinas}},\ }\bibfield  {title}
  {\bibinfo {title} {A broadband superconducting detector suitable for use in
  large arrays},\ }\href {https://doi.org/10.1038/nature02037} {\bibfield
  {journal} {\bibinfo  {journal} {Nature}\ }\textbf {\bibinfo {volume} {425}},\
  \bibinfo {pages} {817} (\bibinfo {year} {2003})}\BibitemShut {NoStop}%
\bibitem [{\citenamefont {Henriques}\ \emph {et~al.}(2019)\citenamefont
  {Henriques}, \citenamefont {Valenti}, \citenamefont {Charpentier},
  \citenamefont {Lagoin}, \citenamefont {Gouriou}, \citenamefont
  {Mart{\'i}nez}, \citenamefont {Cardani}, \citenamefont {Vignati},
  \citenamefont {Gr{\"u}nhaupt}, \citenamefont {Gusenkova}, \citenamefont
  {Ferrero}, \citenamefont {Skacel}, \citenamefont {Wernsdorfer}, \citenamefont
  {Ustinov}, \citenamefont {Catelani}, \citenamefont {Sander},\ and\
  \citenamefont {Pop}}]{doi:10.1063/1.5124967}%
  \BibitemOpen
  \bibfield  {author} {\bibinfo {author} {\bibfnamefont {F.}~\bibnamefont
  {Henriques}}, \bibinfo {author} {\bibfnamefont {F.}~\bibnamefont {Valenti}},
  \bibinfo {author} {\bibfnamefont {T.}~\bibnamefont {Charpentier}}, \bibinfo
  {author} {\bibfnamefont {M.}~\bibnamefont {Lagoin}}, \bibinfo {author}
  {\bibfnamefont {C.}~\bibnamefont {Gouriou}}, \bibinfo {author} {\bibfnamefont
  {M.}~\bibnamefont {Mart{\'i}nez}}, \bibinfo {author} {\bibfnamefont
  {L.}~\bibnamefont {Cardani}}, \bibinfo {author} {\bibfnamefont
  {M.}~\bibnamefont {Vignati}}, \bibinfo {author} {\bibfnamefont
  {L.}~\bibnamefont {Gr{\"u}nhaupt}}, \bibinfo {author} {\bibfnamefont
  {D.}~\bibnamefont {Gusenkova}}, \bibinfo {author} {\bibfnamefont
  {J.}~\bibnamefont {Ferrero}}, \bibinfo {author} {\bibfnamefont {S.~T.}\
  \bibnamefont {Skacel}}, \bibinfo {author} {\bibfnamefont {W.}~\bibnamefont
  {Wernsdorfer}}, \bibinfo {author} {\bibfnamefont {A.~V.}\ \bibnamefont
  {Ustinov}}, \bibinfo {author} {\bibfnamefont {G.}~\bibnamefont {Catelani}},
  \bibinfo {author} {\bibfnamefont {O.}~\bibnamefont {Sander}},\ and\ \bibinfo
  {author} {\bibfnamefont {I.~M.}\ \bibnamefont {Pop}},\ }\bibfield  {title}
  {\bibinfo {title} {Phonon traps reduce the quasiparticle density in
  superconducting circuits},\ }\href {https://doi.org/10.1063/1.5124967}
  {\bibfield  {journal} {\bibinfo  {journal} {Applied Physics Letters}\
  }\textbf {\bibinfo {volume} {115}},\ \bibinfo {pages} {212601} (\bibinfo
  {year} {2019})},\ \Eprint
  {https://arxiv.org/abs/https://doi.org/10.1063/1.5124967}
  {https://doi.org/10.1063/1.5124967} \BibitemShut {NoStop}%
\bibitem [{\citenamefont {Cardani}\ \emph {et~al.}(2021)\citenamefont
  {Cardani}, \citenamefont {Valenti}, \citenamefont {Casali}, \citenamefont
  {Catelani}, \citenamefont {Charpentier}, \citenamefont {Clemenza},
  \citenamefont {Colantoni}, \citenamefont {Cruciani}, \citenamefont
  {D{'}Imperio}, \citenamefont {Gironi}, \citenamefont {Gr{\"{u}}nhaupt},
  \citenamefont {Gusenkova}, \citenamefont {Henriques}, \citenamefont {Lagoin},
  \citenamefont {Martinez}, \citenamefont {Pettinari}, \citenamefont {Rusconi},
  \citenamefont {Sander}, \citenamefont {Tomei}, \citenamefont {Ustinov},
  \citenamefont {Weber}, \citenamefont {Wernsdorfer}, \citenamefont {Vignati},
  \citenamefont {Pirro},\ and\ \citenamefont {Pop}}]{Cardani2021}%
  \BibitemOpen
  \bibfield  {author} {\bibinfo {author} {\bibfnamefont {L.}~\bibnamefont
  {Cardani}}, \bibinfo {author} {\bibfnamefont {F.}~\bibnamefont {Valenti}},
  \bibinfo {author} {\bibfnamefont {N.}~\bibnamefont {Casali}}, \bibinfo
  {author} {\bibfnamefont {G.}~\bibnamefont {Catelani}}, \bibinfo {author}
  {\bibfnamefont {T.}~\bibnamefont {Charpentier}}, \bibinfo {author}
  {\bibfnamefont {M.}~\bibnamefont {Clemenza}}, \bibinfo {author}
  {\bibfnamefont {I.}~\bibnamefont {Colantoni}}, \bibinfo {author}
  {\bibfnamefont {A.}~\bibnamefont {Cruciani}}, \bibinfo {author}
  {\bibfnamefont {G.}~\bibnamefont {D{'}Imperio}}, \bibinfo {author}
  {\bibfnamefont {L.}~\bibnamefont {Gironi}}, \bibinfo {author} {\bibfnamefont
  {L.}~\bibnamefont {Gr{\"{u}}nhaupt}}, \bibinfo {author} {\bibfnamefont
  {D.}~\bibnamefont {Gusenkova}}, \bibinfo {author} {\bibfnamefont
  {F.}~\bibnamefont {Henriques}}, \bibinfo {author} {\bibfnamefont
  {M.}~\bibnamefont {Lagoin}}, \bibinfo {author} {\bibfnamefont
  {M.}~\bibnamefont {Martinez}}, \bibinfo {author} {\bibfnamefont
  {G.}~\bibnamefont {Pettinari}}, \bibinfo {author} {\bibfnamefont
  {C.}~\bibnamefont {Rusconi}}, \bibinfo {author} {\bibfnamefont
  {O.}~\bibnamefont {Sander}}, \bibinfo {author} {\bibfnamefont
  {C.}~\bibnamefont {Tomei}}, \bibinfo {author} {\bibfnamefont {A.~V.}\
  \bibnamefont {Ustinov}}, \bibinfo {author} {\bibfnamefont {M.}~\bibnamefont
  {Weber}}, \bibinfo {author} {\bibfnamefont {W.}~\bibnamefont {Wernsdorfer}},
  \bibinfo {author} {\bibfnamefont {M.}~\bibnamefont {Vignati}}, \bibinfo
  {author} {\bibfnamefont {S.}~\bibnamefont {Pirro}},\ and\ \bibinfo {author}
  {\bibfnamefont {I.~M.}\ \bibnamefont {Pop}},\ }\bibfield  {title} {\bibinfo
  {title} {Reducing the impact of radioactivity on quantum circuits in a
  deep-underground facility},\ }\href
  {https://doi.org/10.1038/s41467-021-23032-z} {\bibfield  {journal} {\bibinfo
  {journal} {Nature Communications}\ }\textbf {\bibinfo {volume} {12}},\
  \bibinfo {pages} {2733} (\bibinfo {year} {2021})}\BibitemShut {NoStop}%
\bibitem [{\citenamefont {Winkel}\ \emph {et~al.}(2020)\citenamefont {Winkel},
  \citenamefont {Borisov}, \citenamefont {Gr\"unhaupt}, \citenamefont {Rieger},
  \citenamefont {Spiecker}, \citenamefont {Valenti}, \citenamefont {Ustinov},
  \citenamefont {Wernsdorfer},\ and\ \citenamefont {Pop}}]{PhysRevX.10.031032}%
  \BibitemOpen
  \bibfield  {author} {\bibinfo {author} {\bibfnamefont {P.}~\bibnamefont
  {Winkel}}, \bibinfo {author} {\bibfnamefont {K.}~\bibnamefont {Borisov}},
  \bibinfo {author} {\bibfnamefont {L.}~\bibnamefont {Gr\"unhaupt}}, \bibinfo
  {author} {\bibfnamefont {D.}~\bibnamefont {Rieger}}, \bibinfo {author}
  {\bibfnamefont {M.}~\bibnamefont {Spiecker}}, \bibinfo {author}
  {\bibfnamefont {F.}~\bibnamefont {Valenti}}, \bibinfo {author} {\bibfnamefont
  {A.~V.}\ \bibnamefont {Ustinov}}, \bibinfo {author} {\bibfnamefont
  {W.}~\bibnamefont {Wernsdorfer}},\ and\ \bibinfo {author} {\bibfnamefont
  {I.~M.}\ \bibnamefont {Pop}},\ }\bibfield  {title} {\bibinfo {title}
  {Implementation of a transmon qubit using superconducting granular
  aluminum},\ }\href {https://doi.org/10.1103/PhysRevX.10.031032} {\bibfield
  {journal} {\bibinfo  {journal} {Phys. Rev. X}\ }\textbf {\bibinfo {volume}
  {10}},\ \bibinfo {pages} {031032} (\bibinfo {year} {2020})}\BibitemShut
  {NoStop}%
\bibitem [{\citenamefont {Wang}\ \emph {et~al.}(2015)\citenamefont {Wang},
  \citenamefont {Axline}, \citenamefont {Gao}, \citenamefont {Brecht},
  \citenamefont {Chu}, \citenamefont {Frunzio}, \citenamefont {Devoret},\ and\
  \citenamefont {Schoelkopf}}]{Wang2015}%
  \BibitemOpen
  \bibfield  {author} {\bibinfo {author} {\bibfnamefont {C.}~\bibnamefont
  {Wang}}, \bibinfo {author} {\bibfnamefont {C.}~\bibnamefont {Axline}},
  \bibinfo {author} {\bibfnamefont {Y.~Y.}\ \bibnamefont {Gao}}, \bibinfo
  {author} {\bibfnamefont {T.}~\bibnamefont {Brecht}}, \bibinfo {author}
  {\bibfnamefont {Y.}~\bibnamefont {Chu}}, \bibinfo {author} {\bibfnamefont
  {L.}~\bibnamefont {Frunzio}}, \bibinfo {author} {\bibfnamefont {M.~H.}\
  \bibnamefont {Devoret}},\ and\ \bibinfo {author} {\bibfnamefont {R.~J.}\
  \bibnamefont {Schoelkopf}},\ }\bibfield  {title} {\bibinfo {title} {Surface
  participation and dielectric loss in superconducting qubits},\ }\href
  {https://doi.org/10.1063/1.4934486} {\bibfield  {journal} {\bibinfo
  {journal} {Applied Physics Letters}\ }\textbf {\bibinfo {volume} {107}},\
  \bibinfo {pages} {162601} (\bibinfo {year} {2015})},\ \Eprint
  {https://arxiv.org/abs/https://doi.org/10.1063/1.4934486}
  {https://doi.org/10.1063/1.4934486} \BibitemShut {NoStop}%
\bibitem [{\citenamefont {Wenner}\ \emph {et~al.}(2011)\citenamefont {Wenner},
  \citenamefont {Barends}, \citenamefont {Bialczak}, \citenamefont {Chen},
  \citenamefont {Kelly}, \citenamefont {Lucero}, \citenamefont {Mariantoni},
  \citenamefont {Megrant}, \citenamefont {O'Malley}, \citenamefont {Sank},
  \citenamefont {Vainsencher}, \citenamefont {Wang}, \citenamefont {White},
  \citenamefont {Yin}, \citenamefont {Zhao}, \citenamefont {Cleland},\ and\
  \citenamefont {Martinis}}]{doi:10.1063/1.3637047}%
  \BibitemOpen
  \bibfield  {author} {\bibinfo {author} {\bibfnamefont {J.}~\bibnamefont
  {Wenner}}, \bibinfo {author} {\bibfnamefont {R.}~\bibnamefont {Barends}},
  \bibinfo {author} {\bibfnamefont {R.~C.}\ \bibnamefont {Bialczak}}, \bibinfo
  {author} {\bibfnamefont {Y.}~\bibnamefont {Chen}}, \bibinfo {author}
  {\bibfnamefont {J.}~\bibnamefont {Kelly}}, \bibinfo {author} {\bibfnamefont
  {E.}~\bibnamefont {Lucero}}, \bibinfo {author} {\bibfnamefont
  {M.}~\bibnamefont {Mariantoni}}, \bibinfo {author} {\bibfnamefont
  {A.}~\bibnamefont {Megrant}}, \bibinfo {author} {\bibfnamefont {P.~J.~J.}\
  \bibnamefont {O'Malley}}, \bibinfo {author} {\bibfnamefont {D.}~\bibnamefont
  {Sank}}, \bibinfo {author} {\bibfnamefont {A.}~\bibnamefont {Vainsencher}},
  \bibinfo {author} {\bibfnamefont {H.}~\bibnamefont {Wang}}, \bibinfo {author}
  {\bibfnamefont {T.~C.}\ \bibnamefont {White}}, \bibinfo {author}
  {\bibfnamefont {Y.}~\bibnamefont {Yin}}, \bibinfo {author} {\bibfnamefont
  {J.}~\bibnamefont {Zhao}}, \bibinfo {author} {\bibfnamefont {A.~N.}\
  \bibnamefont {Cleland}},\ and\ \bibinfo {author} {\bibfnamefont {J.~M.}\
  \bibnamefont {Martinis}},\ }\bibfield  {title} {\bibinfo {title} {Surface
  loss simulations of superconducting coplanar waveguide resonators},\ }\href
  {https://doi.org/10.1063/1.3637047} {\bibfield  {journal} {\bibinfo
  {journal} {Applied Physics Letters}\ }\textbf {\bibinfo {volume} {99}},\
  \bibinfo {pages} {113513} (\bibinfo {year} {2011})},\ \Eprint
  {https://arxiv.org/abs/https://doi.org/10.1063/1.3637047}
  {https://doi.org/10.1063/1.3637047} \BibitemShut {NoStop}%
\bibitem [{\citenamefont {Calusine}\ \emph {et~al.}(2018)\citenamefont
  {Calusine}, \citenamefont {Melville}, \citenamefont {Woods}, \citenamefont
  {Das}, \citenamefont {Stull}, \citenamefont {Bolkhovsky}, \citenamefont
  {Braje}, \citenamefont {Hover}, \citenamefont {Kim}, \citenamefont {Miloshi},
  \citenamefont {Rosenberg}, \citenamefont {Sevi}, \citenamefont {Yoder},
  \citenamefont {Dauler},\ and\ \citenamefont
  {Oliver}}]{doi:10.1063/1.5006888}%
  \BibitemOpen
  \bibfield  {author} {\bibinfo {author} {\bibfnamefont {G.}~\bibnamefont
  {Calusine}}, \bibinfo {author} {\bibfnamefont {A.}~\bibnamefont {Melville}},
  \bibinfo {author} {\bibfnamefont {W.}~\bibnamefont {Woods}}, \bibinfo
  {author} {\bibfnamefont {R.}~\bibnamefont {Das}}, \bibinfo {author}
  {\bibfnamefont {C.}~\bibnamefont {Stull}}, \bibinfo {author} {\bibfnamefont
  {V.}~\bibnamefont {Bolkhovsky}}, \bibinfo {author} {\bibfnamefont
  {D.}~\bibnamefont {Braje}}, \bibinfo {author} {\bibfnamefont
  {D.}~\bibnamefont {Hover}}, \bibinfo {author} {\bibfnamefont {D.~K.}\
  \bibnamefont {Kim}}, \bibinfo {author} {\bibfnamefont {X.}~\bibnamefont
  {Miloshi}}, \bibinfo {author} {\bibfnamefont {D.}~\bibnamefont {Rosenberg}},
  \bibinfo {author} {\bibfnamefont {A.}~\bibnamefont {Sevi}}, \bibinfo {author}
  {\bibfnamefont {J.~L.}\ \bibnamefont {Yoder}}, \bibinfo {author}
  {\bibfnamefont {E.}~\bibnamefont {Dauler}},\ and\ \bibinfo {author}
  {\bibfnamefont {W.~D.}\ \bibnamefont {Oliver}},\ }\bibfield  {title}
  {\bibinfo {title} {Analysis and mitigation of interface losses in trenched
  superconducting coplanar waveguide resonators},\ }\href
  {https://doi.org/10.1063/1.5006888} {\bibfield  {journal} {\bibinfo
  {journal} {Applied Physics Letters}\ }\textbf {\bibinfo {volume} {112}},\
  \bibinfo {pages} {062601} (\bibinfo {year} {2018})},\ \Eprint
  {https://arxiv.org/abs/https://doi.org/10.1063/1.5006888}
  {https://doi.org/10.1063/1.5006888} \BibitemShut {NoStop}%
\bibitem [{\citenamefont {Valenti}\ \emph {et~al.}(2019)\citenamefont
  {Valenti}, \citenamefont {Henriques}, \citenamefont {Catelani}, \citenamefont
  {Maleeva}, \citenamefont {Gr\"unhaupt}, \citenamefont {von L\"upke},
  \citenamefont {Skacel}, \citenamefont {Winkel}, \citenamefont {Bilmes},
  \citenamefont {Ustinov}, \citenamefont {Goupy}, \citenamefont {Calvo},
  \citenamefont {Beno\^{\i}t}, \citenamefont {Levy-Bertrand}, \citenamefont
  {Monfardini},\ and\ \citenamefont {Pop}}]{PhysRevApplied.11.054087}%
  \BibitemOpen
  \bibfield  {author} {\bibinfo {author} {\bibfnamefont {F.}~\bibnamefont
  {Valenti}}, \bibinfo {author} {\bibfnamefont {F.}~\bibnamefont {Henriques}},
  \bibinfo {author} {\bibfnamefont {G.}~\bibnamefont {Catelani}}, \bibinfo
  {author} {\bibfnamefont {N.}~\bibnamefont {Maleeva}}, \bibinfo {author}
  {\bibfnamefont {L.}~\bibnamefont {Gr\"unhaupt}}, \bibinfo {author}
  {\bibfnamefont {U.}~\bibnamefont {von L\"upke}}, \bibinfo {author}
  {\bibfnamefont {S.~T.}\ \bibnamefont {Skacel}}, \bibinfo {author}
  {\bibfnamefont {P.}~\bibnamefont {Winkel}}, \bibinfo {author} {\bibfnamefont
  {A.}~\bibnamefont {Bilmes}}, \bibinfo {author} {\bibfnamefont {A.~V.}\
  \bibnamefont {Ustinov}}, \bibinfo {author} {\bibfnamefont {J.}~\bibnamefont
  {Goupy}}, \bibinfo {author} {\bibfnamefont {M.}~\bibnamefont {Calvo}},
  \bibinfo {author} {\bibfnamefont {A.}~\bibnamefont {Beno\^{\i}t}}, \bibinfo
  {author} {\bibfnamefont {F.}~\bibnamefont {Levy-Bertrand}}, \bibinfo {author}
  {\bibfnamefont {A.}~\bibnamefont {Monfardini}},\ and\ \bibinfo {author}
  {\bibfnamefont {I.~M.}\ \bibnamefont {Pop}},\ }\bibfield  {title} {\bibinfo
  {title} {Interplay between kinetic inductance, nonlinearity, and
  quasiparticle dynamics in granular aluminum microwave kinetic inductance
  detectors},\ }\href {https://doi.org/10.1103/PhysRevApplied.11.054087}
  {\bibfield  {journal} {\bibinfo  {journal} {Phys. Rev. Applied}\ }\textbf
  {\bibinfo {volume} {11}},\ \bibinfo {pages} {054087} (\bibinfo {year}
  {2019})}\BibitemShut {NoStop}%
\bibitem [{\citenamefont {Barends}\ \emph {et~al.}(2011)\citenamefont
  {Barends}, \citenamefont {Wenner}, \citenamefont {Lenander}, \citenamefont
  {Chen}, \citenamefont {Bialczak}, \citenamefont {Kelly}, \citenamefont
  {Lucero}, \citenamefont {O'Malley}, \citenamefont {Mariantoni}, \citenamefont
  {Sank}, \citenamefont {Wang}, \citenamefont {White}, \citenamefont {Yin},
  \citenamefont {Zhao}, \citenamefont {Cleland}, \citenamefont {Martinis},\
  and\ \citenamefont {Baselmans}}]{doi:10.1063/1.3638063}%
  \BibitemOpen
  \bibfield  {author} {\bibinfo {author} {\bibfnamefont {R.}~\bibnamefont
  {Barends}}, \bibinfo {author} {\bibfnamefont {J.}~\bibnamefont {Wenner}},
  \bibinfo {author} {\bibfnamefont {M.}~\bibnamefont {Lenander}}, \bibinfo
  {author} {\bibfnamefont {Y.}~\bibnamefont {Chen}}, \bibinfo {author}
  {\bibfnamefont {R.~C.}\ \bibnamefont {Bialczak}}, \bibinfo {author}
  {\bibfnamefont {J.}~\bibnamefont {Kelly}}, \bibinfo {author} {\bibfnamefont
  {E.}~\bibnamefont {Lucero}}, \bibinfo {author} {\bibfnamefont
  {P.}~\bibnamefont {O'Malley}}, \bibinfo {author} {\bibfnamefont
  {M.}~\bibnamefont {Mariantoni}}, \bibinfo {author} {\bibfnamefont
  {D.}~\bibnamefont {Sank}}, \bibinfo {author} {\bibfnamefont {H.}~\bibnamefont
  {Wang}}, \bibinfo {author} {\bibfnamefont {T.~C.}\ \bibnamefont {White}},
  \bibinfo {author} {\bibfnamefont {Y.}~\bibnamefont {Yin}}, \bibinfo {author}
  {\bibfnamefont {J.}~\bibnamefont {Zhao}}, \bibinfo {author} {\bibfnamefont
  {A.~N.}\ \bibnamefont {Cleland}}, \bibinfo {author} {\bibfnamefont {J.~M.}\
  \bibnamefont {Martinis}},\ and\ \bibinfo {author} {\bibfnamefont {J.~J.~A.}\
  \bibnamefont {Baselmans}},\ }\bibfield  {title} {\bibinfo {title} {Minimizing
  quasiparticle generation from stray infrared light in superconducting quantum
  circuits},\ }\href {https://doi.org/10.1063/1.3638063} {\bibfield  {journal}
  {\bibinfo  {journal} {Applied Physics Letters}\ }\textbf {\bibinfo {volume}
  {99}},\ \bibinfo {pages} {113507} (\bibinfo {year} {2011})},\ \Eprint
  {https://arxiv.org/abs/https://doi.org/10.1063/1.3638063}
  {https://doi.org/10.1063/1.3638063} \BibitemShut {NoStop}%
\bibitem [{\citenamefont {Gao}\ \emph {et~al.}(2008)\citenamefont {Gao},
  \citenamefont {Zmuidzinas}, \citenamefont {Vayonakis}, \citenamefont {Day},
  \citenamefont {Mazin},\ and\ \citenamefont {Leduc}}]{Gao2008}%
  \BibitemOpen
  \bibfield  {author} {\bibinfo {author} {\bibfnamefont {J.}~\bibnamefont
  {Gao}}, \bibinfo {author} {\bibfnamefont {J.}~\bibnamefont {Zmuidzinas}},
  \bibinfo {author} {\bibfnamefont {A.}~\bibnamefont {Vayonakis}}, \bibinfo
  {author} {\bibfnamefont {P.}~\bibnamefont {Day}}, \bibinfo {author}
  {\bibfnamefont {B.}~\bibnamefont {Mazin}},\ and\ \bibinfo {author}
  {\bibfnamefont {H.}~\bibnamefont {Leduc}},\ }\bibfield  {title} {\bibinfo
  {title} {Equivalence of the effects on the complex conductivity of
  superconductor due to temperature change and external pair breaking},\ }\href
  {https://doi.org/10.1007/s10909-007-9688-z} {\bibfield  {journal} {\bibinfo
  {journal} {Journal of Low Temperature Physics}\ }\textbf {\bibinfo {volume}
  {151}},\ \bibinfo {pages} {557} (\bibinfo {year} {2008})}\BibitemShut
  {NoStop}%
\bibitem [{\citenamefont {Pracht}\ \emph {et~al.}(2016)\citenamefont {Pracht},
  \citenamefont {Bachar}, \citenamefont {Benfatto}, \citenamefont {Deutscher},
  \citenamefont {Farber}, \citenamefont {Dressel},\ and\ \citenamefont
  {Scheffler}}]{PhysRevB.93.100503}%
  \BibitemOpen
  \bibfield  {author} {\bibinfo {author} {\bibfnamefont {U.~S.}\ \bibnamefont
  {Pracht}}, \bibinfo {author} {\bibfnamefont {N.}~\bibnamefont {Bachar}},
  \bibinfo {author} {\bibfnamefont {L.}~\bibnamefont {Benfatto}}, \bibinfo
  {author} {\bibfnamefont {G.}~\bibnamefont {Deutscher}}, \bibinfo {author}
  {\bibfnamefont {E.}~\bibnamefont {Farber}}, \bibinfo {author} {\bibfnamefont
  {M.}~\bibnamefont {Dressel}},\ and\ \bibinfo {author} {\bibfnamefont
  {M.}~\bibnamefont {Scheffler}},\ }\bibfield  {title} {\bibinfo {title}
  {Enhanced cooper pairing versus suppressed phase coherence shaping the
  superconducting dome in coupled aluminum nanograins},\ }\href
  {https://doi.org/10.1103/PhysRevB.93.100503} {\bibfield  {journal} {\bibinfo
  {journal} {Phys. Rev. B}\ }\textbf {\bibinfo {volume} {93}},\ \bibinfo
  {pages} {100503} (\bibinfo {year} {2016})}\BibitemShut {NoStop}%
\bibitem [{\citenamefont {Dieterle}\ \emph {et~al.}(2016)\citenamefont
  {Dieterle}, \citenamefont {Kalaee}, \citenamefont {Fink},\ and\ \citenamefont
  {Painter}}]{PhysRevApplied.6.014013}%
  \BibitemOpen
  \bibfield  {author} {\bibinfo {author} {\bibfnamefont {P.~B.}\ \bibnamefont
  {Dieterle}}, \bibinfo {author} {\bibfnamefont {M.}~\bibnamefont {Kalaee}},
  \bibinfo {author} {\bibfnamefont {J.~M.}\ \bibnamefont {Fink}},\ and\
  \bibinfo {author} {\bibfnamefont {O.}~\bibnamefont {Painter}},\ }\bibfield
  {title} {\bibinfo {title} {Superconducting cavity electromechanics on a
  silicon-on-insulator platform},\ }\href
  {https://doi.org/10.1103/PhysRevApplied.6.014013} {\bibfield  {journal}
  {\bibinfo  {journal} {Phys. Rev. Applied}\ }\textbf {\bibinfo {volume} {6}},\
  \bibinfo {pages} {014013} (\bibinfo {year} {2016})}\BibitemShut {NoStop}%
\bibitem [{\citenamefont {Melville}\ \emph {et~al.}(2020)\citenamefont
  {Melville}, \citenamefont {Calusine}, \citenamefont {Woods}, \citenamefont
  {Serniak}, \citenamefont {Golden}, \citenamefont {Niedzielski}, \citenamefont
  {Kim}, \citenamefont {Sevi}, \citenamefont {Yoder}, \citenamefont {Dauler},\
  and\ \citenamefont {Oliver}}]{Melville2020}%
  \BibitemOpen
  \bibfield  {author} {\bibinfo {author} {\bibfnamefont {A.}~\bibnamefont
  {Melville}}, \bibinfo {author} {\bibfnamefont {G.}~\bibnamefont {Calusine}},
  \bibinfo {author} {\bibfnamefont {W.}~\bibnamefont {Woods}}, \bibinfo
  {author} {\bibfnamefont {K.}~\bibnamefont {Serniak}}, \bibinfo {author}
  {\bibfnamefont {E.}~\bibnamefont {Golden}}, \bibinfo {author} {\bibfnamefont
  {B.~M.}\ \bibnamefont {Niedzielski}}, \bibinfo {author} {\bibfnamefont
  {D.~K.}\ \bibnamefont {Kim}}, \bibinfo {author} {\bibfnamefont
  {A.}~\bibnamefont {Sevi}}, \bibinfo {author} {\bibfnamefont {J.~L.}\
  \bibnamefont {Yoder}}, \bibinfo {author} {\bibfnamefont {E.~A.}\ \bibnamefont
  {Dauler}},\ and\ \bibinfo {author} {\bibfnamefont {W.~D.}\ \bibnamefont
  {Oliver}},\ }\bibfield  {title} {\bibinfo {title} {Comparison of dielectric
  loss in titanium nitride and aluminum superconducting resonators},\ }\href
  {https://doi.org/10.1063/5.0021950} {\bibfield  {journal} {\bibinfo
  {journal} {Applied Physics Letters}\ }\textbf {\bibinfo {volume} {117}},\
  \bibinfo {pages} {124004} (\bibinfo {year} {2020})},\ \Eprint
  {https://arxiv.org/abs/https://doi.org/10.1063/5.0021950}
  {https://doi.org/10.1063/5.0021950} \BibitemShut {NoStop}%
\bibitem [{\citenamefont {Tinkham}(2004)}]{tinkham}%
  \BibitemOpen
  \bibfield  {author} {\bibinfo {author} {\bibfnamefont {M.}~\bibnamefont
  {Tinkham}},\ }\href@noop {} {\emph {\bibinfo {title} {Introduction to
  Superconductivity}}},\ \bibinfo {edition} {2nd}\ ed.\ (\bibinfo  {publisher}
  {Courier Corporation},\ \bibinfo {year} {2004})\BibitemShut {NoStop}%
\bibitem [{\citenamefont {Pita-Vidal}\ \emph {et~al.}(2020)\citenamefont
  {Pita-Vidal}, \citenamefont {Bargerbos}, \citenamefont {Yang}, \citenamefont
  {van Woerkom}, \citenamefont {Pfaff}, \citenamefont {Haider}, \citenamefont
  {Krogstrup}, \citenamefont {Kouwenhoven}, \citenamefont {de~Lange},\ and\
  \citenamefont {Kou}}]{PhysRevApplied.14.064038}%
  \BibitemOpen
  \bibfield  {author} {\bibinfo {author} {\bibfnamefont {M.}~\bibnamefont
  {Pita-Vidal}}, \bibinfo {author} {\bibfnamefont {A.}~\bibnamefont
  {Bargerbos}}, \bibinfo {author} {\bibfnamefont {C.-K.}\ \bibnamefont {Yang}},
  \bibinfo {author} {\bibfnamefont {D.~J.}\ \bibnamefont {van Woerkom}},
  \bibinfo {author} {\bibfnamefont {W.}~\bibnamefont {Pfaff}}, \bibinfo
  {author} {\bibfnamefont {N.}~\bibnamefont {Haider}}, \bibinfo {author}
  {\bibfnamefont {P.}~\bibnamefont {Krogstrup}}, \bibinfo {author}
  {\bibfnamefont {L.~P.}\ \bibnamefont {Kouwenhoven}}, \bibinfo {author}
  {\bibfnamefont {G.}~\bibnamefont {de~Lange}},\ and\ \bibinfo {author}
  {\bibfnamefont {A.}~\bibnamefont {Kou}},\ }\bibfield  {title} {\bibinfo
  {title} {Gate-tunable field-compatible fluxonium},\ }\href
  {https://doi.org/10.1103/PhysRevApplied.14.064038} {\bibfield  {journal}
  {\bibinfo  {journal} {Phys. Rev. Applied}\ }\textbf {\bibinfo {volume}
  {14}},\ \bibinfo {pages} {064038} (\bibinfo {year} {2020})}\BibitemShut
  {NoStop}%
\bibitem [{\citenamefont {Brecht}\ \emph {et~al.}(2015)\citenamefont {Brecht},
  \citenamefont {Reagor}, \citenamefont {Chu}, \citenamefont {Pfaff},
  \citenamefont {Wang}, \citenamefont {Frunzio}, \citenamefont {Devoret},\ and\
  \citenamefont {Schoelkopf}}]{doi:10.1063/1.4935541}%
  \BibitemOpen
  \bibfield  {author} {\bibinfo {author} {\bibfnamefont {T.}~\bibnamefont
  {Brecht}}, \bibinfo {author} {\bibfnamefont {M.}~\bibnamefont {Reagor}},
  \bibinfo {author} {\bibfnamefont {Y.}~\bibnamefont {Chu}}, \bibinfo {author}
  {\bibfnamefont {W.}~\bibnamefont {Pfaff}}, \bibinfo {author} {\bibfnamefont
  {C.}~\bibnamefont {Wang}}, \bibinfo {author} {\bibfnamefont {L.}~\bibnamefont
  {Frunzio}}, \bibinfo {author} {\bibfnamefont {M.~H.}\ \bibnamefont
  {Devoret}},\ and\ \bibinfo {author} {\bibfnamefont {R.~J.}\ \bibnamefont
  {Schoelkopf}},\ }\bibfield  {title} {\bibinfo {title} {Demonstration of
  superconducting micromachined cavities},\ }\href
  {https://doi.org/10.1063/1.4935541} {\bibfield  {journal} {\bibinfo
  {journal} {Applied Physics Letters}\ }\textbf {\bibinfo {volume} {107}},\
  \bibinfo {pages} {192603} (\bibinfo {year} {2015})},\ \Eprint
  {https://arxiv.org/abs/https://doi.org/10.1063/1.4935541}
  {https://doi.org/10.1063/1.4935541} \BibitemShut {NoStop}%
\end{thebibliography}%
%\end{thebibliography}%

\end{document}